\documentclass[fleqn,usenatbib]{mnras}

\usepackage{newtxtext,newtxmath}
\usepackage[T1]{fontenc}
\usepackage{ae,aecompl}

\usepackage{graphicx}	% Including figure files
\usepackage{amsmath}	% Advanced maths commands
\usepackage{amssymb}	% Extra maths symbols

\usepackage{gensymb}
\usepackage{subfigure}
\usepackage{bm}
\usepackage{soul}

\providecommand{\sorthelp}[1]{}

\defcitealias{2014PhRvL.113s1303K}{KK14}
\defcitealias{2017A&A...603A..62V}{V17}
\defcitealias{planck2014-a14}{FFP8}
\renewcommand{\vec}[1]{\bm{#1}}

\title[Polarised Dust Anisotropy]{Detection and Removal of B-mode Dust Foregrounds with Signatures of Statistical Anisotropy}

\author[O. H. E. Philcox et al.]{
Oliver H. E. Philcox,$^{1,3}$\thanks{E-mail: ohep2@alumni.cam.ac.uk}
Blake D. Sherwin,$^{2,3}$
Alexander van Engelen$^{4}$
\\
$^{1}$Institute of Astronomy, Madingley Road, Cambridge, CB3 0HA, UK\\
$^{2}$Department of Applied Mathematics and Theoretical Physics, Cambridge, CB3 0WA, UK\\
$^{3}$Kavli Institute for Cosmology Cambridge, Cambridge, CB3 0HA, UK\\
$^{4}$Canadian Institute for Theoretical Astrophysics, 60 St.~George Street, Toronto,  M5S 3H8, Canada
}

\date{Accepted XXX. Received YYY; in original form ZZZ}

\pubyear{2018}

\begin{document}
\label{firstpage}
\pagerange{\pageref{firstpage}--\pageref{lastpage}}
\maketitle

\begin{abstract}
Searches for inflationary gravitational wave signals in the CMB B-mode polarisation are expected to reach unprecedented power over the next decade. A major difficulty in these ongoing searches is that galactic foregrounds such as dust can easily mimic inflationary signals. Though typically foregrounds are separated from primordial signals using the foregrounds' different frequency dependence, in this paper we investigate instead the extent to which the galactic dust B-modes' \emph{statistical anisotropy} can be used to distinguish them from inflationary B-modes, building on the work of Kamionkowski and Kovetz (2014). In our work, we extend existing anisotropy estimators and apply them to simulations of polarised dust to forecast their performance for future experiments. Considering the application of this method as a null-test for dust contamination to CMB-S4, we find that we can detect residual dust levels corresponding to $r\sim0.001$ at $2\sigma$, which implies that statistical anisotropy estimators will be a powerful diagnostic for foreground residuals (though our results show some dependence on the dust simulation used). Finally, considering applications beyond a simple null test, we demonstrate how anisotropy statistics can be used to construct an estimate of the dust B-mode map, which could potentially be used to clean the B-mode sky.
\end{abstract}


%%%%%%%%%%%%%%%%%%%%%%%%%%%%%%%%%%%%%%%%%%%%%%%%%%

%%%%%%%%%%%%%%%%% BODY OF PAPER %%%%%%%%%%%%%%%%%%

\section{Introduction}
Though the cosmic microwave background (CMB) has already provided us with remarkable insights into the properties of our universe, a wealth of cosmological information still remains to be extracted from the polarisation of the CMB. In particular, the CMB B-mode polarisation signal contains much information about weak lensing (e.g.~\citealt{2003PhRvD..68h3002H}) and inflationary gravitational waves (IGWs) (e.g.~\citealt{1997PhRvD..55.7368K}). As direct predictions of many of the simplest models of inflation, B-mode signals induced by IGWs allow physics to be probed at the highest energy scales and are thus of great interest to the fundamental physics community.

The amplitude of any IGW B-modes is expected to be small given current limits, with $r<0.07$ at 95\% confidence \citep{2016PhRvL.116c1302B}. Therefore, precise techniques are likely required to distinguish between these primordial IGW B-modes and galactic foreground B-modes, which are typically much larger.  At large angular scale, the main source of polarised CMB foregrounds is diffuse synchotron emission (at a CMB frequency $\nu\lesssim 100$\,GHz) and thermal dust emission (at higher $\nu$) \citep{planck2014-a12,2016ARA&A..54..227K}. The latter arises from the infrared emission of Galactic dust grains (including silicates, polycyclic aromatic hydrocarbons and carbonaceous grains) which are heated by optical photons \citep{2011piim.book.....D,2017MNRAS.469.2821T}. This emission is polarised due to the preferential alignment of spinning aspherical grains' rotation axes with the Galactic magnetic field (GMF), giving contributions to both CMB E- and B-modes \citep{2009ApJ...696....1D}.  %This picture has recently been supported by magneto-hydrodynamical simulations \citep{2017arXiv171111108K} and with the observation that the neutral interstellar medium, which is coupled to galactic magnetic fields, is correlated with the polarised emission \citep{2015PhRvL.115x1302C}.

The canonical method for distinguishing dust from the underlying CMB signal makes use of the fact that the frequency dependence of the two components differs (e.g.~\citealt{planck2014-XXII}); making use of data from different frequency bands (e.g.~\citealt{planck2016-XLIV}), IGW constraints can thus be obtained despite foreground contamination. However, foreground separation methods can suffer from a number of complications, including incomplete knowledge of the dust frequency dependence in parametric methods, for instance due to multiple components at different temperatures (cf.~\citealt{2017MNRAS.469.2821T}) or decorrelation, potentially arising from different dust structures being probed at different frequencies. In addition, the application of multifrequency foreground separation is more difficult for ground-based CMB experiments, where only a restricted frequency range is available due to atmospheric effects. 

Here we consider the approach proposed by \cite{2014PhRvL.113s1303K} (hereafter \citetalias{2014PhRvL.113s1303K}): to search for foreground-specific signatures of statistical anisotropy in the two-dimensional power-spectra of the B-mode signal, resulting from a roughly coherent Galactic magnetic field (GMF) on few-degree scales, as supported by observations \citep[e.g.\,][]{2015PhRvL.115x1302C,2015ApJ...809..153M,2016ApJ...821..117K}. In this paper, we focus on the use of these statistical anisotropy estimators as a null test to detect dust residuals in a poorly-cleaned CMB map, but also briefly discuss their use in mapping foregrounds and as a tool for dust-removal (`dedusting'). We test the performance of these methods with reference to current and future CMB experiments, using simulated polarised thermal dust emission maps from \citet[hereafter \citetalias{2017A&A...603A..62V}]{2017A&A...603A..62V}, as well as realistic noise and lensing implementations for futuristic CMB experiments such as CMB-S4 \citep{2016arXiv161002743A}. (In some cases, we will also compare with thermal dust simulations from \cite{planck2014-a14}, hereafter \citetalias{planck2014-a14}.) A Python implementation of the code used in this analysis is freely available online.\footnote{A Hexadecapolar Analysis for Dust Estimation in Simulations: \url{http://github.com/oliverphilcox/HADES}}

Although there is a range of previous work regarding the search for non-Gaussianity and statistical anisotropy in dust emission (stretching back to \citealt{2002ApJ...566...19K}), less research has been done on the applications of such a detection of anisotropy; however, the work of \cite{2016JCAP...09..034R} provides one such example. In this work, a bipolar-spherical harmonic basis is used to construct an anisotropy test for Planck 353\,GHz maps, which is then utilised to assess the cleanest regions of the foreground sky.  Here, we use high-resolution simulated data instead of Planck maps to emulate future experiments and focus on the use of anisotropy detection in the context of null-tests. In addition, much of the  literature is concerned with detecting generalised anisotropy, whereas in this paper we consider only a very specific anisotropy signature typical of dust (or synchrotron emission) in a coherent GMF.

The paper is organised as follows. Sec.~\ref{Sec:Maths} describes the mathematical background to the work, including a description of the \citetalias{2014PhRvL.113s1303K} estimators, before the simulations are discussed in Sec.~\ref{Sec:Simulations} and the full methodology is presented in Sec.~\ref{Sec:Methods}. Sec.~\ref{Sec:NullTests} details the application to null tests, and further results are presented in Sec.~\ref{Sec:Results} through \ref{Sec:Dedusting} (discussing spatial distributions, correlations and `dedusting' respectively). Conclusions and a summary are then made in Sec.~\ref{Sec:Conclusion}.

\section{Dust Estimators}\label{Sec:Maths}
First, we describe the B-mode dust model used in this study. Following \citetalias{2014PhRvL.113s1303K}, we assume that, across a sufficiently small section (`tile') of the CMB sky, the GMF should be roughly coherent, leading to the dust having a uniform polarisation direction. This gives rise to a particular \textit{hexadecapole} anisotropy pattern in the two-dimensional B-mode power spectrum of that tile, $|B(\vec{l})|^2$ (for Fourier mode $B(\vec{l})$). For a spectral pixel at position vector $\vec{l}$ and angle $\phi_{\vec{l}}$ from the positive RA axis, the power is expected to take the form;
\begin{eqnarray}\label{Eq:PowModel}
\left|B(\vec{l})\right|^2 = AC_l^\mathrm{fid} [1 - f_c\cos{4\phi_{\vec{l}}} - f_s\sin{4\phi_{\vec{l}}}],
\end{eqnarray}
where the fiducial isotropic $l$ ($=|\vec{l}|$) dependence is given by \cite{planck2014-XXX} as $C_l^\mathrm{fid} = l^{-\gamma}$ with $\gamma = 2.42 \pm 0.02$, obtained by averaging over regions away from the galactic plane with no evidence for departure from this behaviour over 1\% sky regions \citep{pb2015}. 
%Blake comment: Since we are primarily interested in dust in this work, we do not apply the approach of
%\cite{2016PhRvL.116c1302B}, which marginalised over $\gamma \in [2,3]$.} 
The key model parameters in Eq.~\ref{Eq:PowModel} are the monopole dust amplitude $A$ and sine (cosine) hexadecapole coefficients $Af_s$ ($Af_c$) which define the model anisotropy. $f_s$ and $f_c$ denote the fractional hexadecapole power, but will not be directly used in this analysis. From \citetalias{2014PhRvL.113s1303K}, we note that $Af_s$ and $Af_c$ are proportional to $\sin{4\alpha}$ and $\cos{4\alpha}$ respectively, where $\alpha$ is the dust polarisation angle defined by 
\begin{eqnarray}\label{Eq:AlphaQUDefn}
2\alpha=\arctan{\frac{U}{Q}}
\end{eqnarray}
in the IAU polarisation convention.\footnote{We here adopt the IAU polarisation convention used by \texttt{flipper}, in contrast to the COSMO convention of Planck and \texttt{HEALPix}; this corresponds to a sign inversion of the U-map.} This can be found from the measured hexadecapole parameters via
\begin{eqnarray}\label{Eq: AnisotropyAngle}
\tan{4\alpha} = \frac{Af_s}{Af_c}.
\end{eqnarray} 
In this analysis $Af_s$ and $Af_c$ are used rather than the fractional quantities ($f_s$ and $f_c$) due to their higher signal-to-noise ratio (since $f_s$ ($f_c$) is influenced by errors in both $Af_s$ ($Af_c$) and $A$). A generalisation of this to a continuous description that allows for variations in the GMF direction is also presented in \citetalias{2014PhRvL.113s1303K} but we adopt the simpler procedure in this paper.

$A$, $Af_c$ and $Af_s$ may be computed for a specific tile using the following estimators (cf.\,\citetalias{2014PhRvL.113s1303K}, Eq.\,9-10):
\begin{eqnarray}\label{Eq:estimators}
\widehat{A}&=&\frac{\sum_{\vec{l}}\mathcal{B}^2(\vec{l})\Lambda_l^2/{C_l^\mathrm{fid}}}{\sum_{\vec{l}}\Lambda^2_l}\nonumber\\
\nonumber\\
\widehat{Af_c}&=&\frac{\sum_{\vec{l}}\mathcal{B}^2(\vec{l})\Lambda^2_l\cos{4\phi_{\vec{l}}}/{C_l^\mathrm{fid}}}{\sum_{\vec{l}}\left(\Lambda_l\cos{4\phi_{\vec{l}}}\right)^2}\nonumber\\
\nonumber\\
\widehat{Af_s}&=&\frac{\sum_{\vec{l}}\mathcal{B}^2(\vec{l})\Lambda^2_l\sin{4\phi_{\vec{l}}}/{C_l^\mathrm{fid}}}{\sum_{\vec{l}}\left(\Lambda_l\sin{4\phi_{\vec{l}}}\right)^2},
\end{eqnarray}
summing over all pixels at positions $\vec{l}$. $\mathcal{B}^2$ is the expected B-mode power due to dust alone;
\begin{eqnarray}\label{Eq: BmodeDustPower}
\mathcal{B}^2(\vec{l})=|B(\vec{l})|^2-(C_l^\mathrm{lens}+C_l^\mathrm{noise})
\end{eqnarray}
for total 2D measured power $|B(\vec{l})|^2$, accounting for (isotropic) contamination from lensing and noise modes with known profile. This differs from that of \citetalias{2014PhRvL.113s1303K} (who just use $|B(\vec{l})|^2$ directly in the estimators), and is chosen to reduce bias in the low-dust limit. (We note that this is not used for cross-spectra between dust and isotropic Monte Carlo simulations where $\langle{|B(\vec{l})|^2\rangle}=0$ is expected.) 
In Eqs.~\ref{Eq:estimators}, $\Lambda_l$ is the signal-to-noise ratio for each pixel in the 2D power spectrum, chosen here as 
\begin{eqnarray}\label{Eq:SNR}
\Lambda_l = \frac{AC_l^\mathrm{fid}}{AC_l^\mathrm{fid}+C_l^\mathrm{noise}+C_l^\mathrm{lens}},
\end{eqnarray}
\citepalias[cf.][Eq.\,12]{2014PhRvL.113s1303K}, which reduces to $\Lambda_l=1$ in the noiseless limit, affording equal weight to all pixels. The dust-contributions to $C_l$ are included in the denominator to account for errors in the parametrisation. We note that $\Lambda_l$ depends on the monopole amplitude $A$, hence $\widehat{A}$ is computed iteratively, using a first estimate of $A_0 = 1\,\mu\mathrm{K}^2$ and stopping when the estimates for $A$ converge to within 1\% (this generally occurs within a few iterations, even for a far-off initial guess of $A_0=1\mathrm{K}^2$). In addition, once convergence is achieved, the same value of $A$ is used in $\Lambda_l$ for all applications of the estimators to this specific tile.

We note that we have assumed all components aside from dust and lensing to have an isotropic power spectrum. This may not be true in real data, for example if noise is anisotropic due to the survey scan strategy, or if a multifrequency component separation algorithm has been applied that does not preserve isotropy (such as a Needlet Internal Linear Combination algorithm). While the noise problem can easily be solved by relying on cross-spectra, anisotropic effects of component separation would need to be treated by subtracting off a simulated bias, or by a modification of the cleaning procedure. We defer such considerations to future work.

In addition to $A$, $Af_s$ and $Af_c$ we introduce the polarisation angle-independent \textit{hexadecapole amplitude}, $H$, via
\begin{eqnarray}\label{Eq: HexPow}
H^2 = (Af_s)^2+(Af_c)^2.
\end{eqnarray}
This quantity is motivated by the fact that, if we assume no prior knowledge of the dust polarisation direction, $Af_s$ and $Af_c$ are equally likely to be positive or negative; a quadratic statistic is thus needed. To compute the variance of such a measurement of $H^2$, under the null hypothesis of no anisotropy being present, we use a large set of Monte Carlo (MC) simulations created from random Gaussian realisations of the isotropic power spectrum for each tile. A further treatment of this, and the biases to $H^2$ arising from the non-negative nature of the statistic, are discussed in Sec.~\ref{Subsec:MCerrors} and appendices \ref{Subsec:H2Bias} and \ref{Appen:Stats}. 

The debiased fractional anisotropy can also be defined as $\epsilon=\sqrt{f_s^2+f_c^2} = H/A$, although we note that this suffers from the same signal-to-noise limitations as $f_s$ and $f_c$. For convenience, all model parameters and quantities derived from the estimators are summarised in tables \ref{Tab:modelParams} and \ref{Tab:estParams} respectively.

\begin{table*}
\centering
\begin{tabular}{c|l|c}
Symbol & Definition & Fiducial Value\\
\hline
$\gamma$ & Fiducial isotropic map slope (Sec.~\ref{Sec:Maths}) & $2.42$\\
$\Lambda_l$ & Estimator signal-to-noise ratio (Sec.~\ref{Sec:Maths})& Eq.\,\ref{Eq:SNR}\\
$\Delta_\mathrm{P}$ & Noise power (Eq.\,\ref{Eq:noiseModel}) in $\mu$K-arcmin & see tab.\,\ref{Tab:noiseParams}\\
$\theta_\mathrm{FWHM}$ & Noise FWHM (Eq.\,\ref{Eq:noiseModel}) in arcmin & see tab.\,\ref{Tab:noiseParams}\\
$f_\mathrm{lens}$ & Fractional contribution of lensing to $C_l^{BB}$ (Sec.\,\ref{Subsec:LensingNoise}) & see tab.\,\ref{Tab:noiseParams}\\
$\nu$ & Simulated map frequency (Sec.\,\ref{Subsec:DustMaps})& $150\,\text{GHz} $\\
$\beta_\mathrm{d}$ & Dust emissivity spectral index (Sec.\,\ref{Subsec:DustMaps})& 1.53\\
$T_\mathrm{dust}$& Equilibrium dust grain temperature (Sec.\,\ref{Subsec:DustMaps})&19.6\,K\\
$\Delta\theta$ & Tile width in degrees (Sec.\,\ref{Subsec:Cutouts}) & 3\degree\\
$W(\vec{x})$ & Real-space window function & - \\
$R_\mathrm{pad}$ & Ratio of widths for zero padding (Sec.\,\ref{Subsec:PowerMaps})& 2\\
$l_\mathrm{min}$& Lowest $l$ used in statistical estimators (Eq.\,\ref{Eq:estimators})&$360/(R_\mathrm{pad}\Delta\theta{})$\\
$\Delta{l}$ & 1D spectrum binning width (Sec.\,\ref{Subsec:MCerrors}) & $240/R_\mathrm{pad}(\Delta\theta{})$\\
$N_\mathrm{sims}$& Number of Monte Carlo simulations (Sec.\,\ref{Subsec:MCerrors}) & 500\\
$N_\mathrm{bias}$& Number of bias simulations (Sec.\,\ref{Subsec:debiasing}) & 500\\
$f_\mathrm{dust}$& Real-space fractional dust contribution (Sec.\,\ref{Sec:NullTests}) & 1\\
$r_\mathrm{eff}$& Effective tensor ratio for mean patch dust level (Sec.\,\ref{Sec:NullTests}) & Eq.\,\ref{Eq:r_effDefn} 
\end{tabular}
\caption{Model variables used in this study with their default values. For full descriptions see the text.}\label{Tab:modelParams}
\end{table*}

\section{Simulations}\label{Sec:Simulations}
\subsection{Noise \& Lensing B-modes}\label{Subsec:LensingNoise}
To simulate realistic CMB experiments we must consider other (non-dust) sources of B-mode power, arising from noise and weak lensing of large scale structure. (At first, we will neglect contributions from IGWs, which are subdominant to lensing contributions over the $l$-ranges used here, given current constraints on the tensor-to-scalar ratio \citep{2016PhRvL.116c1302B}; however, we explore this further in Sec.~\ref{Subsec:TensorBias}.) 

For instrument noise, we take the commonly used parametrisation \citep[e.g.][]{1995PhRvD..52.4307K,1997ApJ...488....1Z,2002ApJ...574..566H}
\begin{eqnarray}\label{Eq:noiseModel}
C_l^\mathrm{noise} = \Delta_\mathrm{P}^2 \exp\left(\frac{l(l+1)\theta_\mathrm{FWHM}^2}{8\ln{2}}\right)
\end{eqnarray}
for noise power $\Delta_\mathrm{P}$ (in $\mu$K-arcmin) and beam full-width half-maximum (FWHM) $\theta_\mathrm{FWHM}$ (in radians, presented here in arcmin). The fiducial values we assume for each experiment are shown in table \ref{Tab:noiseParams}. This model assumes the noise to be isotropic, unlike the expected B-mode dust signal. 

To simulate maps in the analysis below, noise modes are added separately to each tile cut out from the dust simulations, using Gaussian realisations of the above power spectrum. A more physical approach would be to generate a \textit{full-sky} noise map which could be partitioned and added to the dust modes, taking into account large-scale correlations. However, as shown in appendix \ref{Appen:NoiseBias}, both methods give almost identical results, and hence the tile-based approach will be adopted here for computational efficiency.

\begin{table}
\centering
\begin{tabular}{l|c|c|c}
CMB Experiment & $\Delta_\mathrm{P}$ [$\mu{}$K$\arcmin$] & $\theta_\mathrm{FWHM}$ [\arcmin] & $f_\mathrm{lens}$\\
\hline
Zero-noise & 0 & 0 & 0\\
BICEP2 & 5 & 30 & 1\\
$5\mu{}$K$^{\prime}$-CMB & 5 & 1.8 & 0.4\\
CMB-S4 & 1 & 1.5 & 0.1\\
\end{tabular}
\caption{Assumed noise and lensing parameters (as defined in table \ref{Tab:modelParams}) for current and future CMB experiments. The $5\mu{}$K$^{\prime}$-CMB experiment is included to emulate upcoming experiments with noise parameters similar to that of the Simons Observatory (SO). BICEP2, SO and CMB-S4 parameters are from 
\citet{2014ApJ...792...62B,2016JLTP..184..805S,2016arXiv161002743A} respectively. We note that the $f_\mathrm{lens}$ values are rough approximations only.}\label{Tab:noiseParams}
\end{table}

In this study, we include lensing modes using a lensed CMB map from the Planck FFP10 simulation suite,\footnote{Available from NERSC: \url{https://nim.nersc.gov/}} created with a random realisation of the lensing potentials. Since there is considerably more power in lensed E-modes than B-modes, when we compute B-modes from small cut-outs of the Stokes Q and U maps, there can be significant E-to-B mode leakage. To account for this, the scalar E-modes were removed from the lensed full-sky spherical-harmonic data before the Stokes maps were computed, thus nulling the additional leakage power in our computed B-modes. Although this is unphysical, it can be justified by noting that we are here forecasting experimental performance, and a maximum likelihood approach or better E-B separation method \citep[e.g.\,][]{2009PhRvD..79l3515G} would be able to minimise E-B leakage to a significant extent. (Additionally, a continuous implementation of our method \citepalias[c.f.~][]{2014PhRvL.113s1303K} would obviate the need for E-B decomposition on small tiles.)  

Each individual lensing event is expected to have a small quadrupolar B-mode anisotropy signature \citep[e.g.][]{2016ARA&A..54..227K}; however, the power derives from a sum over many lensing modes, leading to the power spectrum becoming almost isotropic when averaged over the $3\degree$ regions used here. To approximate the process of `delensing' \citep[e.g.][]{2002PhRvL..89a1303K,2002PhRvL..89a1304K,2012JCAP...06..014S,2017ApJ...846...45M}, we rescale the lensing power-spectrum by a factor $f_\mathrm{lens}$, where $f_\mathrm{lens}=0.1$ would signify 90\% delensing for example. Fiducial values for this are given in table \ref{Tab:noiseParams}.

\begin{figure*}
\centering
\includegraphics[width=0.9\linewidth]{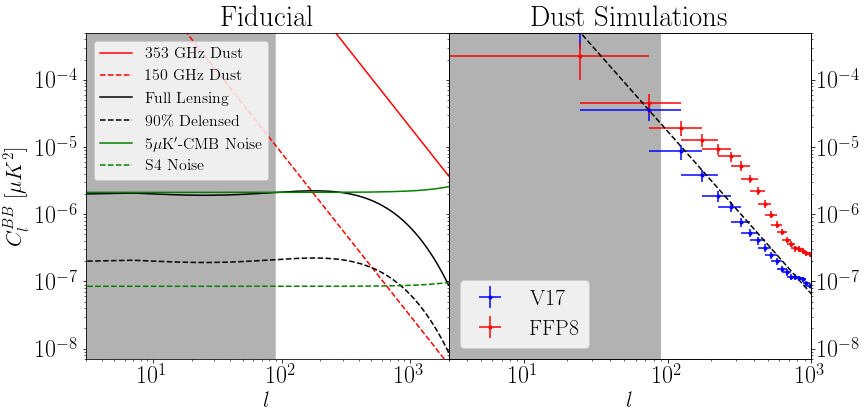}
\caption{Contributions to the B-mode power spectrum, $C_l^{BB}$. \textbf{Left:} A fiducial $l^{-2.42}$ dust power spectrum (red) at 353\,GHz (full lines) and 150\,GHz (dashed lines) with frequency scaling obtained via the Planck dust spectral energy distribution \citep{2012AdAst2012E..41T,2018arXiv180104945P}. The amplitude is taken from a representative tile in the 5\% sky patch (Sec.~\ref{Subsec:Cutouts}) using the \citet[\citetalias{2017A&A...603A..62V}]{2017A&A...603A..62V} simulation. Also shown is the lensing B-mode spectrum (black, taken from a full-sky Planck FFP10 lensed scalar simulation) assuming 0\% (full lines) and 90\% delensing (dashed lines) efficiency. Noise modes (green) are displayed using the fiducial noise model, (Eq.\,\ref{Eq:noiseModel}, e.g. \citealt{1995PhRvD..52.4307K}) with 5$\mu$K$^\prime$-CMB (full-lines) and CMB-S4 (dashed lines) experimental parameters (table \ref{Tab:noiseParams}). Regions with too low $l$ to be investigated via our methods (using $3\degree$ tile-widths) are shown in grey. \textbf{Right:} Binned B-mode power spectra of the \citetalias{2017A&A...603A..62V} and \citetalias{planck2014-a14} simulations extrapolated to 150~GHz, computed over the 5\% sky mask for $l\in[50,2000]$ using the \texttt{PolSpice} pseudo-$C_l$ estimator \citep{2004MNRAS.350..914C}, with a maximum angle of $50\degree$ and binned with $\Delta{}l=50$. No mask-deconvolution is performed here, which is not expected to have a large impact due to the wide mask used. A dashed line indicates the fiducial $l$-dependence (here calibrated to the \citetalias{2017A&A...603A..62V} simulations at $l=80$).}\label{Fig:ClBBContributions}
\end{figure*}

Fig.~\ref{Fig:ClBBContributions} (left panel) shows the contributions to the isotropic $C_l^{BB}$ power spectrum from the relevant components using two sets of noise parameters and delensing efficiencies. It is apparent, for the plotted dust-level, that a significant detection of anisotropy at $\nu = 150$\,GHz in the $l>90$ region available to this test (Sec.~\ref{Subsec:PowerMaps}) will require low noise and, ideally, a substantially delensed map.

\subsection{Dust Maps}\label{Subsec:DustMaps}
The primary thermal-dust simulation used in this paper is from \citet[][\citetalias{2017A&A...603A..62V}]{2017A&A...603A..62V}. However, we also compare it to the simulation of \citet[][\citetalias{planck2014-a14}]{planck2014-a14}. Both simulations cover the full-sky and make use of Planck $\mathrm{D}_{353}$ dust intensity maps at 353 GHz (where the Stokes maps are dominated by thermally emitting dust). Each is given to a HEALPix\footnote{\url{http://healpix.sf.net}} \citep{2005ApJ...622..759G} resolution  of \texttt{NSIDE}~$=2048$; we use data up to $l = 2000$ in our analysis. B-mode power spectra for the two simulations are shown in Fig.~\ref{Fig:ClBBContributions} (right panel), computed for the 5\% sky region used via the pseudo-$C_l$ estimator \texttt{PolSpice}\footnote{\url{http://www2.iap.fr/users/hivon/software/PolSpice/}} \citep{2004MNRAS.350..914C} which takes into account the effects of masking, using a apodisation width and maximum angle of $50\degree$ (matching the patch-width).

We will briefly review the construction of these two simulations. The \citetalias{2017A&A...603A..62V} simulations proceed by assuming that the GMF may be approximated by a combination of uniform and turbulent components. From Gaussian realisations of the turbulent GMF, a set of polarisation angles are computed that give a theoretical model for the Q/I and U/I ratio maps, which are then combined with the (Planck PR2) $\mathrm{D}_{353}$ map to estimate Q and U. Hyperparameters of the method are optimised using Planck data on low-dust regions, which ensures that observational trends such as the ratio of E- to B-mode power are well represented (e.g. \citealt{planck2014-XXX,planck2016-XLIV}). The approximation of constant background GMF becomes invalid too close to the Galactic centre; therefore, when considering full-sky scenarios, we mask the central galactic regions with the Planck GAL80 mask.\footnote{\url{http://pla.esac.esa.int/pla/}} From Fig.~\ref{Fig:ClBBContributions} (right panel), we note that the $l$ dependence of this simulation is well approximated by the fiducial $l^{-2.42}$ model. 

In contrast, the \citetalias{planck2014-a14} simulations are computed from observational (component-separated) Planck PR1 Q/I and U/I maps smoothed to a resolution of 0.5\degree, with small-scale data added by analytically continuing the $C_l$ power spectra. This is then multiplied by full-resolution Planck $\mathrm{D}_{353}$ maps to compute Q and U. From Fig.~\ref{Fig:ClBBContributions}, it is clear that the small-scale (high-$l$) power differs significantly between the simulations. Because of the greater physical motivation behind the construction of the \citetalias{2017A&A...603A..62V} simulations (and better apparent fit to observational trends), we will centre our analyses around them.

Since a multifrequency analysis is beyond the scope of this work, we generate simulations at  150~GHz, close to where the CMB is brightest, by rescaling the dust simulations, which are provided at 353~GHz. This is achieved by treating the polarised dust frequency spectrum as a modified blackbody \citep{2012AdAst2012E..41T,2017MNRAS.469.2821T} with temperature and spectral indices ($T_\mathrm{dust}$ \& $\beta_\mathrm{d}$, table \ref{Tab:modelParams}) taken from \cite{planck2014-XIX,2018arXiv180104945P}. We additionally use the colour and unit conversions described in \cite{planck2013-p03d,2015PhRvD..91h1303K}, giving an overall reduction in $C_l^{BB}$ of $0.0416^2\approx 1/580$ \citep[cf.][]{planck2014-XXX,2016JCAP...09..034R}. 

This simple scaling may not be perfectly accurate, since we neglect the existence of multi-temperature dust components and the (potentially frequency-dependent) spatial variation in $T_\mathrm{dust}$ and $\beta_\mathrm{d}$ \citep{2017MNRAS.469.2821T}. Whilst these are an important consideration for multifrequency dust-cleaning techniques where high precision is needed, here we require only an estimate of the frequency dependence to generate simulations and hence forecast the applicability of our techniques to future experiments. 

The simple foreground model leading to our anisotropy estimators (Eqs.\,\ref{Eq:estimators}) assumes similar effects in both E- and B-modes and hence that the power spectra are equal; $C_l^{EE}=C_l^{BB}$.  This contradicts the measurements of \citet{planck2014-XXX}, which have $C_l^{EE}\approx2C_l^{BB}$. The difference between E- and B-mode power is believed to arise from coupling of the filamentary structure of dust with the magnetic field, affecting both polarisation direction and level of intensity fluctuations \citep{2015PhRvL.115x1302C,planck2015-XXXVIII,2017ApJ...839...91C}. This effect is included in \citetalias{2017A&A...603A..62V} via their anisotropic realisations of the turbulent filamentary GMF with hyperparameters tuned to match \textit{Planck} data, but not in the earlier \citetalias{planck2014-a14} simulations, since they pre-date the \textit{Planck} discovery of asymmetry.

In the context of this work, since our primary \citetalias{2017A&A...603A..62V} simulations include these E-mode and B-mode effects in a physically motivated manner, the results presented below already take the asymmetry into account, with differences resulting in only a potential loss of significance due to the estimator being somewhat sub-optimal. By relying on more physically correct models of the dust B-modes in future work, one could potentially develop an improved (yet similar) estimator that would be even more sensitive to dust contamination.

\section{Implementation}\label{Sec:Methods}
Below we describe the computation of the hexadecapole parameters (Sec.~\ref{Sec:Maths}) from the full-sky simulated Stokes maps. 

\subsection{Cutting out Tiles}\label{Subsec:Cutouts}
To match the simulated data to futuristic experiments, for the majority of this analysis we consider only parts of the entire sky, hereafter denoted `patches'. A simulation region of size 2150\,$\mathrm{deg}^2$ is used primarily, covering 5\% of the sky with Galactic co-ordinates [-110\degree, -20\degree] (RA), [-85\degree, -45\degree] (dec). This was chosen since it (a) encloses the BICEP region, and (b) is a region of minimal dust, as is desirable to search for IGWs. This mask was smoothed with a $2\degree$~FWHM Gaussian kernel in \texttt{HEALPix} to avoid edge effects, and is the default patch used in the analysis. For the Simons Observatory (SO), we note that the actual experimental region may be as large as $4000$\,$\mathrm{deg}^2$ \citep{2016JLTP..184..805S}, which would boost the significance of any detections of hexadecapolar anisotropy compared to that found by our mock $5\mu{}$K$^{\prime}$-CMB experiment.

The patch is then partitioned into tiles of width $\Delta\theta$ (where $\Delta\theta$ is small to ensure GMF coherency) by projecting each section of the relevant whole-sky (dust or lensing) Stokes map onto a flat grid in galactic co-ordinates. 
%Blake comment, with an angular resolution of 30$''$ (corresponding to $l_\mathrm{max} \approx 20000$, significantly above the maximum $l$ used here to avoid information loss). 
This partitioning is also applied to the mask map, which, when multiplied by a half-cosine apodisation window of width $\Delta\theta/10$ (to remove edge discontinuity effects), provides a window function for each tile. A discussion of the potential biases given by the projection onto the flat-sky is given in appendix \ref{Appen:ProjectionDistortions}. We note that there is some small information loss due to the apodisation, since the unapodised window functions do not overlap (to avoid double counting modes), thus we lose some data at the edges of each tile. This may slightly reduce the significances of anisotropy detection.

Here, square tiles of $\Delta\theta = 3\degree$ are used, giving a sky fraction of $0.02\%$ per tile.\footnote{Whilst larger tiles allow lower $l$ to be probed (where there are greater dust contributions), this reduces the coherency of the GMF necessary to observe the hexadecapole structure. $\Delta\theta=3\degree$ was found to give the best performance in initial studies.} For studies investigating the spatial distribution of the hexadecapole parameters $\alpha$ and $H^2$ across the patch (Sec.~\ref{Sec:Results}), the tiles are chosen to be overlapping, with centres separated by $0.5\degree$, to give increased resolution for display in figures. However, the overlapping tiling is not used to calculate any detection significances (Sec.~\ref{Subsec:PatchAnisotropy}).

\subsection{Mock Data and Power Maps}\label{Subsec:PowerMaps}
From the cut-out Q and U lensing and dust tiles, we compute two-dimensional Fourier-space B-mode maps using \texttt{flipper}\footnote{\url{https://github.com/sudeepdas/flipper}} \citep{2009PhRvD..79h3008D} and \texttt{flipperPol}\footnote{\url{https://github.com/amaurea/flipperpol}} \citep{2013MNRAS.435.2040L}, via the latter's `hybrid' method, which uses spatial derivatives of the window function to avoid dust E- to B-mode leakage. These are used to create a combined Fourier-space map (hereafter referred to as the `mock data');
\begin{eqnarray}\label{Eq:ContaminationMap}
B_\mathrm{tot}(\vec{l}) = f_\mathrm{dust}B_\mathrm{dust}(\vec{l})+\sqrt{f_\mathrm{lens}}B_\mathrm{lens}(\vec{l})+B_\mathrm{noise}(\vec{l})
\end{eqnarray}
where $f_\mathrm{lens}$ is included to represent delensing and we generate a Fourier-space noise map, $B_\mathrm{noise}(\vec{l})$, for each tile separately from a Gaussian realisation of the 1D noise spectrum (Eq.~\ref{Eq:noiseModel}). A scaling factor $f_\mathrm{dust}$, corresponding to the level of dust residuals, is included to emulate a null test after incomplete foreground subtraction. The 2D power spectrum is given by $|B_\mathrm{tot}(\vec{l})|^2/\langle{W^2}\rangle$, where the angle bracket indicates a spatial average over the window function $W^2(\mathbf{x})$ \citep{2014MNRAS.438.1507N}. We note that this procedure is only approximate, and a more complete treatment would perform full inversion of the mode-coupling matrix.

Since we only consider small tiles, there exists a limited resolution of the $l$-space pixels constraining the minimum $l$ that can be used in our estimators (given by $l_\mathrm{min}\approx360\degree/\Delta\theta$). To ameliorate this, we applied \textit{zero-padding} to increase the effective resolution of the maps, using a padding ratio, $R_\mathrm{pad}$ (the factor by which each tile width is increased), of 2. 

Following this, the anisotropy estimators are applied to the combined power map, using $l\in[l_\mathrm{min},2000]$, with $l_\mathrm{min}=360/(R_\mathrm{pad}\Delta\theta)=90$. It can be shown that there is a small additional bias caused by the square shape of the individual pixels, with, for example, the central nine pixels all satisfying $\sin{4\phi_{\vec{l}}}=0$ (independently of the padding). This gives unphysical contributions to $Af_s$ and $Af_c$, implying that the estimated quantities are not invariant under rotation of the pixel-grid of the power-map. To account for this, we performed 20 uniformly spaced rotations of the map, averaging over the rotationally corrected estimated quantities. Further details of this and zero-padding may be found in appendix \ref{Appen:Pixellation}.

\subsection{Monte Carlo Errors}\label{Subsec:MCerrors}
To compute the variances in the hexadecapole parameters, we adopt a Monte Carlo (MC) procedure, based on the null hypothesis of there being no anisotropy present. We must therefore generate MC simulations which are isotropic, but have the same 1D power spectrum as is present on each tile. For this purpose, first the 1D power spectrum of the combined 2D power-space map is computed by binning in annuli with a width of $\Delta{l}=\frac{2}{3}l_\mathrm{min}=60$ (equal to the pixel width). 
% This uses the \textit{unpadded} 2D power spectra here (with the same Gaussian noise and lensing realisation as their padded analogues) to avoid bias, since the padding introduces a smoothing to the $C_l$ curve, further enhanced by zero-padding of the MC simulations, making them not directly comparable to the initial data. \alex{I am confused about why we ever want to pad in that case.}\oliver{a good point: although fitting the MC sims to padded/unpadded data doesn't actually make any significant difference (although padding as a whole does)}  
The $l$ co-ordinate for each annular bin was estimated via a simple analytic model of the expected 1D spectrum (using the fiducial $C_l$ and the computed monopole amplitude $A$), giving greater accuracy than assuming it to be in the centre of the $l$-range for each bin. The binned (logarithmic) data was then fit to a univariate spline curve, which well represents the data.% in \texttt{Scipy}\footnote{\url{https://docs.scipy.org/doc/scipy/reference/}}, which well represents the data. 

Gaussian random field maps drawn from this 1D power spectrum are generated as for the noise spectrum, applying zero-padding as previously. (The 2D power spectra of these maps will be isotropic on average, but, due to the stochastic nature of the limited-resolution maps, will still exhibit some random fluctuations of anisotropy.) The hexadecapole parameters are computed for $N_\mathrm{sims}$ MC realisations, giving estimates for the variances and distributions of $A$, $Af_s$ and $Af_c$. 

Both the isotropic simulations and the pipeline were tested by comparing the estimated anisotropy parameters to those of input \textit{isotropic}, dust-free maps and verifying that a null response was obtained. This was satisfied for sufficiently large $N_\mathrm{sims}$, here set to 500.

\subsection{Realisation Dependent Debiasing}\label{Subsec:debiasing}
Due to the non-negative nature of the $H^2$ statistic, there will be a significant biasing contribution from random fluctuations of isotropic maps, giving $\langle{H^2_\mathrm{MC}}\rangle \neq 0$ even for isotropic MC maps. Since $H^2$ is a four-point correlation function in $B$, this bias can be understood as a Gaussian or disconnected contribution to $H^2$. Na\"ively, we may debias by subtracting $\langle{H^2_\mathrm{MC}}\rangle$ from the mock data-derived $H^2$ estimate, but we use a more robust ``realisation-dependent'' approach here, which can correct for first-order errors in the estimation of the 1D power spectra. (A similar approach is typically used in measurements of the lensing power spectrum, cf.~\citealt{2014MNRAS.438.1507N}). Utilising estimates of $H^2$ obtained from power spectra of the MC Fourier-space maps (denoted SS) and cross-spectra of these with the mock data maps (denoted DS), we compute an \textit{isotropic bias} term
\begin{eqnarray}\label{H2Bias}
\langle{H^2\rangle}_\mathrm{bias,iso}= 4\langle{H^2\rangle}_\mathrm{DS}-\langle{H^2\rangle}_\mathrm{SS}
\end{eqnarray}
%\langle{H^2\rangle}_\mathrm{bias,iso}=
%\begin{array}{@{}ll@{}}
%    4\langle{H^2\rangle}_\mathrm{DS}-%\langle{H^2\rangle}_\mathrm{SS} & \text{(Mock %Data)}
% \end{array}
%.
%\end{eqnarray}
as derived and explained in appendix \ref{Subsec:H2Bias}. (For computational efficiency, we approximate this full expression for the bias as $ \langle{H^2}\rangle_\mathrm{bias,iso}=\langle{H^2}\rangle_{\mathrm{SS}} $ when using Monte Carlo simulations to obtain error bars and significances.) The \textit{debiased hexadecapole statistic} is thus 
\begin{eqnarray}\label{Eq:DebiasedH2}
\mathcal{H}^2=H^2 - \langle{H^2\rangle}_\mathrm{bias,iso}
\end{eqnarray}
which can be used to compare data and simulations, since it should (a) have zero mean for an isotropic map, and (b) be robust to small errors in the assumed power spectra of MC simulations. Here we use $N_\mathrm{bias}=500$ independent MC simulations for each tile (distinct from the previous set of simulations) to estimate a expectation value for this bias.

Since $\mathcal{H}^2$ derives from four copies of the B-mode Fourier map, we normalise it by dividing by $\sqrt{\langle{W^4}\rangle}$ rather than $\langle{W^2}\rangle$ \citep{2014MNRAS.438.1507N}. 

To compute the significance of hexadecapole detection for a single tile, we must consider the statistics of $\mathcal{H}^2$ for the isotropic MC simulations. Given a cumulative distribution function, $p(\mathcal{H}^2)$, we can compute the percentile value for the data-derived $\mathcal{H}^2$ estimate, allowing us to assess whether the data is indeed anisotropic (for $p\rightarrow1$) or whether it is characteristic of random fluctuations of an isotropic spectrum (for $p\sim0.5$). An analytic (exponential) CDF is derived in appendix \ref{Appen:Stats}; here we adopt a simpler procedure, computing the statistical percentile from the $N_\mathrm{sims}=500$ MC simulations alone.

In addition, we define the \textit{equivalent significance}, $\mathcal{P}_\sigma(\mathcal{H}^2)$ of a detection of anisotropy, by converting the statistical percentile $p(\mathcal{H}^2)$ into an associated $\sigma$ value according to a standard Gaussian distribution CDF (e.g. $p=95\%$ corresponds to $\mathcal{P}_\sigma=2$). This is used for greater clarity with $\langle{\mathcal{P}_\sigma\rangle}=0$ for isotropic tiles, and large $\mathcal{P}_\sigma$ representing significant anisotropy. 

\subsection{Patch Anisotropy}\label{Subsec:PatchAnisotropy}
To assess the likelihood of anisotropy over an entire patch of the sky, we define the \textit{patch anisotropy}, $\Xi$, whose estimator is given by the mean of the hexadecapole power $\mathcal{H}^2$;
\begin{eqnarray}\label{Eq:PatchAnisotropy}
\widehat{\Xi} = \frac{1}{N}\sum_{i=1}^{N}\widehat{\mathcal{H}}^2,
\end{eqnarray}
%blake comment: hat on the hs? - yes
where $N$ is the number of non-overlapping tiles. Using the MC simulations, we can construct a probability distribution in $\Xi$ for isotropic dust realisations, which is well approximated as a Gaussian, $\mathcal{N}(\mu_\Xi,\sigma^2_\Xi)$, from the Central Limit Theorem for large $N$ (appropriate for the 5\% sky patch, with $N = 245$).\footnote{Due to long-wavelength fluctuations of lensing modes and large-scale power in dust, there will be some correlations in $\mathcal{H}^2$ between tiles even in the absence of anisotropy, meaning the tiles are not strictly independent (as assumed by the Central Limit Theorem). The effects of this are limited since the lowest measurable mode has $l\approx90$, and the distribution was found to be well fit by a Gaussian.} 

The mock data patch-hexadecapole value, $\Xi_0$, can thus be compared to the MC distribution via its \textit{significance}: 
\begin{eqnarray}\label{Eq:Significance}
\mathcal{S}=\Xi_0/\sigma_\Xi-\langle{\mathcal{S}_\mathrm{bias}\rangle},
\end{eqnarray}
indicating how well anisotropy can be detected in a particular experiment. We include a bias term to account for the small hexadecapole signature from lensing, though it is very small, corresponding to $|\langle{\mathcal{S}_\mathrm{bias}}\rangle|<1.5\sigma_\Xi$ in all cases. (No bias is found when Gaussian isotropic realisations of the $C_l^\mathrm{lens}$ spectrum are used instead of the FFP10 maps). %Blake comment: i dont get this - this is just saying that the bias is just from the fact that lensing isn't isotropic, so if instead we use an isotropic realisation we get no bias.
%This is negligible on single-tile scales, and may be simply removed via MC simulations of the expected lensing potential.

% Here, we note that there is a small additional bias present inherent in the MC simulations (due to a non-zero lensing hexadecapole signature) creating a non-zero $\mu_\Xi$, meaning that significance of a detection of $\Xi$ is non-zero even in simulations with no dust included. This is a very minor effect, giving a patch-averaged bias magnitude within $1.5\sigma$ in all cases. This is negligible on a tile-only level, hence was not previously discussed, but can be important when probing low dust levels. To rectify this, we compute the bias by running the analysis on the same patch with $f_\mathrm{dust} = 0$ and computing $\langle\mathcal{S}_\mathrm{bias}\rangle$. This is simply subtracted from the significance to produce a final significance free from MC bias, and could be emulated in experimental data using MC simulations of only lensing and noise. 
% \blake{it might be good to have just one overall equation, rather than describing steps in the text. Also, got to here thus far.}

\begin{table}
\centering
\begin{tabular}{l|l}
Symbol & Definition\\
\hline
$A$ & Monopole dust B-mode amplitude (Eq.\,\ref{Eq:PowModel})\\
$Af_s$, $Af_c$ & Sine \& cosine hexadecapole coefficients\\
$f_s$, $f_c$ & Fractional hexadecapole coefficients\\
$H^2$ & Hexadecapole strength (Eq.\,\ref{Eq: HexPow})\\
$\mathcal{H}^2$ & Debiased Hexadecapole strength (eq.\,\ref{Eq:DebiasedH2})\\
$\epsilon$ & Anisotropy Fraction ($H/A$)\\
$\alpha$ & Anisotropy angle (Eq.\,\ref{Eq: AnisotropyAngle})\\
$\sigma_{Af}$ & MC error (mean of $\sigma_{Af_s}$ and $\sigma_{Af_c}$)\\
$p(\mathcal{H}^2)$ & Anisotropy Likelihood (Eq.\,\ref{Eq:Chi2Prob})\\
$\mathcal{P}_\sigma(\mathcal{H}^2)$ & Equivalent anisotropy sigificance (Sec.\,\ref{Subsec:MCerrors})\\
$\Xi$ & Patch Hexadecapole (Eq.\,\ref{Eq:PatchAnisotropy})\\
$\mathcal{S}$ & Significance of $\Xi$ (Eq.\,\ref{Eq:Significance})
\end{tabular}
\caption{Estimated quantities derived from the hexadecapole and other estimators (Eq.\,\ref{Eq:estimators})}
\end{table}\label{Tab:estParams}

\section{Null Tests}\label{Sec:NullTests}
\subsection{Quantifying the Power of Statistical Anisotropy Null Tests for Future Experiments}
An important application of these dust anisotropy estimation techniques is to perform \textit{null tests} to assess how well the hexadecapole estimators can detect residual dust after dust-subtraction using standard (multi-frequency) methods \citep[e.g.][]{1996MNRAS.281.1297T,2008A&A...491..597L}. 
We imagine an experiment which has performed dust cleaning using maps at multiple frequencies, uniformly reducing the real-space dust level to a fraction $f_\mathrm{dust}$. How significantly can we then detect a dust residual in the 5\% sky region? Here, this is achieved by rescaling the \citetalias{2017A&A...603A..62V} and \citetalias{planck2014-a14} dust maps by $f_\mathrm{dust}$, and computing the associated significance of anisotropy detection ($\mathcal{S}$), as described above. 

To provide a physical scale to $f_\mathrm{dust}$ (following \citealt{planck2014-XXX}) we define the \textit{effective tensor-to-scalar ratio}, $r_\mathrm{eff}$, as the ratio of dust monopole amplitude to the expected B-mode power from tensor modes at $r = 1$. $r_\mathrm{eff}$ is thus a useful quantity, since it corresponds to the level of a spurious $r$-detection that (potentially) could be investigated with our null test. We evaluate the dust amplitude $A$ at $l = 80$ (approximating the peak sensitivity of future ground-based experiments), giving
\begin{eqnarray}\label{Eq:r_effDefn}
r_\mathrm{eff}(f_\mathrm{dust})=\frac{\langle{A_{80}}\rangle}{C_{80}^\mathrm{tensor}(r=1)}f_\mathrm{dust}^2.
\end{eqnarray}
for expected tensor contribution $C_{80}^\mathrm{tensor}(r=1) = 7.31\times10^{-5} \mu{}\mathrm{K}^2$ (from \texttt{CAMB}, \citealt{2000ApJ...538..473L}), and mean dust power at $l=80$ $\langle{A_{80}}\rangle$ (averaged over all 245 tiles with no lensing or noise modes included). Using the 5\% sky region, $\langle{}{A_{80}}\rangle$ is given by $3.9\times{}10^{-5}$\,$\mu{}\mathrm{K}^2$ ($1.5\times{}10^{-4}$\,$\mu{}\mathrm{K}^2$) in the \citetalias{2017A&A...603A..62V} (\citetalias{planck2014-a14}) simulation.

\begin{figure*}
\includegraphics[width=0.9\linewidth]{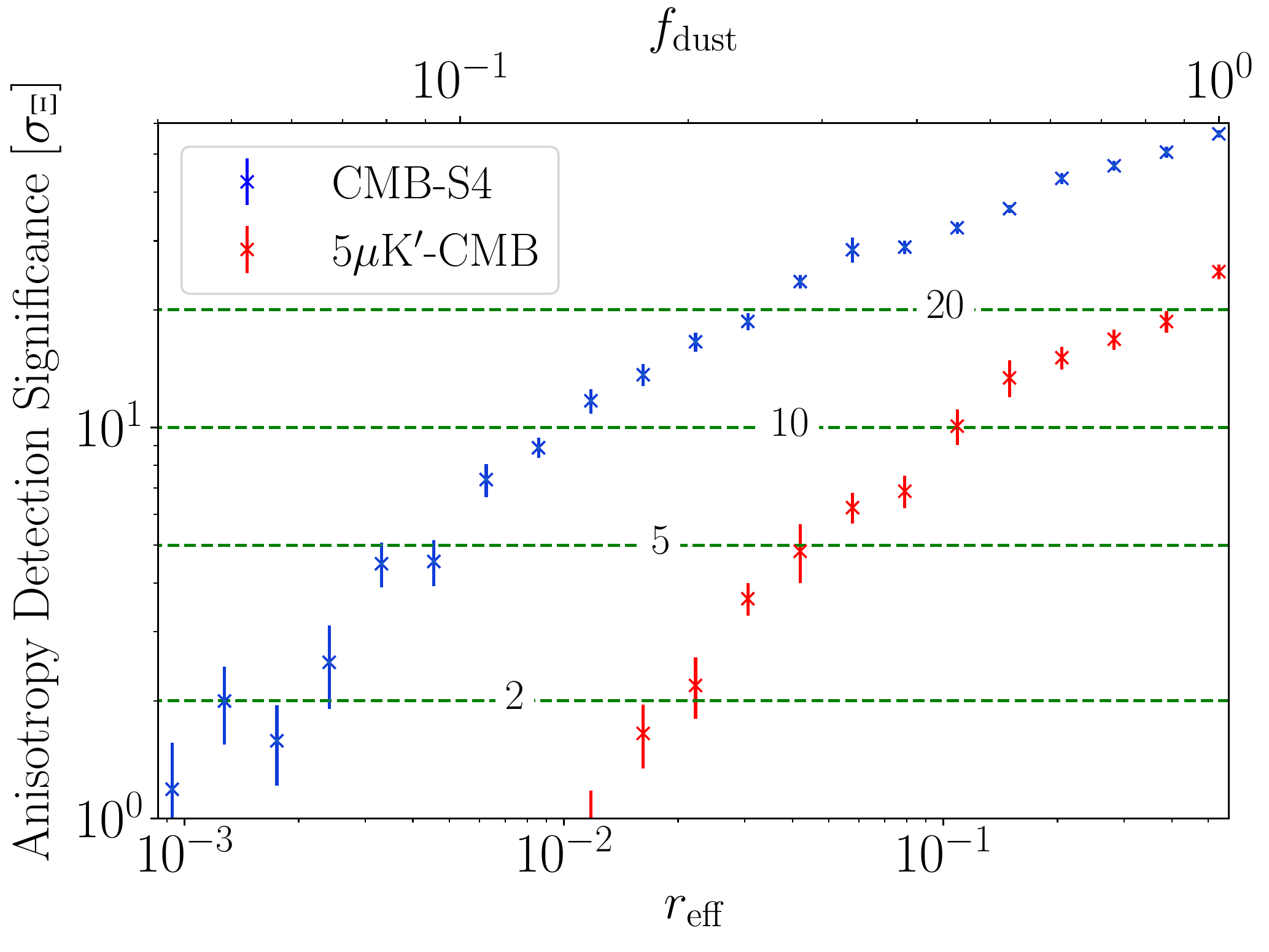}
\caption{Significance, $\mathcal{S}$ (in units of $\sigma_\Xi$), of a detection of dust statistical anisotropy via the patch-averaged hexadecapole, $\Xi$. This is given as a function of residual dust fraction, $f_\mathrm{dust}$ and equivalent effective $r$ (Eq.\,\ref{Eq:r_effDefn}) with green dashed lines representing 2, 5, 10 and 20$\sigma_\Xi$ significances. The large values of $\mathcal{S}$ imply that this would be a valuable null test for future experiments. This uses the \citetalias{2017A&A...603A..62V} dust simulation with noise and lensing parameters appropriate for the $5\mu{}$K$^{\prime}$-CMB and CMB-S4 experiments (tab.\,\ref{Tab:noiseParams}). Mock data were obtained from 245 $3\degree$ width tiles of the 5\% sky region, using $N_\mathrm{bias}=N_\mathrm{sims}=500$ and error-bars represent the standard error obtained by re-runnning the analysis 10 times for each data-point. The significances are corrected for the small intrinsic MC error from lensing ($\sim$\,$1\sigma_\Xi$) as described in Sec.~\ref{Subsec:PatchAnisotropy}.}\label{Fig:NullPlot}
\end{figure*}

Fig.~\ref{Fig:NullPlot} shows the results of this for a range of values of $f_\mathrm{dust}$, using \citetalias{2017A&A...603A..62V} simulations with noise and lensing parameters appropriate for the $5\mu{}$K$^{\prime}$-CMB and CMB-S4 experiments (see table \ref{Tab:noiseParams}). The significance is plotted in units of $\sigma_\Xi$, showing the mean and standard error of $\mathcal{S}$ for each $f_\mathrm{dust}$ value obtained from 10 iterations of the test. (We expect some variation in $\mathcal{S}$ due to the stochastic nature of the noise contributions to the mock data.) There is a similar shape to the plots for both experiments with greater $\mathcal{S}$ for CMB-S4 due to the smaller noise levels and larger assumed delensing efficiency. As expected, $\mathcal{S}$ rises monotonically with $r_\mathrm{eff}$ due to the increasing dominance of dust over the noise and lensing modes in $C_l^{BB}$. We report a peak significance of $25\pm1\sigma_\Xi$ ($56\pm1\sigma_\Xi$) using $5\mu{}$K$^{\prime}$-CMB (CMB-S4) parameters. 

\begin{table}
\centering
\begin{tabular}{c||c|c}
$\mathcal{S}$ [$\sigma_\Xi$] & $5\mu{}$K$^{\prime}$-CMB & CMB-S4\\\hline
2 & 0.02 & 0.001\\
3 & 0.03 & 0.002\\
5 & 0.05 & 0.004\\
10 & 0.1 & 0.01\\
\end{tabular}
\caption{Projected upper bounds on $r_\mathrm{eff}$ (Eq.\,\ref{Eq:r_effDefn}) for given significance levels, $\mathcal{S}$, for two forthcoming CMB experiments, using the \citetalias{2017A&A...603A..62V} simulations across a 5\% sky region split into $3\degree$ tiles with a padding ratio, $R_\mathrm{pad}$, of 2.}\label{Tab:rEffLimits}
\end{table}

From this analysis, we can place bounds on the dust fraction detectable using hexadecapolar anisotropies at a given significance level and the associated effective tensor-to-scalar ratio as shown in table \ref{Tab:rEffLimits}.

We attribute the large difference in $\mathcal{S}$ between $5\mu{}$K$^{\prime}$-CMB and CMB-S4 to the much larger lensing contributions in the former case (we assume $f_\mathrm{lens}=0.4$ for 5$\mu{}$K$^{\prime}$-CMB and 0.1 for CMB-S4). Also, we note that this analysis uses the 5\% sky region to allow direct comparison, although this may not reflect forthcoming experiments such as the Simons Observatory, where the experimental region may be as large as 10\%. Comparing these constraints to the current limits on IGWs ($r\lesssim0.07$; \citealt{planck2013-p17,2016PhRvL.116c1302B}), it is clear that the technique can be used to test for the presence of dust-residuals even for substantially cleaned maps and is useful for verifying a detection of even a small value of $r$. We note a marked improvement using the lower CMB-S4 lensing and noise parameters. 
%In addition, the hexadecapole significances would be greatly boosted at higher frequencies, due to the significant power reduction in transforming to the 150\,GHz used here.

\begin{figure}
\includegraphics[width=\linewidth]{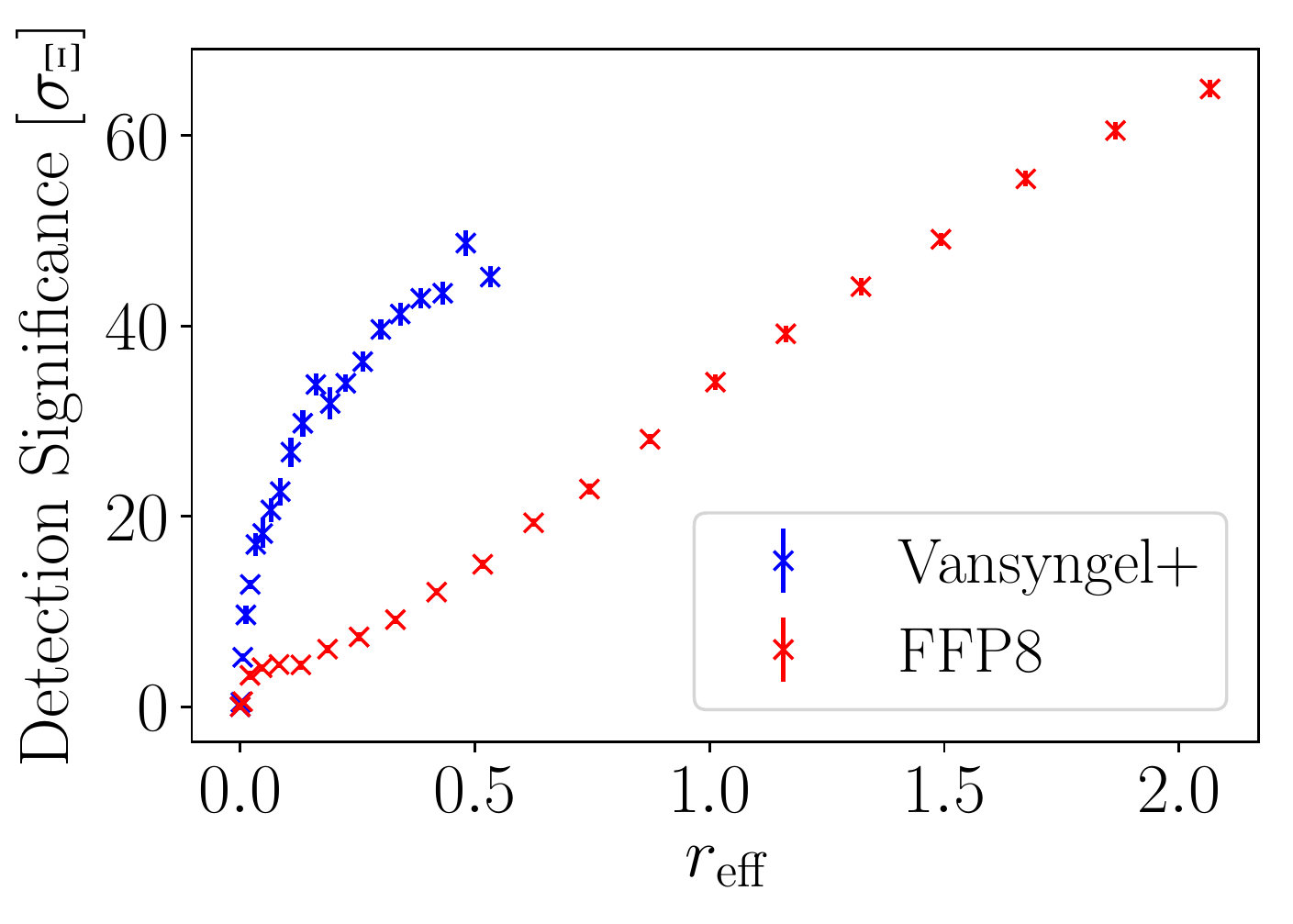}
\caption{As Fig.~\ref{Fig:NullPlot}, but comparing both \citetalias{2017A&A...603A..62V} and \citetalias{planck2014-a14} simulations for CMB-S4 noise and lensing parameters. The $r_\mathrm{eff}$ scale is calibrated separately for each simulation using the relevant mean dust amplitude $\langle{A_{80}}\rangle$ (Eq.\,\ref{Eq:r_effDefn}) and we use linear $r_\mathrm{eff}$ scales for clarity. The curves look more similar when plotted in terms of $f_\mathrm{dust}$.}\label{Fig:V17vsFFP8Plot}
\end{figure}

Fig.~\ref{Fig:V17vsFFP8Plot} compares the results from the \citetalias{planck2014-a14} and \citetalias{2017A&A...603A..62V} simulations using CMB-S4 noise and lensing parameters, and we note markedly (and surprisingly) different behaviour between the two dust models. (The $f_\mathrm{dust}$ axis is not included here since $r_\mathrm{eff}$ is calibrated with a different $\langle{A_{80}}\rangle$ for each simulation.) 

First, we note that the maximum $r_\mathrm{eff}$ (at $f_\mathrm{dust}=1$) is much greater for \citetalias{planck2014-a14}; a consequence of a larger mean dust amplitude in the 5\% sky region, with $\langle{A_{80}}\rangle_\mathrm{FFP8}/\langle{A_{80}}\rangle_\mathrm{V17} \approx 3.9$. In addition, $\mathcal{S}$ is much reduced across the range of $r_\mathrm{eff}$ tested, with a $2\sigma$ anisotropy detection corresponding to $r_\mathrm{eff}\approx0.02$ in this case. This is due to the different assumptions made in the creation of the two simulations; the older \citetalias{planck2014-a14} maps have small-scale power added from an extension of the smoothed  experimental Q/I and U/I power spectra, whereas \citetalias{2017A&A...603A..62V} utilises newer Planck maps and a method for generating high-$l$ power via realisations of the turbulent GMF, with hyperparameters optimised to reproduce \cite{planck2014-XXX,planck2016-XLIV} results. However, it is not immediately clear how the physical differences in the generation of the two simulations result in the trends found in Fig.~\ref{Fig:V17vsFFP8Plot}. The different simulation choices also lead to the different $C_l^{BB}$ spectra as noted in Fig.~\ref{Fig:ClBBContributions}.

Due to the greater physical motivation in the mid-$l$ regions probed in this paper from the inclusion of a GMF model, we affix greater credibility to the \citetalias{2017A&A...603A..62V} simulation. The differences illustrate that there is significant theoretical uncertainty in our analyses that should be explored with future simulations and experimental data. In addition, it is pertinent to note that plotting $\mathcal{S}$ as a function of $f_\mathrm{dust}$ instead of $r_\mathrm{eff}$ gives considerably more similar results with $\mathcal{S}\approx65$ at $f_\mathrm{dust}=1$ for \citetalias{planck2014-a14}. The results are thus comparable if we are concerned purely with the factor by which dust power can be reduced.

\subsection{The Null Test in the Presence of Tensor Modes}\label{Subsec:TensorBias}
An important check of our null test methodology is whether it avoids false positives when tensor modes are present instead of dust. As previously noted, IGW-induced spectra are expected to be isotropic \cite[e.g.][]{2010Sci...328..989K}; therefore we do not expect them to contribute to the debiased hexadecapole statistic, $\mathcal{H}^2$, on which our null test is based (as any lower-order multipoles should vanish in the $e^{4i\phi_{\vec{l}}}$-weighted summations of Eqs.~\ref{Eq:estimators}). 

To investigate our null test in the presence of tensor modes, we consider a full-sky Stokes map realisation of the $r = 0.1$ tensor spectrum (motivated by the current constraints on $r$, \cite{2016PhRvL.116c1302B}). This map is obtained from a rescaled Planck FFP10 unlensed tensor mode simulation.\footnote{Also available from NERSC.} The full analysis pipeline of Sec.~\ref{Sec:Methods} was reapplied to this map (instead of the \citetalias{2017A&A...603A..62V} simulation) after the addition of noise and lensing modes, and the patch anisotropy $\Xi$ was computed for the 5\% sky region. (We use a whole-sky realisation of this spectrum rather than generating spectra individually for each tile to fully test for any additional sources of bias between the full-sky maps and the MC debiasing procedure.)

Computing $\Xi$ using $\Delta\theta=3\degree$ with $N_\mathrm{bias}=N_\mathrm{sims}=500$ for the predicted CMB-S4 noise and lensing parameters gives a significance of 
\begin{eqnarray}\label{Eq:TensorBias}
\mathcal{S}_\mathrm{bias, tensor}^{r=0.1} = -0.04\pm0.58\,\sigma_\Xi
\end{eqnarray}
where the error is obtained by rerunning the analysis 10 times with independent MC debiasing and noise modes. This is clearly consistent with zero, and far below the $10\,\sigma_\Xi$ detection found for 150~GHz \citetalias{2017A&A...603A..62V} dust at this amplitude level. We hence conclude that our method is robust and capable of providing a null test for dust without false positives in the presence of tensor modes.

\subsection{Anisotropy Detection Significances for Different Experimental Configurations}\label{Subsec:NoiseParamsXi}
\begin{figure}\centering
\subfigure[Full Lensing, $f_\mathrm{lens}=1$]{\includegraphics[width=\linewidth]{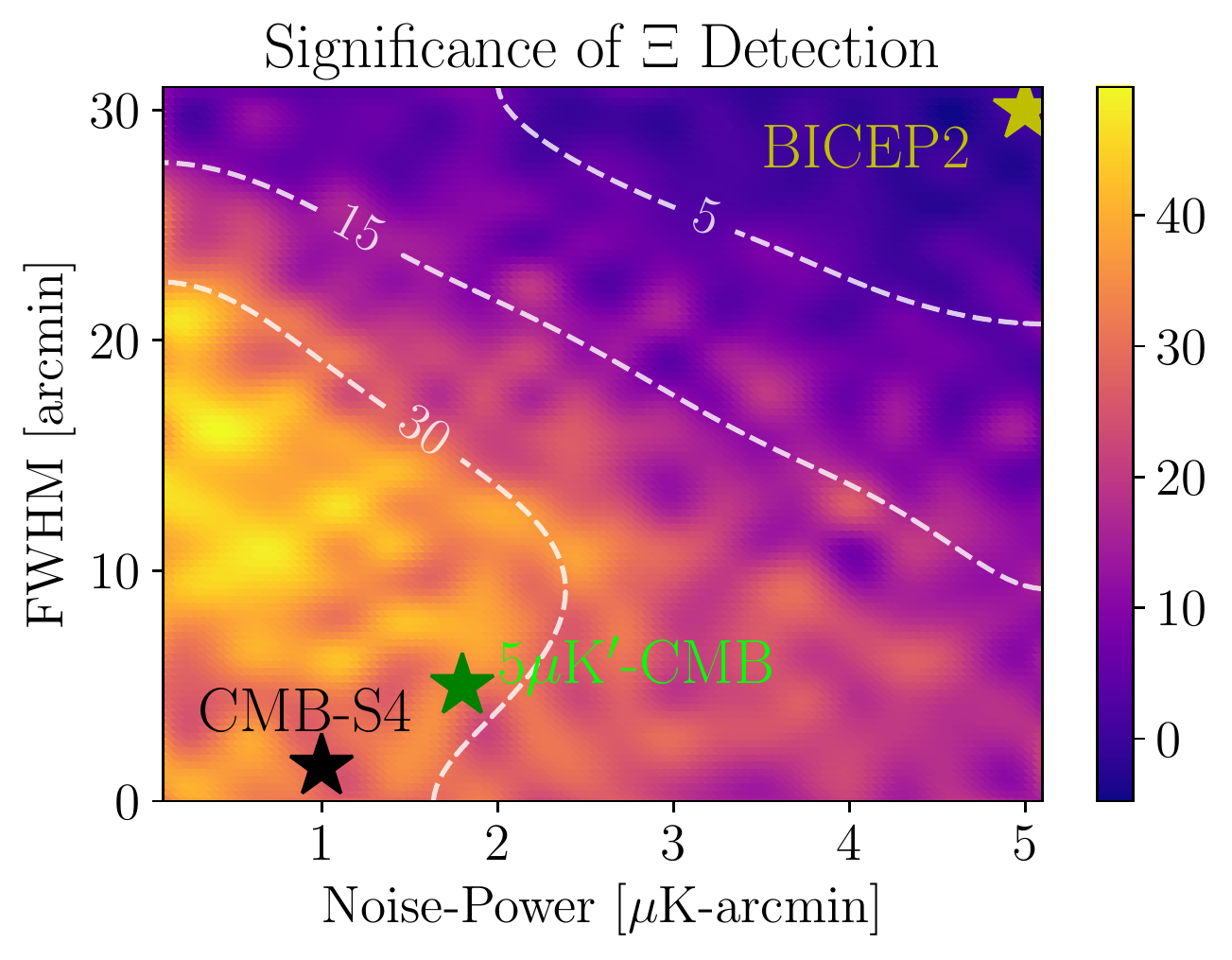}}
\subfigure[90\% Delensing, $f_\mathrm{lens}=0.1$]{
\includegraphics[width=\linewidth]{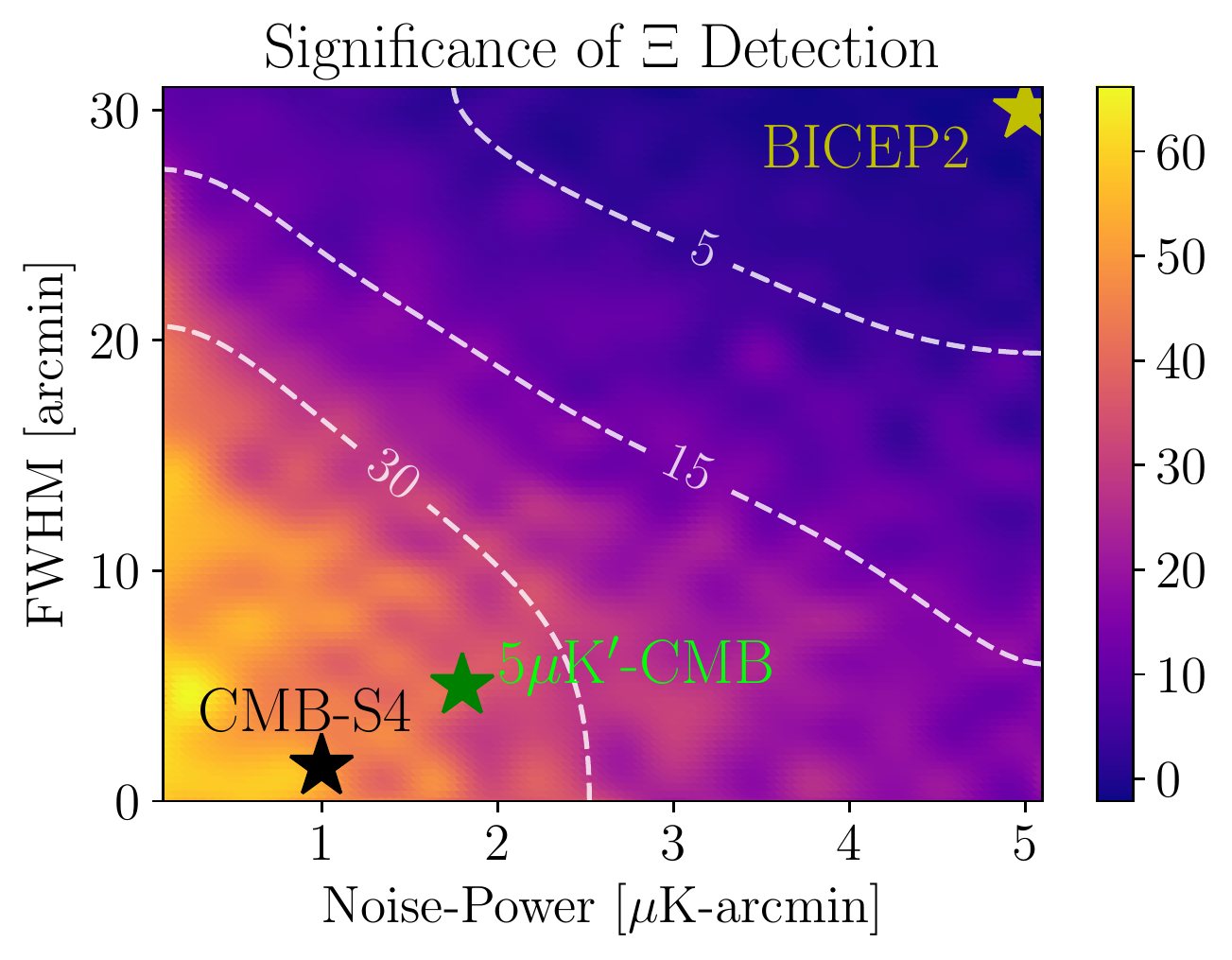}}
\caption{Significance of a detection of patch anisotropy, $\Xi$, (eq\,.\ref{Eq:PatchAnisotropy}) in units of $\sigma_\Xi$ for the 5\% sky region using a range of noise parameters (noise power $\Delta_\mathrm{P}$ and FWHM $\theta_\mathrm{FWHM}$). The panels assume (a) full lensing contributions to $C_l^{BB}$ and (b) 90\% delensing efficiency (i.e. $0.1C_l^\mathrm{lens}$). The plots are generated via interpolation of 400 individual tests with the small MC bias subtracted and visualised using \texttt{Scipy}. Stars show the location of the BICEP2 (yellow), $5\mu{}$K$^{\prime}$-CMB (green) and CMB-S4 (black) experiments and significance contours are also marked. $\Xi$ was computed using $\Delta\theta = 3\degree$ with $N_\mathrm{sims}=N_\mathrm{bias}=500$.}\label{Fig:WidePatchNoiseParams}
\end{figure}

We will now briefly consider how the ability of our test to detect dust depends on the experimental configuration. To simplify the discussion, we will not simultaneously vary $f_\mathrm{dust}$, experiment parameters and sky areas, but will instead fix $f_\mathrm{dust}=1$ and consistently analyse the 5\% sky region. As detailed previously, we compute the patch-averaged hexadecapole, $\Xi$, and derive its detection significance, $\mathcal{S}$, for different experimental noise levels. We thus obtain Fig.~\ref{Fig:WidePatchNoiseParams}, in which we show the dust detection significance $\mathcal{S}$ as a function of noise parameters $\Delta_\mathrm{P}$ and $\theta_\mathrm{FWHM}$ for both $f_\mathrm{lens}$ = 1 and 0.1 (i.e. zero and 90\% delensing efficiency). These plots use computations of $\mathcal{S}$ at 400 pairs of noise parameters.%, applying a bivariate spline smoothing using \texttt{Scipy} for visualisation.

The form of the plots is as expected, with high significances for low noise, which fall as $\Delta_\mathrm{P}$ and $\theta_\mathrm{FWHM}$ increase and obscure the dust signal. The scatter seen in adjacent regions may be attributed to the fact that each data-point uses a finite number of different realisations of the noise. Comparing lensed to delensed plots, the significances are higher in the latter case for low-noise, but there is no significant difference for high noise parameters (where the significances are consistent with zero), since the lensing modes become subdominant.

For both cases, when using BICEP2 noise parameters \citep{2014ApJ...792...62B} on the 5\% sky region, $\mathcal{S}$ is negligible, which is as expected since the BICEP2 experiment detected the dust monopole only at moderate significance, and the hexadecapole should be smaller still. Also marked are the assumed locations of the $5\mu{}$K$^{\prime}$-CMB and CMB-S4 experiments, which have much higher detection significances; around 50$\sigma$ (35$\sigma$) for CMB-S4 assuming $f_\mathrm{lens} = 0.1$ ($f_\mathrm{lens} = 1$). (We expect $f_\mathrm{lens}\approx0.4$ for $5\mu{}$K$^{\prime}$-CMB, thus the delensed significances cannot be directly obtained from these plots). These levels of significance confirm that dust statistical anisotropy is clearly detectable for a range of future experiments. 

\section{Visualising Dust Anisotropy}\label{Sec:Results}

\subsection{Spatial Distributions}\label{Subsec:WideResults}
\begin{figure*}\centering
\subfigure[Monopole amplitude $\log_{10}(A\,{[\mathrm{K}^2]})$]{\includegraphics[width=0.49\linewidth]{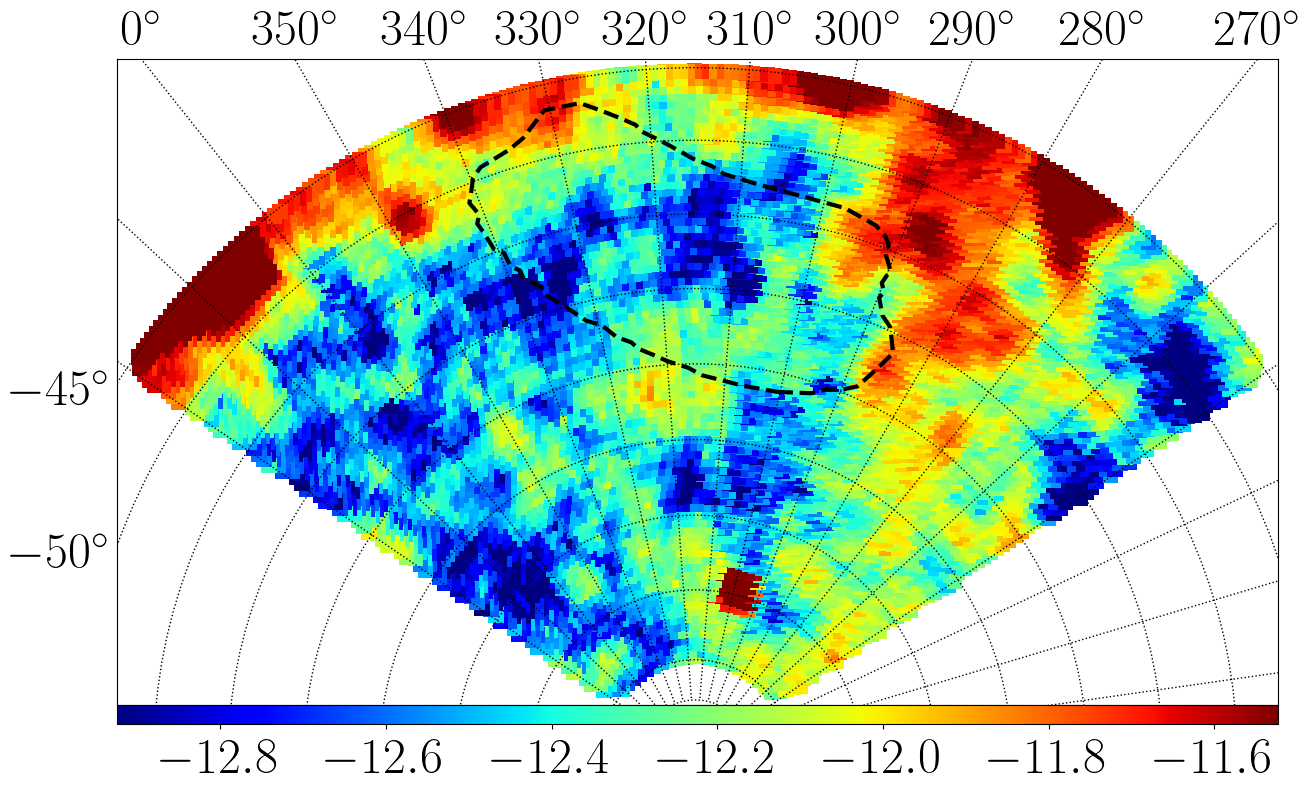}}
\subfigure[Debiased Hexadecapole Power $\mathcal{H}^2 {[\mathrm{K}^4]}$]{\includegraphics[width=0.49\linewidth]{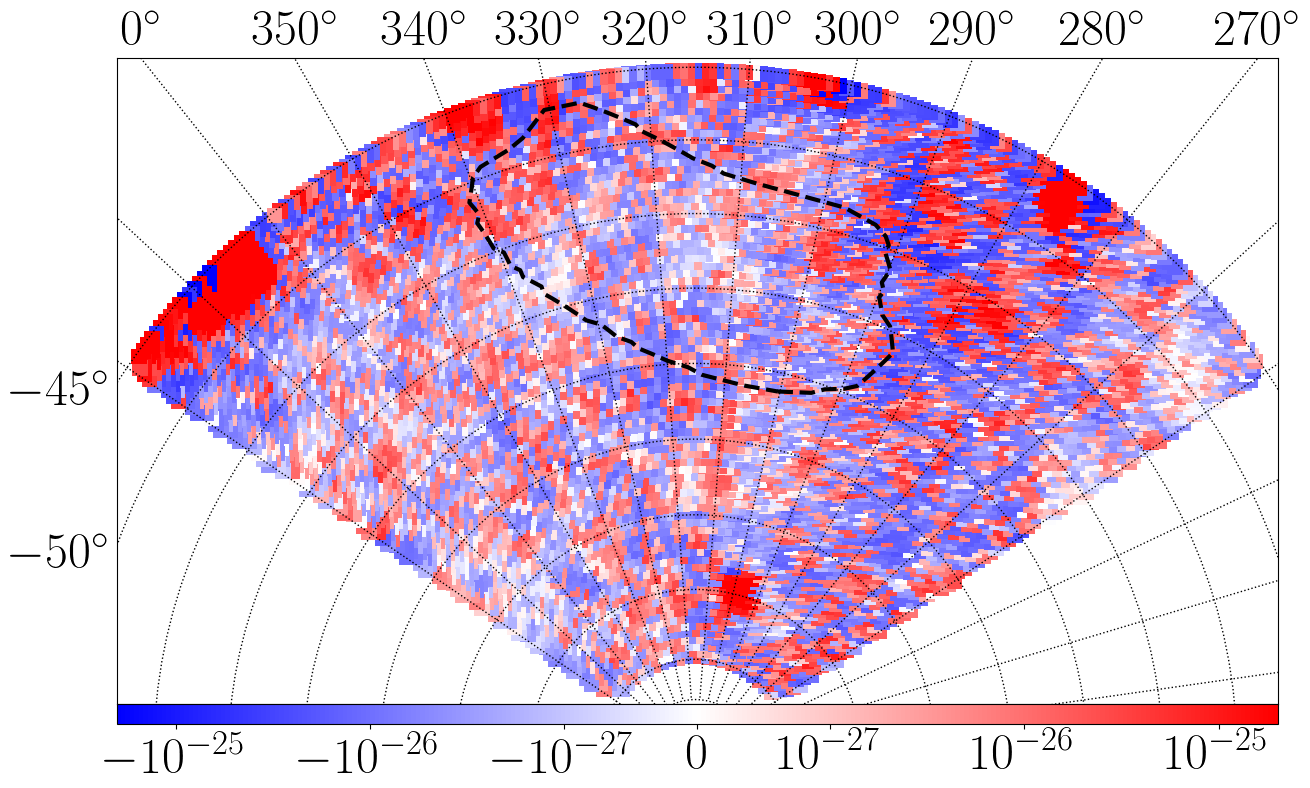}}
\subfigure[Hexadecapole Fraction $\epsilon = H/A$]
{\includegraphics[width=0.49\linewidth]{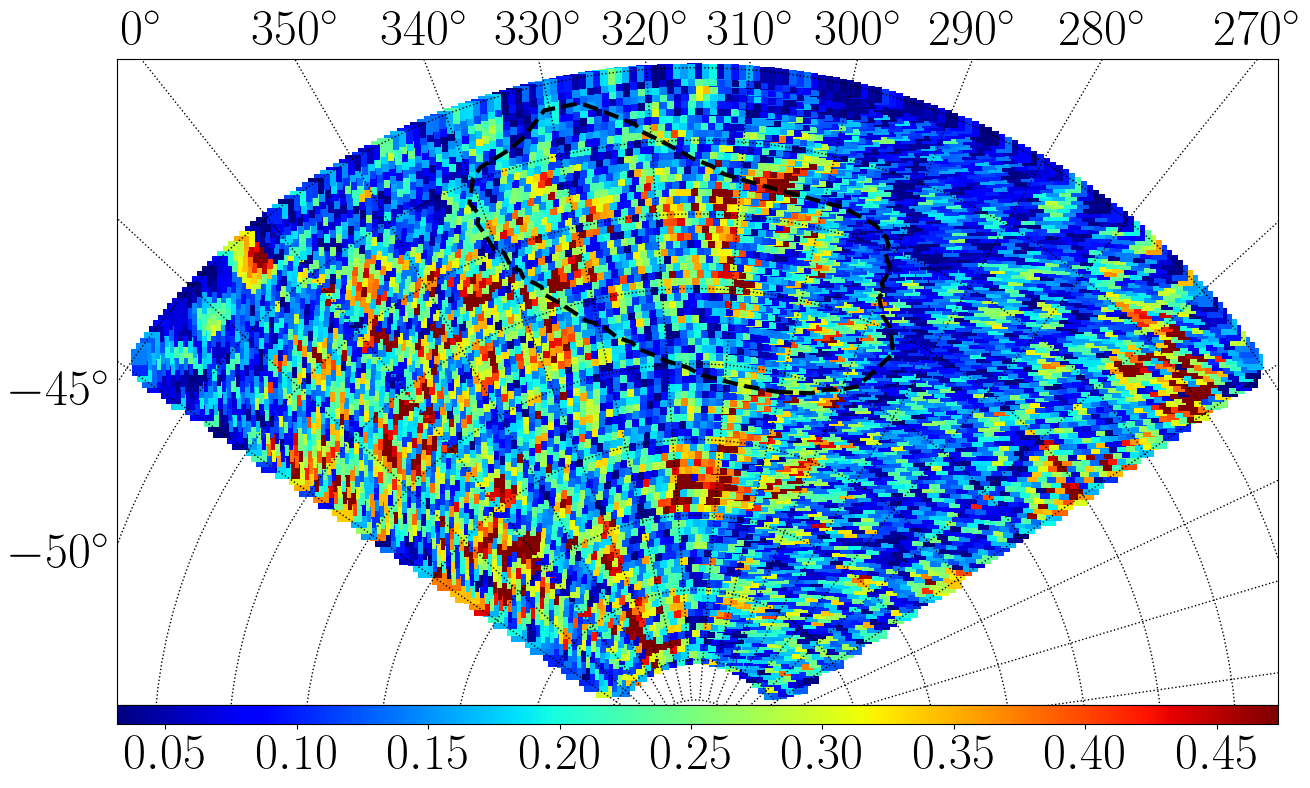}}
\subfigure[Hexadecapole Angle $\alpha {[\degree]}$]{\includegraphics[width=0.49\linewidth]{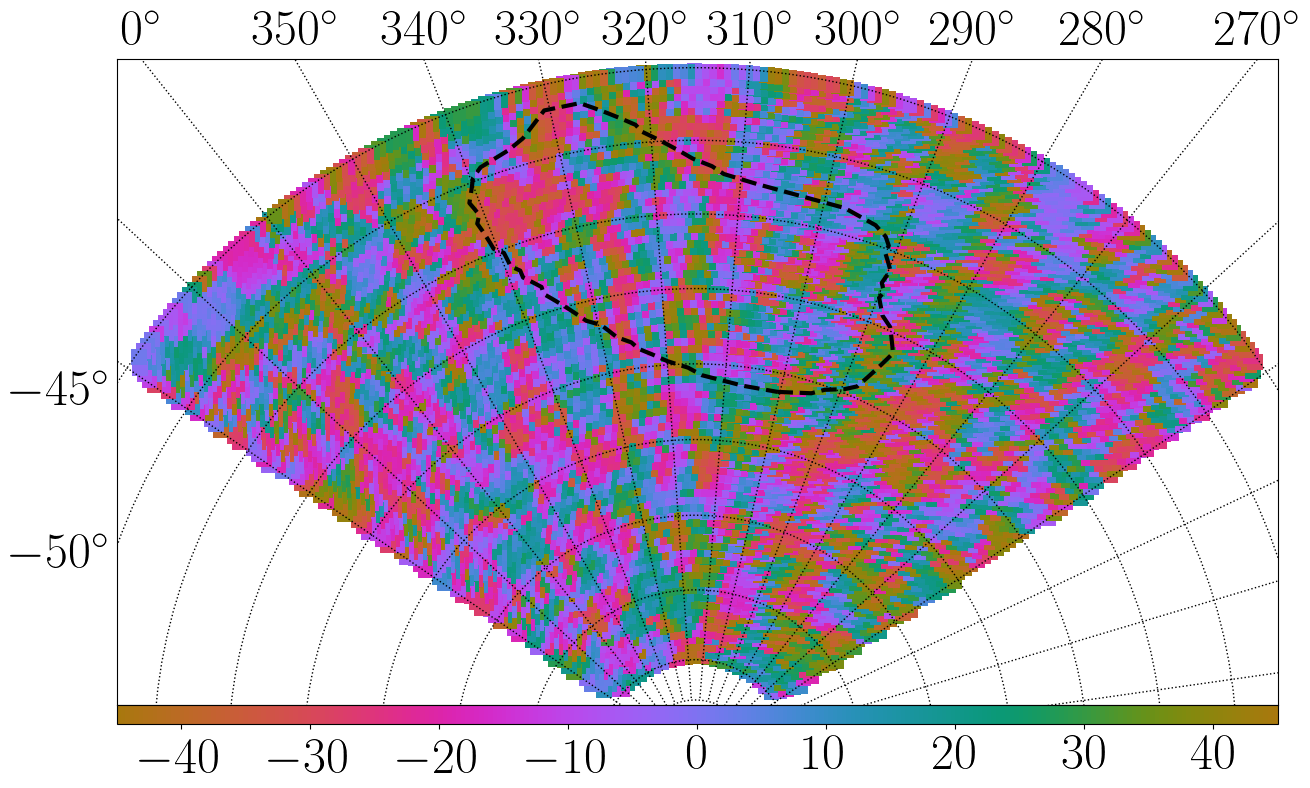}}
\subfigure[Equivalent Anisotropy Significance $\mathcal{P}_\sigma(\mathcal{H}^2)$]{\includegraphics[width=0.5\linewidth]{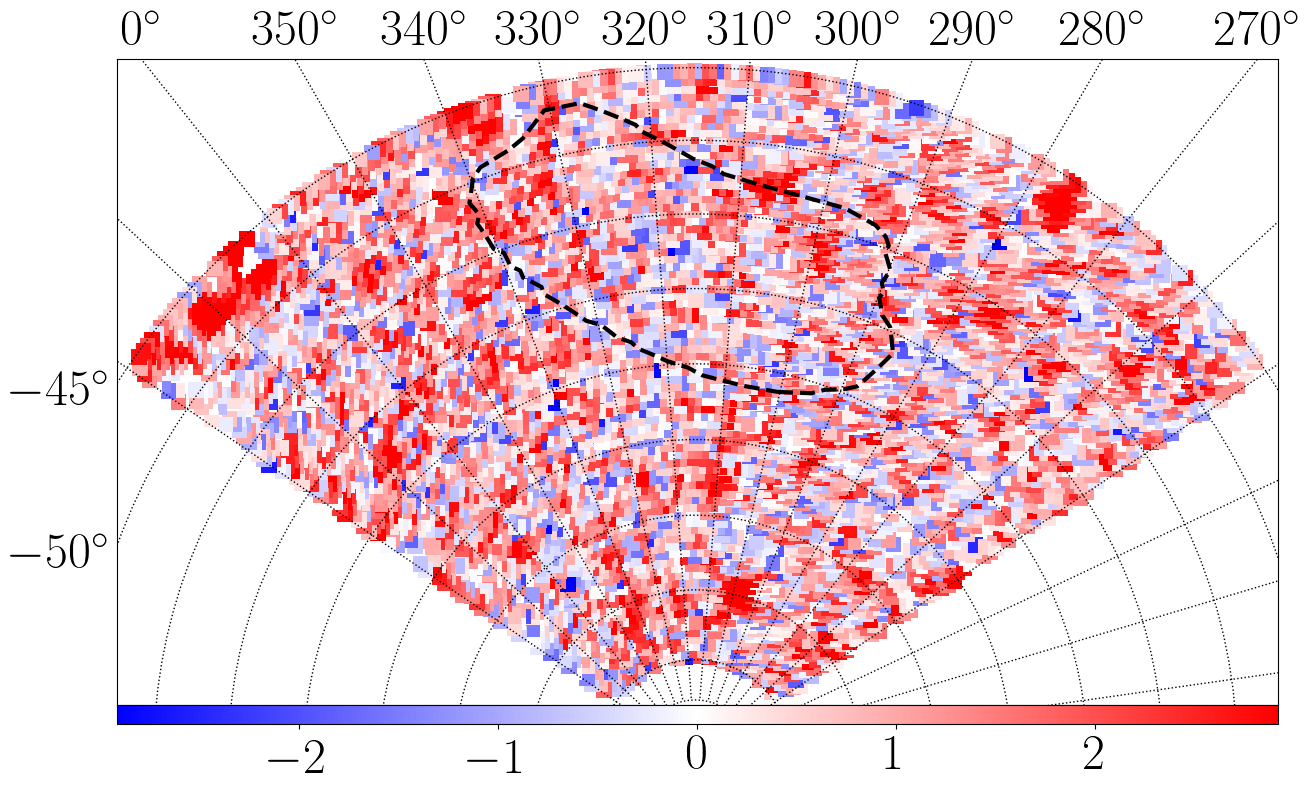}}
\caption{Plots showing the estimated anisotropy quantities (see tab.\,\ref{Tab:estParams}) for the 5\% (2150\,$\mathrm{deg}^2$) sky patch, using CMB-S4 noise parameters (tab.\,\ref{Tab:noiseParams}) on $\sim9000$ $3\degree$ width tiles with centres separated by $0.5\degree$. In (b), the debiased hexadecapole statistic is plotted on a symmetric logarithmic scale (using a logarithmic scale for both positive and negative data, with a small linear region encompassing zero) to include random negative fluctuations below the isotropic mean, with red regions indicating considerable anisotropy. For (c), we note that $\epsilon$ exhibits some bias, since it depends on the biased quantity $H^2$ rather than $\mathcal{H}^2$. 
Plot (e) gives the significance of a hexadecapole detection for a single tile, compared to isotropic MC simulations. This is converted into equivalent Gaussian significances, with zero (white, $\mathcal{P}_\sigma\approx{}0$) and large (red) values representing isotropy and significant anisotropy respectively. RA (top) and declination (left) in Galactic co-ordinates are shown on the plots, and the BICEP region is indicated with a black dashed line. The panels are plotted as Albers equal area conic projections.}\label{Fig:WidePatchPlots}
\end{figure*}

Following the methods of Sec.~\ref{Sec:Methods}, we may also compute spatially-resolved maps of the hexadecapole parameters across the 5\% sky patch, as shown in Fig.~\ref{Fig:WidePatchPlots} (henceforth setting $f_\mathrm{dust}=1$). This uses lensing and noise parameters appropriate for CMB-S4 (table \ref{Tab:noiseParams}), and plots are given as Albers equal area conic projections, created using the Python \texttt{skymapper} package.\footnote{\url{https://github.com/pmelchior/skymapper}} Each individual pixel represents a $3\degree$ width tile, whose centres are separated by $0.5\degree$ to improve visibility, thus violating their independence.

The monopole amplitude $A$ (Fig.~\ref{Fig:WidePatchPlots}a) exhibits substantial variation across the map, with high amplitudes mostly concentrated towards lower latitudes, and hence closer to the galactic centre. We note that these amplitudes are still several orders of magnitude smaller than those closer to the Galactic plane. The region used by the BICEP experiment (shown by the dotted line) has low dust amplitude, as expected, though we note some higher amplitude regions, particularly towards smaller RA. Notably, the form of $A$ in the BICEP region is heuristically similar to that found in \citet[fig.\,7]{2016JCAP...09..034R}.

Due to the low dust amplitudes and considerable contributions from the stochastic noise spectra, we expect some scatter in $\mathcal{H}^2$ around its true value, leading to $\mathcal{H}^2<0$ for some tiles. This can be partially ameliorated by using a larger $N_\mathrm{bias}$ to better constrain the bias term, although this adds considerable computation time. In Fig.~\ref{Fig:WidePatchPlots}b, $\mathcal{H}^2$ is displayed on a symmetric logarithmic scale (utilising a logarithmic scale for the majority of the data, with a linear region about zero to avoid infinities), thus including negative values, but we note that the \textit{biased} quantity, $H^2$, is used in Fig.~\ref{Fig:WidePatchPlots}c to define the anisotropy fraction $\epsilon=H/A$, else this is undefined for $\mathcal{H}^2<0$. 

By eye, a correlation between $\mathcal{H}^2$ and $A$ is apparent, as confirmed by the values of the hexadecapole fraction $\epsilon$, which varies predominantly between 10\% and 45\% across the map. This will be further discussed in Sec.~\ref{Sec:Correlations}. In addition, we note that the anisotropy angle (Fig.~\ref{Fig:WidePatchPlots}d) seems to be coherent on scales larger than the pixel-separation (0.5$\degree$), as was originally assumed in construction of the estimators. (Even though the dust-contributions of overlapping tiles are not independent, the noise realisations are generated separately, thus we would not expect coherency on scales above $0.5\degree$ if the angle was significantly biased by noise).

The hexadecapole `equivalent significance' (Fig.~\ref{Fig:WidePatchPlots}e, as defined in Sec.~\ref{Subsec:debiasing}) shows that we detect anisotropy at levels exceeding $2\sigma$ for a number of regions across the patch, with highly anisotropic regions correlating with regions of high $A$. For most tiles however, we have small $\mathcal{P}_\sigma$, implying largely insignificant detections of anisotropy on a single-tile level.

\subsection{Full Sky Correlations}\label{Sec:Correlations}
Using $A$ and $\mathcal{H}^2$ estimates from $3\degree$ tiles across the majority of the sphere, we can probe correlations over a range of angular scales as well as different levels of dust intensity. As noted in Sec.~\ref{Sec:Simulations}, the \citetalias{2017A&A...603A..62V} simulations do not accurately predict the dust polarisation spectra close to the galactic plane, thus we use the Planck \texttt{GAL80} unapodised mask to remove central galactic regions.\footnote{\url{http://pla.esac.esa.int/pla/}} Regions with declinations above $85\degree$ suffer from slight distortions due to the projection (as shown in appendix \ref{Appen:ProjectionDistortions}) thus these are also excluded, and the overall mask is apodised with a 3$\degree$ FWHM Gaussian kernel to avoid edge effects.

$A$ and $\mathcal{H}^2$ are then computed across the remainder of the sky using tiles of $\Delta\theta=3\degree$ separated by $1\degree$ to increase resolution, using CMB-S4 noise parameters and $N_\mathrm{bias}=100$ (to keep the computation time tractable). This gives $\sim40,000$ measurements which are mapped into a \texttt{HEALPix} $\texttt{NSIDE}=64$ (50,000 pixels) map, using linear interpolation to compute the parameters at the centre of the \texttt{HEALPix} pixels. (We note that the two pixellation routines are not directly compatible since our cut-out technique natively takes square patches along the RA axis unlike the \texttt{HEALPix} tiling).

\begin{figure*}
\centering
\subfigure[Monopole Amplitude]{\includegraphics[width=0.53\linewidth]{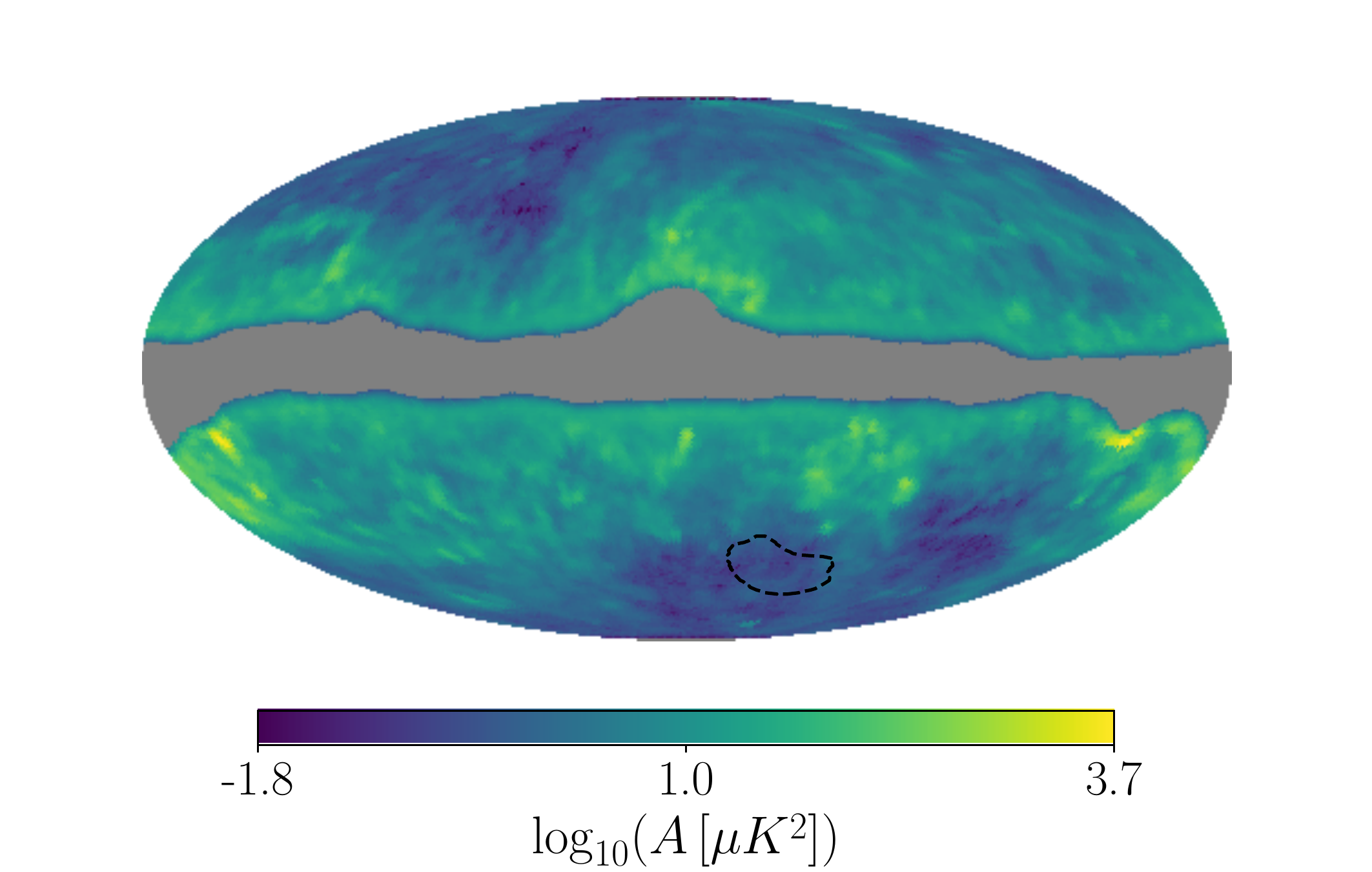}}
\subfigure[Debiased Hexadecapole Power]{\includegraphics[width=0.45\linewidth]{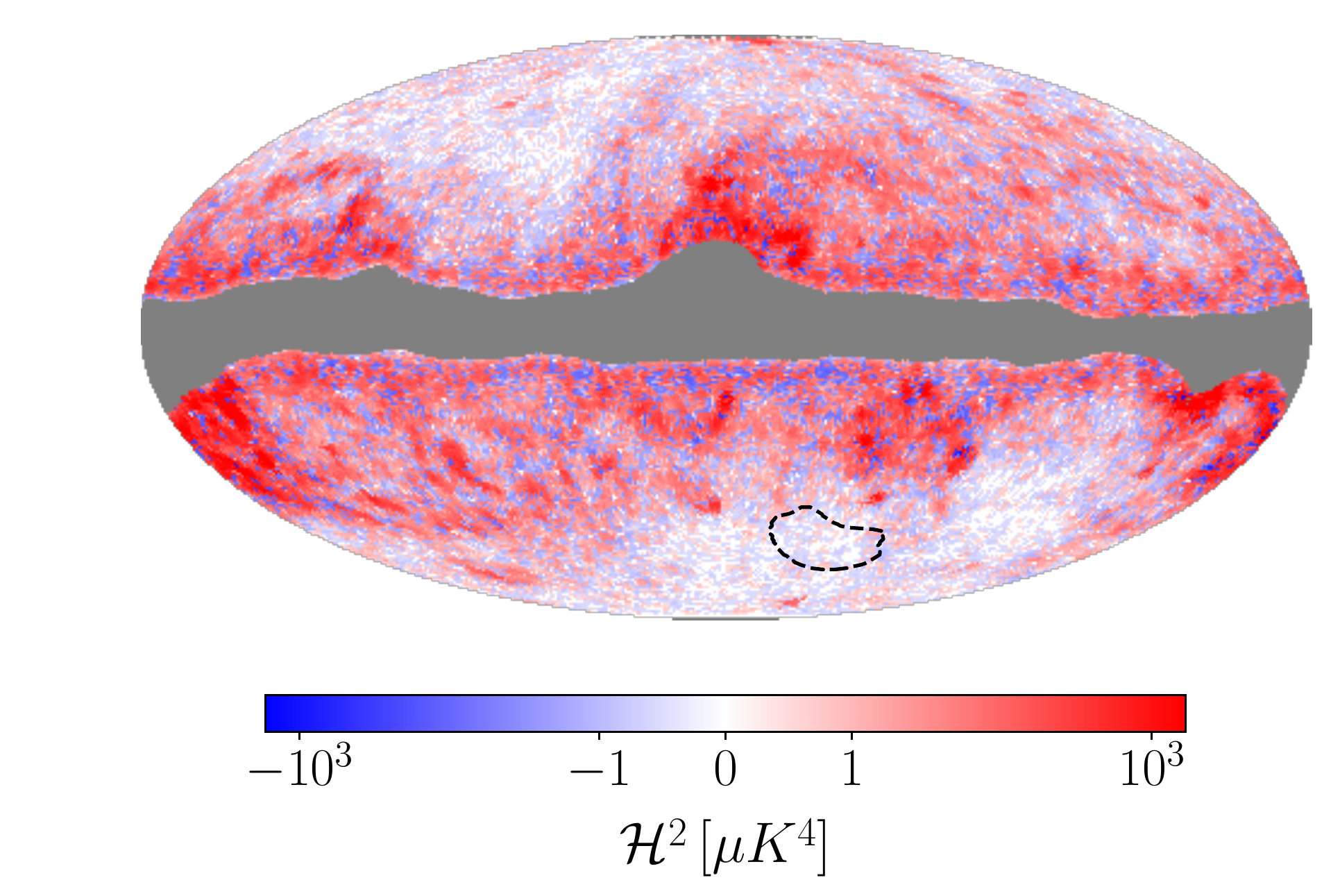}}
% \subfigure[Hexadecapole Angle, $\alpha$]{\includegraphics[width=\linewidth]{fullAng.pdf}}
\caption{Full sky Mollweide projections of the dust monopole amplitude and hexadecapole power, computed at a spatial resolution of $1\degree$ using CMB-S4 noise and lensing parameters with the \citetalias{2017A&A...603A..62V} simulation. Data is interpolated linearly onto the \texttt{HEALPix} grid from $\sim$40,000 parameter estimates, applying the Planck \texttt{GAL80} mask and additional masking at declinations $\delta$ satisfying $|\delta|>85\degree$. $\mathcal{H}^2$ is plotted on a symmetric logarithmic scale (to include negative fluctuations) and exhibits clear correlations with $A$. The BICEP region is shown as a dotted line.}\label{Fig:FullSkyGrids}
\end{figure*}

Fig.~\ref{Fig:FullSkyGrids} shows Mollweide projections of the computed parameters, and we note that the majority of the $\mathcal{H}^2$ values are positive, indicating significant anisotropy detections. The monopole and hexadecapole plots are very similar in form across the scales probed here, especially in high-dust regions, thus it is instructive to consider their correlation coefficient. 

\texttt{PolSpice} \citep{2004MNRAS.350..914C} is used to compute the necessary power-spectra, treating the \texttt{HEALPix} maps of $A$ and $\mathcal{H}^2$ as spin-zero fields, weighted by the apodised masks. From the output pseudo-$C_l$ spectra we can assess the correlation of $A$ and $\mathcal{H}^2$ on different scales via the cross-correlation; 
\begin{eqnarray}\label{Eq:CrossCorrelation}
\rho_{A\mathcal{H}^2}(l)=\frac{C_l^{A\mathcal{H}^2}}{\sqrt{C_l^{AA}C_l^{\mathcal{H}^2\mathcal{H}^2}}}.
\end{eqnarray}
Here $\rho_{A\mathcal{H}^2}\rightarrow1$ represents full correlation between variables.

\begin{figure}%\label{Fig:FullSkyCorrelationCoeff}
\includegraphics[width=\linewidth]{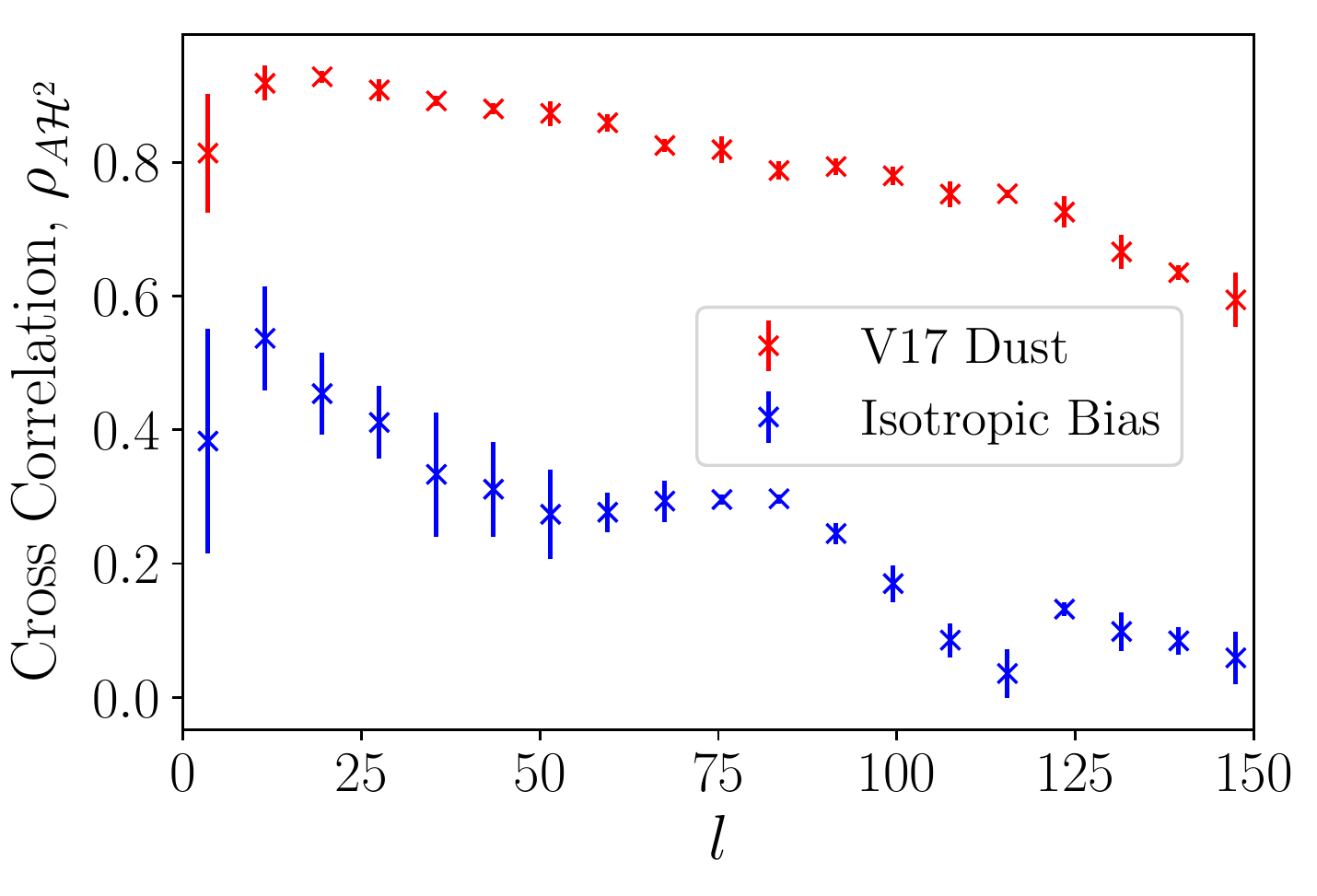}
\caption{Correlation coefficient (Eq.\,\ref{Eq:CrossCorrelation}) between the monopole amplitude $A$ and debiased hexadecapole coefficient $\mathcal{H}^2$ for $\sim$40,000 $3\degree$ width tiles on the sphere (as for Fig.~\ref{Fig:FullSkyGrids}). Data is given for $\mathcal{H}^2$ estimates from the \citetalias{2017A&A...603A..62V} dust simulation (red, assuming CMB-S4 noise and lensing) and a debiased Gaussian realisation of the isotropic $C_l$ spectrum in each tile (blue). We note strong correlations between $\mathcal{H}^2$ and $A$ for the dust (with $\rho\rightarrow1$ indicating full correlation), far in excess of those due to the isotropic bias. This uses the \texttt{PolSpice} emulator \citep{2004MNRAS.350..914C} to compute pseudo-$C_l$ spectra, with errors obtained from the variances of the binned data.}\label{Fig:FullSkyCorrelationCoeff}
\end{figure}

This is shown in Fig.~\ref{Fig:FullSkyCorrelationCoeff}, with errors taken from the variances of the binned data. For the \citetalias{2017A&A...603A..62V} mock data, there are clearly very strong correlations on large angular scales (small $l$) with $\rho_{A\mathcal{H}^2}\approx0.9$ for $l\in[5,50]$ and $\rho_{A\mathcal{H}^2}>0.7$ for $l<150$. (We note that extremely low $l$ modes have considerable uncertainty due to cosmic variance and the application of the mask). 
At larger $l$, the correlation is reduced, most likely due to the small-scale fluctuations in $\mathcal{H}^2$ as a result of noise contamination. In addition, we cannot probe smaller scales than $l\sim150$ with this dataset, since $A$ and $\mathcal{H}^2$ are only sampled at $1\degree$ resolution. 

To account for spurious correlations due to the possibility of bias terms in $\mathcal{H}^2$ correlating with $A$, we also plot $\rho_{A\mathcal{H}^2}$ using $\mathcal{H}^2$ estimates obtained from a single MC realisation of the isotropic $C_l$ spectrum for each tile. The correlation coefficient is clearly much smaller than that of the anisotropic data, thus the strong correlations seen for dust are (to a large extent) a real effect and not just an artefact of noise in $\mathcal{H}^2$.

The significant correlations between the monopole and hexadecapole powers confirm that by simply tracing $\mathcal{H}^2$ we can obtain clear bounds on $A$ and hence the level of dust. In addition, this correlation motivates us to consider a rudimentary method for cleaning the B-mode sky using statistical anisotropy estimators.

\section{Dedusting}\label{Sec:Dedusting}
\subsection{Methodology}
Here we consider the possibility of using the above correlations to `dedust' the B-mode sky; i.e.~to construct an estimate of the dust B-mode map that can be subtracted off. In particular, from the measured hexadecapolar anisotropy, we can obtain both the polarisation angle and a proxy for the polarisation amplitude and thus generate such a template of dust B-modes; this can then be subtracted from the observed B-mode map in order to isolate any IGW signature. This technique of generating a template of large-scale B-modes based on the measured higher-order statistics on small scales is analogous to that used in delensing (e.g.~\citealt{2002PhRvL..89a1304K}, \citealt{2012JCAP...06..014S}, \citealt{2015PhRvD..92d3005S}). If it proves feasible, the approach will in principle be possible at only a single frequency, given data with sufficiently low noise and high delensing efficacy. 

Throughout the above analysis we have assumed that the polarisation angle $\alpha$ is constant over small scales, allowing us to measure $Af_s$ and $Af_c$ (proportional to $\sin{4\alpha}$ and $\cos{4\alpha}$ respectively), which encode information on the ratio of the $Q$ and $U$ maps via Eq.~\ref{Eq:AlphaQUDefn}. Whilst we still assume $\alpha$ to be constant over each tile, on larger scales, we promote $\alpha$ to be a function of spatial position,  %, with each tile-based measurement taken %as the value of $2\alpha(\vec{x}_0)$ 
which we denote $\alpha(\vec{r})$. Here $\vec{r}$ denotes the position of the centre of each tile. In addition, we assume that the position dependence of $\sqrt{Q^2+U^2}$ may be written on these large scales as $R(\vec{r})I(\vec{r})$, where $I(\vec{r})$ is the familiar Stokes intensity map (also available at high resolution), and $R(\vec{r})$ is a function closely related to the polarisation fraction that is also taken to be constant over each tile. Thus%.  We can hence write
\begin{eqnarray}\label{Eq:QUdedust}
\widehat{U}(\vec{r})&\propto{}&I(\vec{r})R(\vec{r})\sin(2\alpha(\vec{r}))\nonumber\\
\widehat{Q}(\vec{r})&\propto{}&I(\vec{r})R(\vec{r})\cos(2\alpha(\vec{r})).
\end{eqnarray}
%with a scaling factor $R(\vec{x}_0)$ which we take to be constant over each tile.  A simple parametrisation would assume that the amplitude variation of $Q$ and $U$ is purely proportional to $I$, %(i.e.~$R(\vec{x})=1\,\forall\vec{x}$)
We allow $R(\vec{r})$ to vary across the sky following the empirical results of \cite{planck2014-XXX}. Here, we approximate this via the rescaling factor
\begin{eqnarray}\label{Eq:DedustingScaling}
R(\vec{r})=\left(\frac{\mathcal{H}^2(\vec{r})}{\langle{I^4\rangle}_\mathrm{tile}(\vec{r})}\right)^{1/4}
\end{eqnarray}
where the average intensity is taken over $3\degree$ width tiles (as for $\mathcal{H}^2$) and $\langle{I^4}\rangle$ is used since $\mathcal{H}^2$ is a 4-field quantity. The hexadecapole amplitude is a good proxy for the B-mode power (and hence the polarisation level), since it is highly correlated with the monopole amplitude (Sec.~\ref{Sec:Correlations}) and free from bias from other sources including IGWs (Sec.~\ref{Subsec:TensorBias}) and lensing.

We may obtain measurements of $\sin{2\alpha(\vec{r})}$ and $\cos{2\alpha(\vec{r})}$ on each tile from the measured parameters $Af_s$ and $Af_c$ via the relations
\begin{eqnarray}\label{Eq:sin2alphaDedust}
\sin{2\alpha}&=&\pm\frac{\overline{f_s}}{\sqrt{2(\overline{f_c}+1)}}\nonumber\\
\cos{2\alpha}&=&\pm\sqrt{\frac{\overline{f_c}+1}{2}},
\end{eqnarray}
where both expressions have the same unknown sign, and $\overline{f_s}$ and $\overline{f_c}$ represent $Af_s$ and $Af_c$ normalised by $\sqrt{(Af_s)^2+(Af_c)^2}$ (these are equal to $\sin{4\alpha(\vec{r})}$ and $\cos{4\alpha(\vec{r})}$ respectively). Note that the unknown sign of $\overline{f_s}$ and $\overline{f_c}$ is from information loss due to our estimators only being sensitive to the angle $4\alpha_\mathrm{tile}$. This (spatially dependent) ambiguity can be resolved by measuring a cross-correlation with the sky B-mode for each tile. We detail this procedure in App.~\ref{App:dedustAngle}.

With expressions for $R$, $I$ and $2\alpha$ at each $\vec{r}$, we may now construct $Q(\vec{r})$ and $U(\vec{r})$ from Eq.~\ref{Eq:QUdedust} and hence can obtain a dust B-mode map, $\eta \widehat{ B}$, up to an overall constant of proportionality, $\eta$. (Note that before converting to a final B-mode map, we convert our variables from the coarse co-ordinate grid $\vec{r}$ defined on tile centres to a high-resolution co-ordinate system $\vec{x}$; this is detailed in the following subsection.)

We now consider the application and expected performance of this technique. To construct a cleaned B-mode map $B_C(\vec{l})=B(\vec{l})-a_c(\vec{l}) \widehat{B}(\vec{l})$ for true dust B-mode Fourier map $B(\vec{l})$, we must first determine the cleaning coefficient $a_c$ which minimises the cleaned (`dedusted') power-spectrum. This factor $a_c$ will be equal to $\eta$ in the limit of no noise. The power spectrum of the cleaned B-modes will be 
\begin{eqnarray}\label{Eq:CleanPowerDedust}
C_l^{B_CB_C}=C_l^{BB}-2 a_c {}C_l^{B\widehat{B}}+a_c^2C_l^{\widehat{B}\widehat{B}}.
\end{eqnarray}
Minimising $C_l^{B_CB_C}$ with respect to the unknown $a_c$, we obtain $a_c(l) = C_l^{\widehat{B}B}/C_l^{\widehat{B}\widehat{B}}$. With a measurement of the relevant cross- and auto-power spectra, we can thus determine $a_c$ and hence remove a significant fraction of the dust B-modes. Using the standard correlation coefficient $\rho_{B\widehat{B}}(l)$ between fields $B(\vec{l})$ and $\widehat{B}(\vec{l})$ (cf.~Eq.\,\ref{Eq:CrossCorrelation}), the residual dust power spectrum of the cleaned dust map becomes
\begin{eqnarray}\label{Eq:CleanedBMap}
C_l^{B_CB_C}=C_l^{BB}[1-\rho^2_{B\widehat{B}}(l)].
\end{eqnarray}
The reduction in dust B-mode power expected from our cleaning is thus entirely specified by the correlation coefficient, with $\rho_{B\widehat{B}}\rightarrow1$ indicating perfect cleaning. 

\subsection{Application to Simulated Data}
We now test this idea with simulated data. First, we compute the (phase-corrected) angle $2\alpha$ and ratio $R$ as described previously for $3\degree$ width tiles in the 5\% sky region, using CMB-S4 noise and delensing parameters.\footnote{For simplicity, we assume the processes of dedusting and delensing are independent here, i.e. we may dedust an already delensed map.} This gives a set of values of $2\alpha(\vec{r})$ and $R(\vec{r})$ (from tiles separated by $0.5\degree$), which are mapped via linear interpolation onto a \texttt{HEALPix} \texttt{NSIDE} = 128 partial sky grid (as in Sec.~\ref{Sec:Correlations}). The resolutions of these maps are upgraded to \texttt{NSIDE} = 1024 (on a higher resolution $\vec{x}$ coordinate grid, matching the intensity map), and smoothed with a Gaussian kernel of $0.5\degree$ FWHM. (To correctly smooth angular data, we must apply Gaussian smoothing to $\sin{2\alpha}$ and $\cos{2\alpha}$ maps rather than a map of $2\alpha$.) We note that $R(\vec{x})$ is broadly consistent across the region, except for two isolated patches, which can be shown to be high-dust regions. 

Here we apply the dedusting analysis primarily to a 1\% sky region (RA $\in [-60\degree,-20\degree]$, dec $\in [-70\degree, -45\degree]$ in Galactic co-ordinates), which excludes the high-$R$ regions and is small enough that the flat-sky limit is appropriate. From $2\alpha(\vec{x})$ and $R(\vec{x})$, we construct estimates for the Stokes maps $\widehat{Q}(\vec{x})$ and $\widehat{U}(\vec{x})$ from the \citetalias{2017A&A...603A..62V} high resolution intensity map $I(\vec{x})$ (via Eq.~\ref{Eq:QUdedust}), and hence $C_l^{\widehat{B}\widehat{B}}$ using \texttt{flipperPol}'s `hybrid' B-mode estimator. 

\begin{figure*}
\subfigure[Estimated B-mode dust map, $\eta\widehat{B}(\vec{x})\,{[\mathrm{K}]}$]{\includegraphics[width=0.9\linewidth]{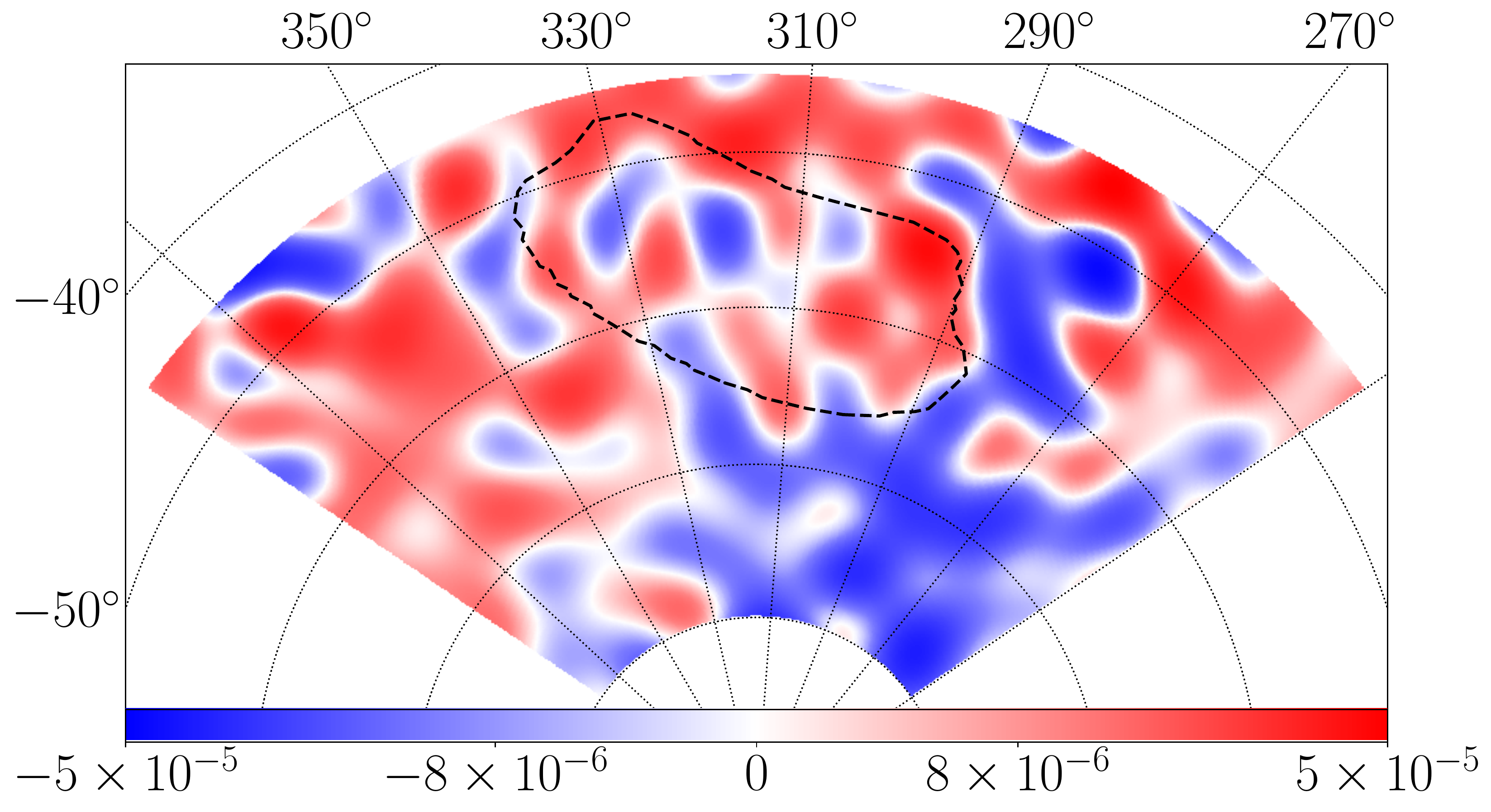}}
\subfigure[True B-mode dust map, $B(\vec{x})\,{[\mathrm{K}]}$]
{\includegraphics[width=0.9\linewidth]{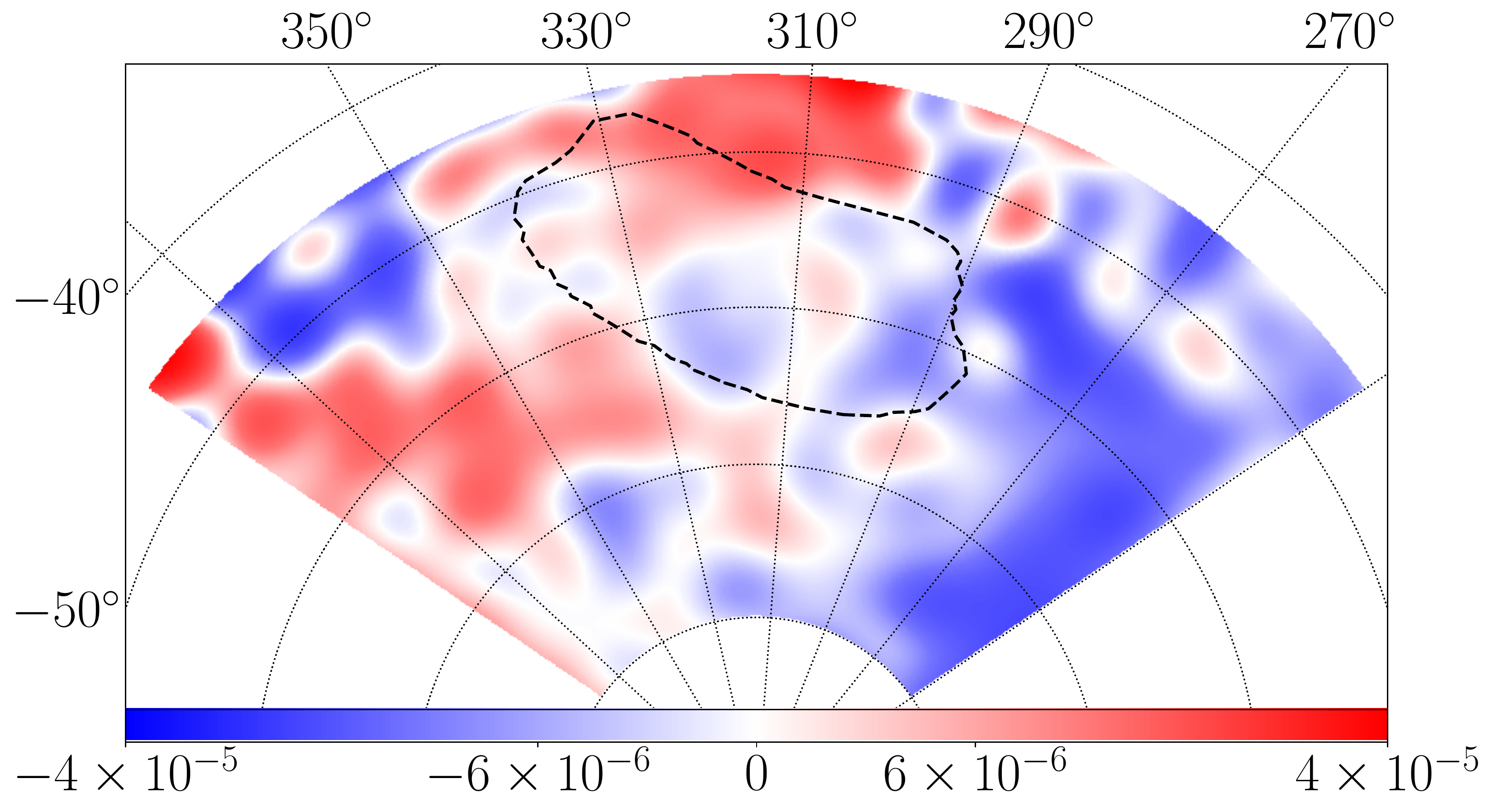}}
\caption{Real-space maps of the (a) estimated and (b) true B-modes from dust, with estimates obtained solely from statistical anisotropy signatures (the hexadecapole parameters) as well as the dust intensity map (Eq.\,\ref{Eq:QUdedust} \& \ref{Eq:DedustingScaling}). $\widehat{B}$ is defined up to an overall scaling factor $\eta$ which can be determined empirically.
The plot uses symmetric logarithmic colour-bars and an Albers equal area conic projection; we filter out modes with $l>60$. The 5\% sky region is shown, with the BICEP patch indicated by a dotted line. We note marked similarities between the estimated and true maps on large scales, demonstrating that we have successfully estimated the dust B-modes from anisotropy statistics alone.}\label{Fig:DedustingMaps}
\end{figure*}

\begin{figure}
\includegraphics[width=\linewidth]{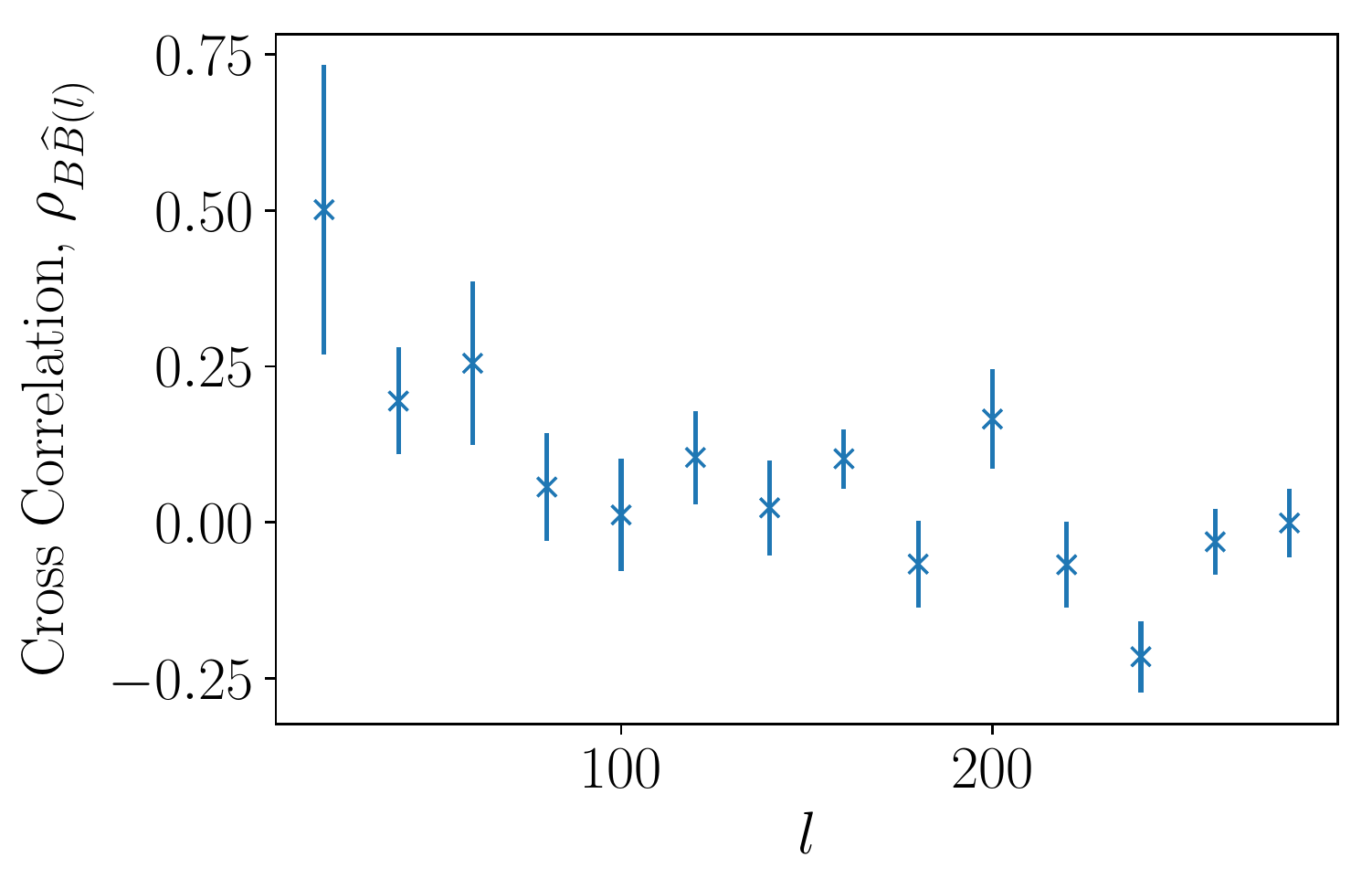}
\caption{Correlation coefficient $\rho_{B\widehat{B}}(l)$ (Eq.\,\ref{Eq:CrossCorrelation}) between estimated ($\widehat{B}$) and true ($B$) dust B-mode maps for the 1\% flat-sky region described in the text. Errors are from the variation in $\rho$ across annular bins in the 2D power-spectra.}\label{Fig:DedustingEstimate}
\end{figure}

A comparison between the estimated and true B-mode dust maps is shown for the larger 5\% region in Fig.~\ref{Fig:DedustingMaps}, with $B(\vec{x})$ computed from the $\widehat{B}_{lm}$ and $B_{lm}$ spherical harmonic coefficients, showing only modes with $l\leq60$. In addition, comparison between the true and estimated B-mode power spectra and the cross-spectrum gives the correlation coefficient $\rho_{B\widehat{B}}$, as plotted in Fig.~\ref{Fig:DedustingEstimate} for the 1\% region.

From the real-space maps, we observe clear correlations between $\widehat{B}(\vec{x})$ and $B(\vec{x})$ on large scales, in agreement with the correlation coefficient, which is markedly non-zero for $l\lesssim80$, with a peak of 0.5 for $l\sim{}20$. Recalling that the cleaned power-spectrum is proportional to $(1-\rho^2_{B\widehat{B}})$, this clearly demonstrates that the dedusting technique has partially worked in this case and would allow the removal of a significant fraction of low-$l$ dust.

It is important to note that this method is very approximate, since we use a constant $2\alpha$ for each tile (an assumption partially alleviated by using overlapping tiles) and assume that $Q$ and $U$ have $\vec{l}$ dependencies represented only by $I(\vec{l})$ and a simple position dependent factor $R(\vec{x})$. The method could be significantly improved through additional constraints from EB and EE power spectra or through a method for computing a continuous angle map (cf.\,\citetalias{2014PhRvL.113s1303K}) instead of the discretely sampled technique used here. However, the fact that we can achieve distinctly non-zero correlation on large scales using our simple method suggests that, with further development, this technique could become useful for the subtraction of dust-induced B-modes for futuristic experiments, complementing usual approaches by using only single-frequency data. An advantage of our method is that no understanding of the dust frequency dependence is required. More generally, detailed knowledge of the dust properties is not needed for our method to work: adding our low-$l$ B-mode template to the bank of multifrequency data used by ILC techniques should yield improved dust mitigation, as long as the correlation coefficient between our B-mode template and the actual B modes is high.

\section{Conclusions}\label{Sec:Conclusion}
Using measurements of the B-mode polarisation, future CMB experiments will be able to place strong bounds on the level of inflationary gravitational waves (IGWs) or detect them for the first time. However, robust measurements of IGWs require the true signal to be separated from thermal dust contamination and other foregrounds. While the canonical method for diagnosing and subtracting dust uses observations at different frequencies for foreground separation, in this paper we have explored complementary single-frequency methods.

We developed and tested methods for detecting dust via B-mode statistical anisotropy, following the suggestions of \cite{2014PhRvL.113s1303K}. Assuming a coherent Galactic magnetic field on few-degree scales, we expect a `hexadecapole' pattern in B-mode power spectra, which we searched for with specialised estimators and Monte Carlo simulations. The technique was applied to realistic simulations of thermal dust emission (predominantly \citealt{2017A&A...603A..62V}) to construct (a) a null-test for the presence of dust contamination on large scales, (b) spatial maps of the dust anisotropy, and (c) a proof-of-concept for a rudimentary technique for dust removal.

% Our methods work by assuming that the Galactic magnetic field is coherent on few-degree scales, such that the polarisation angle is constant across a sufficiently small region of the sky (here taken as $3\degree$). This gives a `hexadecapole' pattern in the 2D power spectra of such regions, which we constructed estimators to detect. These are intrinsically biased estimates, necessitating the creation of isotropic Monte Carlo simulations to calculate the measurement bias. This allows computation of the hexadecapole strength and angle across the full-sky, and was found to be independent of the inclusion of tensor modes.\blake{perhaps shorten /cut - don't need to explain details in conclusions?}

Applying our null test methodology to simulations, we found that such null tests can be a powerful tool for future experiments, giving the ability to detect dust residuals with an amplitude corresponding to a tensor-to-scalar ratio of only 0.001 at a 95\% confidence level, assuming a CMB-S4-like survey. We caution that this result depends on the theoretical assumptions in the dust simulations, as seen in the significant differences when another type of simulation was tested. However, improved future simulation efforts and measurements of polarised emission from dust on small scales are expected to reduce the uncertainty arising from simulation inaccuracy.

We also found strong correlations between the dust monopole and anisotropy fields, with correlation coefficients approaching unity on small scales. This motivated an application of dust anisotropy statistics to the problem of single-frequency dust-subtraction: we derived and tested a rudimentary method for estimating the dust B-mode map from the computed hexadecapole angles and strengths. In simulations, this statistical-anisotropy-derived dust B-mode map was found to match the true B-mode map well on large scales, with significant correlation found.

There is much work to be done in this area in order to increase both the significances of anisotropy detection and the effectiveness of the new dust removal methods. The inclusion of E-mode data may contribute to these efforts, with the extra information that can be extracted from $C_l^{EB}$ and $C_l^{EE}$ spectra potentially allowing lower dust levels to be probed. Furthermore, the discrete sampling of angles used here could be generalised to a continuous field, which would allow greater resolution and the probing of higher multipoles in the 2D power spectra (as suggested by \citealt{2014PhRvL.113s1303K}). Additional work could also involve combining these anisotropy tests with conventional multi-frequency approaches to obtain tighter constraints on residual dust levels and more efficient dust removal.  Finally, although our work has focused on thermal dust emission, very similar anisotropy methods could also be constructed and implemented for synchrotron foreground emission - in fact, our method would presumably capture several types of statistical foreground anisotropy simultaneously.

The techniques presented here give a promising avenue into the detection and potential removal of polarised foregrounds via their anisotropy signature, and could, even in their current form, be used as a powerful null test in forthcoming CMB experiments. 


\section*{Acknowledgements}
The authors would like to thank David Alonso, Anthony Challinor, Colin Hill, Kevin Huffenberger, and Peter Martin for useful feedback. We also thank the anonymous referee for their insightful comments. AvE was supported by the Beatrice and Vincent Tremaine Fellowship at CITA. BDS was supported by an STFC Ernest Rutherford Fellowship and an Isaac Newton Trust Early Career Grant. Some of the results in this paper have been derived using the \texttt{HEALPix} package \citep{2005ApJ...622..759G}. This research used data stored at the National Energy Research Scientific Computing Center, which is supported by the Office of Science of the U.S. Department of Energy.

%%%%%%%%%%%%%%%%%%%%%%%%%%%%%%%%%%%%%%%%%%%%%%%%%%



\bibliographystyle{mnras}
\bibliography{adslib,Planck_bib} % if your bibtex file is called example.bib

\begin{thebibliography}{}
\makeatletter
\relax
\def\mn@urlcharsother{\let\do\@makeother \do\$\do\&\do\#\do\^\do\_\do\%\do\~}
\def\mn@doi{\begingroup\mn@urlcharsother \@ifnextchar [ {\mn@doi@}
  {\mn@doi@[]}}
\def\mn@doi@[#1]#2{\def\@tempa{#1}\ifx\@tempa\@empty \href
  {http://dx.doi.org/#2} {doi:#2}\else \href {http://dx.doi.org/#2} {#1}\fi
  \endgroup}
\def\mn@eprint#1#2{\mn@eprint@#1:#2::\@nil}
\def\mn@eprint@arXiv#1{\href {http://arxiv.org/abs/#1} {{\tt arXiv:#1}}}
\def\mn@eprint@dblp#1{\href {http://dblp.uni-trier.de/rec/bibtex/#1.xml}
  {dblp:#1}}
\def\mn@eprint@#1:#2:#3:#4\@nil{\def\@tempa {#1}\def\@tempb {#2}\def\@tempc
  {#3}\ifx \@tempc \@empty \let \@tempc \@tempb \let \@tempb \@tempa \fi \ifx
  \@tempb \@empty \def\@tempb {arXiv}\fi \@ifundefined
  {mn@eprint@\@tempb}{\@tempb:\@tempc}{\expandafter \expandafter \csname
  mn@eprint@\@tempb\endcsname \expandafter{\@tempc}}}

\bibitem[\protect\citeauthoryear{{Abazajian} et~al.,}{{Abazajian}
  et~al.}{2016}]{2016arXiv161002743A}
{Abazajian} K.~N.,  et~al., 2016, preprint, \href
  {https://ui.adsabs.harvard.edu/#abs/2016arXiv161002743A} {} (\mn@eprint
  {arXiv} {1610.02743})

\bibitem[\protect\citeauthoryear{{BICEP2 Collaboration}}{{BICEP2
  Collaboration}}{2014}]{2014ApJ...792...62B}
{BICEP2 Collaboration} 2014, \mn@doi [\apj] {10.1088/0004-637X/792/1/62}, \href
  {https://ui.adsabs.harvard.edu/#abs/2014ApJ...792...62B} {792}

\bibitem[\protect\citeauthoryear{{BICEP2 Collaboration}}{{BICEP2
  Collaboration}}{2016}]{2016PhRvL.116c1302B}
{BICEP2 Collaboration} 2016, \mn@doi [\prl] {10.1103/PhysRevLett.116.031302},
  \href {https://ui.adsabs.harvard.edu/#abs/2016PhRvL.116c1302B} {116}

\bibitem[\protect\citeauthoryear{{BICEP2/Keck Array and Planck
  Collaborations}}{{BICEP2/Keck Array and Planck
  Collaborations}}{2015}]{pb2015}
{BICEP2/Keck Array and Planck Collaborations} 2015, \mn@doi [\prl]
  {10.1103/PhysRevLett.114.101301}, 114, 101301

\bibitem[\protect\citeauthoryear{{Caldwell}, {Hirata}  \&
  {Kamionkowski}}{{Caldwell} et~al.}{2017}]{2017ApJ...839...91C}
{Caldwell} R.~R.,  {Hirata} C.,   {Kamionkowski} M.,  2017, \mn@doi [\apj]
  {10.3847/1538-4357/aa679c}, \href
  {https://ui.adsabs.harvard.edu/#abs/2017ApJ...839...91C} {839, 91}

\bibitem[\protect\citeauthoryear{{Chon}, {Challinor}, {Prunet}, {Hivon}  \&
  {Szapudi}}{{Chon} et~al.}{2004}]{2004MNRAS.350..914C}
{Chon} G.,  {Challinor} A.,  {Prunet} S.,  {Hivon} E.,   {Szapudi} I.,  2004,
  \mn@doi [\mnras] {10.1111/j.1365-2966.2004.07737.x}, \href
  {https://ui.adsabs.harvard.edu/#abs/2004MNRAS.350..914C} {350, 914}

\bibitem[\protect\citeauthoryear{{Clark}, {Hill}, {Peek}, {Putman}  \&
  {Babler}}{{Clark} et~al.}{2015}]{2015PhRvL.115x1302C}
{Clark} S.~E.,  {Hill} J.~C.,  {Peek} J.~E.~G.,  {Putman} M.~E.,   {Babler}
  B.~L.,  2015, \mn@doi [Physical Review Letters]
  {10.1103/PhysRevLett.115.241302}, \href
  {http://adsabs.harvard.edu/abs/2015PhRvL.115x1302C} {115, 241302}

\bibitem[\protect\citeauthoryear{{Das}, {Hajian}  \& {Spergel}}{{Das}
  et~al.}{2009}]{2009PhRvD..79h3008D}
{Das} S.,  {Hajian} A.,   {Spergel} D.~N.,  2009, \mn@doi [\prd]
  {10.1103/PhysRevD.79.083008}, \href
  {https://ui.adsabs.harvard.edu/#abs/2009PhRvD..79h3008D} {79}

\bibitem[\protect\citeauthoryear{{Draine}}{{Draine}}{2011}]{2011piim.book.....D}
{Draine} B.~T.,  2011, {Physics of the Interstellar and Intergalactic Medium}

\bibitem[\protect\citeauthoryear{{Draine} \& {Fraisse}}{{Draine} \&
  {Fraisse}}{2009}]{2009ApJ...696....1D}
{Draine} B.~T.,  {Fraisse} A.~A.,  2009, \mn@doi [\apj]
  {10.1088/0004-637X/696/1/1}, \href
  {https://ui.adsabs.harvard.edu/#abs/2009ApJ...696....1D} {696}

\bibitem[\protect\citeauthoryear{{G{\'o}rski}, {Hivon}, {Banday}, {Wandelt},
  {Hansen}, {Reinecke}  \& {Bartelmann}}{{G{\'o}rski}
  et~al.}{2005}]{2005ApJ...622..759G}
{G{\'o}rski} K.~M.,  {Hivon} E.,  {Banday} A.~J.,  {Wandelt} B.~D.,  {Hansen}
  F.~K.,  {Reinecke} M.,   {Bartelmann} M.,  2005, \mn@doi [\apj]
  {10.1086/427976}, \href
  {https://ui.adsabs.harvard.edu/#abs/2005ApJ...622..759G} {622, 759}

\bibitem[\protect\citeauthoryear{{Grain}, {Tristram}  \& {Stompor}}{{Grain}
  et~al.}{2009}]{2009PhRvD..79l3515G}
{Grain} J.,  {Tristram} M.,   {Stompor} R.,  2009, \mn@doi [\prd]
  {10.1103/PhysRevD.79.123515}, \href
  {http://adsabs.harvard.edu/abs/2009PhRvD..79l3515G} {79, 123515}

\bibitem[\protect\citeauthoryear{{Hirata} \& {Seljak}}{{Hirata} \&
  {Seljak}}{2003}]{2003PhRvD..68h3002H}
{Hirata} C.~M.,  {Seljak} U.,  2003, \mn@doi [\prd]
  {10.1103/PhysRevD.68.083002}, \href
  {https://ui.adsabs.harvard.edu/#abs/2003PhRvD..68h3002H} {68}

\bibitem[\protect\citeauthoryear{{Hu} \& {Okamoto}}{{Hu} \&
  {Okamoto}}{2002}]{2002ApJ...574..566H}
{Hu} W.,  {Okamoto} T.,  2002, \mn@doi [\apj] {10.1086/341110}, \href
  {https://ui.adsabs.harvard.edu/#abs/2002ApJ...574..566H} {574, 566}

\bibitem[\protect\citeauthoryear{{Kalberla}, {Kerp}, {Haud}, {Winkel}, {Ben
  Bekhti}, {Fl{\"o}er}  \& {Lenz}}{{Kalberla}
  et~al.}{2016}]{2016ApJ...821..117K}
{Kalberla} P.~M.~W.,  {Kerp} J.,  {Haud} U.,  {Winkel} B.,  {Ben Bekhti} N.,
  {Fl{\"o}er} L.,   {Lenz} D.,  2016, \mn@doi [\apj]
  {10.3847/0004-637X/821/2/117}, \href
  {https://ui.adsabs.harvard.edu/#abs/2016ApJ...821..117K} {821, 117}

\bibitem[\protect\citeauthoryear{{Kamionkowski} \& {Kovetz}}{{Kamionkowski} \&
  {Kovetz}}{2014}]{2014PhRvL.113s1303K}
{Kamionkowski} M.,  {Kovetz} E.~D.,  2014, \mn@doi [\prl]
  {10.1103/PhysRevLett.113.191303}, \href
  {https://ui.adsabs.harvard.edu/#abs/2014PhRvL.113s1303K} {113}

\bibitem[\protect\citeauthoryear{{Kamionkowski} \& {Kovetz}}{{Kamionkowski} \&
  {Kovetz}}{2016}]{2016ARA&A..54..227K}
{Kamionkowski} M.,  {Kovetz} E.~D.,  2016, \mn@doi [Annual Review of Astronomy
  and Astrophysics] {10.1146/annurev-astro-081915-023433}, \href
  {https://ui.adsabs.harvard.edu/#abs/2016ARA&A..54..227K} {54, 227}

\bibitem[\protect\citeauthoryear{{Kamionkowski}, {Kosowsky}  \&
  {Stebbins}}{{Kamionkowski} et~al.}{1997}]{1997PhRvD..55.7368K}
{Kamionkowski} M.,  {Kosowsky} A.,   {Stebbins} A.,  1997, \mn@doi [\prd]
  {10.1103/PhysRevD.55.7368}, \href
  {https://ui.adsabs.harvard.edu/#abs/1997PhRvD..55.7368K} {55, 7368}

\bibitem[\protect\citeauthoryear{{Kesden}, {Cooray}  \&
  {Kamionkowski}}{{Kesden} et~al.}{2002}]{2002PhRvL..89a1304K}
{Kesden} M.,  {Cooray} A.,   {Kamionkowski} M.,  2002, \mn@doi [\prl]
  {10.1103/PhysRevLett.89.011304}, \href
  {https://ui.adsabs.harvard.edu/#abs/2002PhRvL..89a1304K} {89}

\bibitem[\protect\citeauthoryear{{Knox}}{{Knox}}{1995}]{1995PhRvD..52.4307K}
{Knox} L.,  1995, \mn@doi [\prd] {10.1103/PhysRevD.52.4307}, \href
  {https://ui.adsabs.harvard.edu/#abs/1995PhRvD..52.4307K} {52, 4307}

\bibitem[\protect\citeauthoryear{{Knox} \& {Song}}{{Knox} \&
  {Song}}{2002}]{2002PhRvL..89a1303K}
{Knox} L.,  {Song} Y.-S.,  2002, \mn@doi [\prl]
  {10.1103/PhysRevLett.89.011303}, \href
  {https://ui.adsabs.harvard.edu/#abs/2002PhRvL..89a1303K} {89}

\bibitem[\protect\citeauthoryear{{Komatsu}, {Wandelt}, {Spergel}, {Banday}  \&
  {G{\'o}rski}}{{Komatsu} et~al.}{2002}]{2002ApJ...566...19K}
{Komatsu} E.,  {Wandelt} B.~D.,  {Spergel} D.~N.,  {Banday} A.~J.,
  {G{\'o}rski} K.~M.,  2002, \mn@doi [\apj] {10.1086/337963}, \href
  {https://ui.adsabs.harvard.edu/#abs/2002ApJ...566...19K} {566, 19}

\bibitem[\protect\citeauthoryear{{Kovetz} \& {Kamionkowski}}{{Kovetz} \&
  {Kamionkowski}}{2015}]{2015PhRvD..91h1303K}
{Kovetz} E.~D.,  {Kamionkowski} M.,  2015, \mn@doi [\prd]
  {10.1103/PhysRevD.91.081303}, \href
  {https://ui.adsabs.harvard.edu/#abs/2015PhRvD..91h1303K} {91}

\bibitem[\protect\citeauthoryear{{Krauss}, {Dodelson}  \& {Meyer}}{{Krauss}
  et~al.}{2010}]{2010Sci...328..989K}
{Krauss} L.~M.,  {Dodelson} S.,   {Meyer} S.,  2010, \mn@doi [Science]
  {10.1126/science.1179541}, \href
  {https://ui.adsabs.harvard.edu/#abs/2010Sci...328..989K} {328, 989}

\bibitem[\protect\citeauthoryear{{Leach} et~al.,}{{Leach}
  et~al.}{2008}]{2008A&A...491..597L}
{Leach} S.~M.,  et~al., 2008, \mn@doi [\aap] {10.1051/0004-6361:200810116},
  \href {https://ui.adsabs.harvard.edu/#abs/2008A&A...491..597L} {491, 597}

\bibitem[\protect\citeauthoryear{{Lewis}, {Challinor}  \& {Lasenby}}{{Lewis}
  et~al.}{2000}]{2000ApJ...538..473L}
{Lewis} A.,  {Challinor} A.,   {Lasenby} A.,  2000, \mn@doi [\apj]
  {10.1086/309179}, \href
  {https://ui.adsabs.harvard.edu/#abs/2000ApJ...538..473L} {538, 473}

\bibitem[\protect\citeauthoryear{{Louis}, {N{\ae}ss}, {Das}, {Dunkley}  \&
  {Sherwin}}{{Louis} et~al.}{2013}]{2013MNRAS.435.2040L}
{Louis} T.,  {N{\ae}ss} S.,  {Das} S.,  {Dunkley} J.,   {Sherwin} B.,  2013,
  \mn@doi [\mnras] {10.1093/mnras/stt1421}, \href
  {https://ui.adsabs.harvard.edu/#abs/2013MNRAS.435.2040L} {435, 2040}

\bibitem[\protect\citeauthoryear{{Manzotti} et~al.,}{{Manzotti}
  et~al.}{2017}]{2017ApJ...846...45M}
{Manzotti} A.,  et~al., 2017, \mn@doi [\apj] {10.3847/1538-4357/aa82bb}, \href
  {https://ui.adsabs.harvard.edu/#abs/2017ApJ...846...45M} {846}

\bibitem[\protect\citeauthoryear{{Martin}, {Blagrave}, {Lockman}, {Pinheiro
  Gon{\c{c}}alves}, {Boothroyd}, {Joncas}, {Miville-Desch{\^e}nes}  \&
  {Stephan}}{{Martin} et~al.}{2015}]{2015ApJ...809..153M}
{Martin} P.~G.,  {Blagrave} K.~P.~M.,  {Lockman} F.~J.,  {Pinheiro
  Gon{\c{c}}alves} D.,  {Boothroyd} A.~I.,  {Joncas} G.,
  {Miville-Desch{\^e}nes} M.~A.,   {Stephan} G.,  2015, \mn@doi [\apj]
  {10.1088/0004-637X/809/2/153}, \href
  {https://ui.adsabs.harvard.edu/#abs/2015ApJ...809..153M} {809, 153}

\bibitem[\protect\citeauthoryear{{Namikawa} \& {Takahashi}}{{Namikawa} \&
  {Takahashi}}{2014}]{2014MNRAS.438.1507N}
{Namikawa} T.,  {Takahashi} R.,  2014, \mn@doi [\mnras]
  {10.1093/mnras/stt2290}, \href
  {https://ui.adsabs.harvard.edu/#abs/2014MNRAS.438.1507N} {438, 1507}

\bibitem[\protect\citeauthoryear{{Planck Collaboration}}{{Planck
  Collaboration}}{2018}]{2018arXiv180104945P}
{Planck Collaboration} 2018, preprint, \href
  {https://ui.adsabs.harvard.edu/#abs/2018arXiv180104945P} {} (\mn@eprint
  {arXiv} {1801.04945})

\bibitem[\protect\citeauthoryear{{\sorthelp{Planck Collaboration 2014I}}{Planck
  Collaboration IX}}{{\sorthelp{Planck Collaboration 2014I}}{Planck
  Collaboration IX}}{2014}]{planck2013-p03d}
{\sorthelp{Planck Collaboration 2014I}}{Planck Collaboration IX} 2014, \mn@doi
  [\aap] {10.1051/0004-6361/201321531}, 571, A9

\bibitem[\protect\citeauthoryear{{\sorthelp{Planck Collaboration 2014V}}{Planck
  Collaboration XXII}}{{\sorthelp{Planck Collaboration 2014V}}{Planck
  Collaboration XXII}}{2014}]{planck2013-p17}
{\sorthelp{Planck Collaboration 2014V}}{Planck Collaboration XXII} 2014,
  \mn@doi [\aap] {10.1051/0004-6361/201321569}, 571, A22

\bibitem[\protect\citeauthoryear{{\sorthelp{Planck Collaboration 2015J}}{Planck
  Collaboration X}}{{\sorthelp{Planck Collaboration 2015J}}{Planck
  Collaboration X}}{2016}]{planck2014-a12}
{\sorthelp{Planck Collaboration 2015J}}{Planck Collaboration X} 2016, \mn@doi
  [\aap] {10.1051/0004-6361/201525967}, 594, A10

\bibitem[\protect\citeauthoryear{{\sorthelp{Planck Collaboration 2015L}}{Planck
  Collaboration XII}}{{\sorthelp{Planck Collaboration 2015L}}{Planck
  Collaboration XII}}{2016}]{planck2014-a14}
{\sorthelp{Planck Collaboration 2015L}}{Planck Collaboration XII} 2016, \mn@doi
  [\aap] {10.1051/0004-6361/201527103}, 594, A12

\bibitem[\protect\citeauthoryear{{\sorthelp{Planck Collaboration IntS}}{Planck
  Collaboration Int. XIX}}{{\sorthelp{Planck Collaboration IntS}}{Planck
  Collaboration Int. XIX}}{2015}]{planck2014-XIX}
{\sorthelp{Planck Collaboration IntS}}{Planck Collaboration Int. XIX} 2015,
  \mn@doi [\aap] {10.1051/0004-6361/201424082}, 576, A104

\bibitem[\protect\citeauthoryear{{\sorthelp{Planck Collaboration IntV}}{Planck
  Collaboration Int. XXII}}{{\sorthelp{Planck Collaboration IntV}}{Planck
  Collaboration Int. XXII}}{2015}]{planck2014-XXII}
{\sorthelp{Planck Collaboration IntV}}{Planck Collaboration Int. XXII} 2015,
  \mn@doi [\aap, submitted] {10.1051/0004-6361/201424088}, 576, A107

\bibitem[\protect\citeauthoryear{{\sorthelp{Planck Collaboration IntZE}}{Planck
  Collaboration Int. XXX}}{{\sorthelp{Planck Collaboration IntZE}}{Planck
  Collaboration Int. XXX}}{2016}]{planck2014-XXX}
{\sorthelp{Planck Collaboration IntZE}}{Planck Collaboration Int. XXX} 2016,
  \mn@doi [\aap] {10.1051/0004-6361/201425034}, 586, A133

\bibitem[\protect\citeauthoryear{{\sorthelp{Planck Collaboration IntZM}}{Planck
  Collaboration Int. XXXVIII}}{{\sorthelp{Planck Collaboration IntZM}}{Planck
  Collaboration Int. XXXVIII}}{2016}]{planck2015-XXXVIII}
{\sorthelp{Planck Collaboration IntZM}}{Planck Collaboration Int. XXXVIII}
  2016, \mn@doi [\aap] {10.1051/0004-6361/201526506}, 586, A141

\bibitem[\protect\citeauthoryear{{\sorthelp{Planck Collaboration IntZS}}{Planck
  Collaboration Int. XLIV}}{{\sorthelp{Planck Collaboration IntZS}}{Planck
  Collaboration Int. XLIV}}{2016}]{planck2016-XLIV}
{\sorthelp{Planck Collaboration IntZS}}{Planck Collaboration Int. XLIV} 2016,
  \mn@doi [\aap] {10.1051/0004-6361/201628636}, 596, A105

\bibitem[\protect\citeauthoryear{{\sorthelp{Planck Collaboration IntZW}}{Planck
  Collaboration Int. XLVIII}}{{\sorthelp{Planck Collaboration IntZW}}{Planck
  Collaboration Int. XLVIII}}{2016}]{planck2016-XLVIII}
{\sorthelp{Planck Collaboration IntZW}}{Planck Collaboration Int. XLVIII} 2016,
  \mn@doi [\aap] {10.1051/0004-6361/201629022}, 596, A109

\bibitem[\protect\citeauthoryear{{Rotti} \& {Huffenberger}}{{Rotti} \&
  {Huffenberger}}{2016}]{2016JCAP...09..034R}
{Rotti} A.,  {Huffenberger} K.,  2016, \mn@doi [Journal of Cosmology and
  Astro-Particle Physics] {10.1088/1475-7516/2016/09/034}, \href
  {https://ui.adsabs.harvard.edu/#abs/2016JCAP...09..034R} {2016}

\bibitem[\protect\citeauthoryear{{Sherwin} \& {Schmittfull}}{{Sherwin} \&
  {Schmittfull}}{2015}]{2015PhRvD..92d3005S}
{Sherwin} B.~D.,  {Schmittfull} M.,  2015, \mn@doi [\prd]
  {10.1103/PhysRevD.92.043005}, \href
  {https://ui.adsabs.harvard.edu/#abs/2015PhRvD..92d3005S} {92}

\bibitem[\protect\citeauthoryear{{Smith}, {Hanson}, {LoVerde}, {Hirata}  \&
  {Zahn}}{{Smith} et~al.}{2012}]{2012JCAP...06..014S}
{Smith} K.~M.,  {Hanson} D.,  {LoVerde} M.,  {Hirata} C.~M.,   {Zahn} O.,
  2012, \mn@doi [Journal of Cosmology and Astro-Particle Physics]
  {10.1088/1475-7516/2012/06/014}, \href
  {https://ui.adsabs.harvard.edu/#abs/2012JCAP...06..014S} {2012}

\bibitem[\protect\citeauthoryear{{Suzuki} et~al.,}{{Suzuki}
  et~al.}{2016}]{2016JLTP..184..805S}
{Suzuki} A.,  et~al., 2016, \mn@doi [Journal of Low Temperature Physics]
  {10.1007/s10909-015-1425-4}, \href
  {https://ui.adsabs.harvard.edu/#abs/2016JLTP..184..805S} {184, 805}

\bibitem[\protect\citeauthoryear{{Tegmark} \& {Efstathiou}}{{Tegmark} \&
  {Efstathiou}}{1996}]{1996MNRAS.281.1297T}
{Tegmark} M.,  {Efstathiou} G.,  1996, \mn@doi [\mnras]
  {10.1093/mnras/281.4.1297}, \href
  {https://ui.adsabs.harvard.edu/#abs/1996MNRAS.281.1297T} {281, 1297}

\bibitem[\protect\citeauthoryear{{Thorne}, {Dunkley}, {Alonso}  \&
  {N{\ae}ss}}{{Thorne} et~al.}{2017}]{2017MNRAS.469.2821T}
{Thorne} B.,  {Dunkley} J.,  {Alonso} D.,   {N{\ae}ss} S.,  2017, \mn@doi
  [\mnras] {10.1093/mnras/stx949}, \href
  {https://ui.adsabs.harvard.edu/#abs/2017MNRAS.469.2821T} {469, 2821}

\bibitem[\protect\citeauthoryear{{Tibbs}, {Paladini}  \& {Dickinson}}{{Tibbs}
  et~al.}{2012}]{2012AdAst2012E..41T}
{Tibbs} C.~T.,  {Paladini} R.,   {Dickinson} C.,  2012, \mn@doi [Advances in
  Astronomy] {10.1155/2012/124931}, \href
  {https://ui.adsabs.harvard.edu/#abs/2012AdAst2012E..41T} {2012}

\bibitem[\protect\citeauthoryear{{Vansyngel} et~al.,}{{Vansyngel}
  et~al.}{2017}]{2017A&A...603A..62V}
{Vansyngel} F.,  et~al., 2017, \mn@doi [\aap] {10.1051/0004-6361/201629992},
  \href {https://ui.adsabs.harvard.edu/#abs/2017A&A...603A..62V} {603}

\bibitem[\protect\citeauthoryear{{Zaldarriaga}, {Spergel}  \&
  {Seljak}}{{Zaldarriaga} et~al.}{1997}]{1997ApJ...488....1Z}
{Zaldarriaga} M.,  {Spergel} D.~N.,   {Seljak} U.,  1997, \mn@doi [\apj]
  {10.1086/304692}, \href
  {https://ui.adsabs.harvard.edu/#abs/1997ApJ...488....1Z} {488, 1}

\makeatother
\end{thebibliography}
%%%%%%%%%%%%%%%%%%%%%%%%%%%%%%%%%%%%%%%%%%%%%%%%%%

%%%%%%%%%%%%%%%%% APPENDICES %%%%%%%%%%%%%%%%%%%%%

\appendix
\section{Correcting for the Intrinsic Bias of $H^2$}\label{Subsec:H2Bias}
\newcommand{\mean}[1]{\left\langle{#1}\right\rangle}
We here derive the $H^2$ bias term used in the computation of $\mathcal{H}^2$ in Eq.~\ref{Eq:DebiasedH2}. To do this, we first consider the generalised estimator for $H^2$ obtained from two B-mode maps $X$ and $Y$. This will allow us to compute the realisation-dependent bias using $H^2$ estimates obtained both from MC maps and from their cross-spectra with the mock data.

\subsection{Deriving a Generalised Estimator for $H^2$}
Consider Gaussian Random Fields (GRFs) $x^{X}(\vec{l})$ where $x$ represents the Fourier-space component of the B-mode CMB map with label $X$ representing either MC simulations or mock data ($x$ is treated as a real field for simplicity). From the definition of the power spectra, $\mean{|x^{X}(\vec{l}_i)x^{Y}(\vec{l}_j)|} = C_l^{XY}\delta_{ij}$ for Kronecker delta $\delta_{ij}$. For the cross spectrum $X\bigotimes{}Y$, we have the generalised estimator for $Af_s$ given two maps $X$ and $Y$:
\begin{eqnarray}
\left(\widehat{Af_s}\right)_{XY}=\frac{1}{\mathcal{N}}\sum_{\vec{l}}\frac{|x_{\vec{l}}^{X}x_{\vec{l}}^{Y}|}{C_l^\mathrm{fid}}\Lambda^2_l\sin{4\phi_{\vec{l}}}
\end{eqnarray}
 (cf.\,Eq.\,\ref{Eq:estimators}) where the normalisation factor is $\mathcal{N}=\sum_{\vec{l}}\Lambda_l^2\sin^2{4\phi_{\vec{l}}}$. ($\widehat{Af_c}$ folllows by symmetry.) To compute the \textit{isotropic} power, $H^2_\mathrm{iso}$, from the two maps we use
 \begin{eqnarray}
 \widehat{H^2}_{\mathrm{iso}\,,XY}=(\widehat{Af_s})_{XY}^2+(\widehat{Af_c})^2_{XY}
 \end{eqnarray}
(cf.\,Eq.\,\ref{Eq: HexPow}) and note that $\mean{Af_s}=\mean{Af_c}=0$. By rotational invariance, the mean values satisfy $\mean{Af_s^2}_{XY}=\mean{Af_c^2}_{XY}$, thus giving the expectation value for $H^2_{\mathrm{iso},XY}$;
\begin{eqnarray}
\mean{H^2_\mathrm{iso}}_{XY}=\frac{2}{\mathcal{N}^2}\sum_{\vec{l}_1}\sum_{\vec{l}_2}\frac{\mean{|x_1^Xx_1^Yx_2^Xx_2^Y|}}{(C_l^\mathrm{fid})^2}\Lambda^2_1\Lambda^2_2\sin{4\phi_1}\sin{4\phi_2}
\end{eqnarray}
where a subscript $i$ is understood to mean $\vec{l}_i$. We proceed via Wick's theorem to expand the GRF term;
\begin{eqnarray}
\mean{|x_1^Xx_1^Yx_2^Xx_2^Y|}&=&\mean{|x_1^Xx_1^Y|}\mean{|x_2^Xx_2^Y|}+\mean{|x_1^Xx_2^X|}\mean{|x_1^Yx_2^Y|}\nonumber\\
&+&\mean{|x_1^Xx_2^Y|}\mean{|x_1^Yx_2^X|}\nonumber\\
&=&C_{l_1}^{XY}C_{l_2}^{XY} + \delta_{12}C_{l_1}^{XX}C_{l_1}^{YY}+\nonumber\\
&+&\delta_{12}C_{l_1}^{XY}C_{l_1}^{XY}.
\end{eqnarray}

Inserting these into the expression for $\mean{H^2_\mathrm{iso}}_{XY}$ we note that the first term vanishes, since the summations are separable and each involves an angular sum over $\sin{4\phi_{\vec{l}}}$, which is zero by isotropy. Applying the Kronecker delta, we are left with a general expression for the isotropic bias;
\begin{eqnarray}\label{Eq:BiasExpression}
\mean{H^2_\mathrm{iso}}_{XY}=\frac{2}{\mathcal{N}^2}\sum_{\vec{l}}\frac{[C_l^{XX}C_l^{YY}+(C_l^{XY})^2]}{(C_l^f)^2}\Lambda_l^4\sin^2{4\phi_{\vec{l}}}.
\end{eqnarray}

\subsection{Application to Mock Data and Simulated Maps}
We now derive expressions for the expected value of $\mean{H^2_\mathrm{iso}}$ for the isotropic part of the mock data (D), the simulations (S) and for a cross-correlation of data and simulations (D $\bigotimes$ S). 

Given a data power spectrum $C_l^{DD}$ we may assume that our MC simulations are created to a reasonable level of accuracy and can thus be expressed as $C_l^{SS}$=$C_l^{DD}(1+\Delta_l)$ where $\Delta_l<1$ quantifies the departure from the true 1D spectrum. Using the generalised expectation expression of Equation \ref{Eq:BiasExpression} to first order in $\Delta_l$, the mean hexadecapole powers become
\begin{eqnarray}
\mean{H^2_\mathrm{iso}}_{DD}&=&\frac{4}{\mathcal{N}^2}\sum_{\vec{l}}\left(\frac{C_l^{DD}}{C_l^\mathrm{fid}}\right)^2\Lambda_l^4\sin^2{4\phi_{\vec{l}}}\nonumber\\
\mean{H^2_\mathrm{iso}}_{SS}&=&\frac{4}{\mathcal{N}^2}\sum_{\vec{l}}(1+2\Delta_l)\left(\frac{C_l^{DD}}{C_l^\mathrm{fid}}\right)^2\Lambda_l^4\sin^2{4\phi_{\vec{l}}}+\mathcal{O}(\Delta_l^2)\nonumber\\
\mean{H^2_\mathrm{iso}}_{DS}&=&\frac{2}{\mathcal{N}^2}\sum_{\vec{l}}\frac{[(1+\Delta_l)(C_l^{DD})^2+(C_l^{DS})^2]}{(C_l^\mathrm{fid})^2}\Lambda_l^4\sin^2{4\phi_{\vec{l}}}.
\end{eqnarray}
Furthermore, since the data and simulations are created using different Gaussian realisations of approximately the same spectrum, they will be uncorrelated on average (since their random phases will lead to cancellation), hence $C_l^{DS} = 0$. Combining the expressions for $\mean{H^2_\mathrm{iso}}_{SS}$ and $\mean{H^2_\mathrm{iso}}_{DS}$ we can eliminate the $\mathcal{O}(\Delta_l)$ term and are hence left with the expression;
\begin{eqnarray}
\mean{H^2_\mathrm{iso}}_{DD} = 4\mean{H^2_\mathrm{iso}}_{DS}-\mean{H^2_\mathrm{iso}}_{SS} + \mathcal{O}(\Delta_l^2).
\end{eqnarray}
This gives us an expression for the mean isotropic hexadecapole bias which is free from first order errors due to a poor implementation of the MC model. To compute further corrections, we would need to compute higher-order correlation functions, which is not necessary here. 

\section{Probability Distributions for the Hexadecapole Parameters}\label{Appen:Stats}
\subsection{Analytic Distributions}\label{Appen:AnalyticPDFs}
For an isotropic map, we expect $A$, $Af_s$ and $Af_c$ to be Gaussian distributed with the latter variables having zero mean and, by symmetry, equal variances (denoted $\sigma_{Af}$). This follows from the central limit theorem, since they are computed from an appropriately weighted sum over many stochastic pixels (whose Fourier space components are $\chi^2$-distributed). Since $Af_s$ and $Af_c$ describe the B-mode power with respect to two orthogonal basis functions ($\sin{4\phi_{\vec{l}}}$ and $\cos{4\phi_{\vec{l}}}$), they are here assumed to be independently distributed. From this, we note that the angle $\alpha$ obtained from MC simulations is expected to be uniformly distributed in $(-45,45]$ degrees. Its expected error, $\sigma_\alpha$, may thus be computed via
\begin{eqnarray}\label{Eq: AngleError}
\centering
\sigma_\alpha &=& \left[\left(\frac{\partial{}\alpha}{\partial{}Af_s}\sigma_{Af_s}\right)^2+\left(\frac{\partial{}\alpha}{\partial{}Af_c}\sigma_{Af_c}\right)^2\right]^{1/2}\nonumber\\
&=& \frac{\sigma_{Af}}{4\sqrt{Af_s^2+Af_c^2}}\frac{180}{\pi},
\end{eqnarray}
where the factor of $180/\pi$ is to convert the angle to degrees. 

To derive the statistics of $\mathcal{H}^2$ for the MC simulations, first consider
\begin{eqnarray}\label{Eq:Chi2Param}
Q = \frac{\mathcal{H}^2+\langle{H^2}\rangle_\mathrm{bias,\,iso}}{\sigma_{Af}^2 W_\mathrm{corr}} = \left(\frac{Af_s}{\sigma_{Af_s}}\right)^2+\left(\frac{Af_c}{\sigma_{Af_c}}\right)^2,
\end{eqnarray}
where $W_\mathrm{corr}$ is the window-correction factor $\langle{W^2}\rangle^2/\langle{W^4}\rangle$ and $\langle{H^2}\rangle_\mathrm{bias,\,iso}$ is the isotropic simulation bias. Being the sum of two independent uniform Gaussians, $Q$ is distributed according to the $\chi^2$-distribution with two degrees of freedom and hence has probability density function (PDF) $f(Q) = \frac{1}{2}e^{-Q/2}$. This implies that the PDF and cumulative density function (CDF) for a measurement of $\mathcal{H}^2$ are
\begin{eqnarray}\label{Eq:Chi2Prob}
f(\mathcal{H}^2)&=&\frac{1}{2\sigma^2_{Af}W_\mathrm{corr}}\exp\left(-\frac{\mathcal{H}^2+\langle{H^2}\rangle_\mathrm{bias,\,iso}}{2\sigma_{Af}^2W_\mathrm{corr}}\right) \nonumber\\
p(\mathcal{H}^2) &=& 1 - \exp\left(-\frac{\mathcal{H}^2+\langle{H^2}\rangle_\mathrm{bias,\,iso}}{2\sigma_{Af}^2W_\mathrm{corr}}\right),
\end{eqnarray}
respectively. The analytic distribution $p(\mathcal{H}^2)$ can be used to parametrise the probability that a particular tile is anisotropic for runs where it is not feasible to create many simulations, since only $\sigma_{Af}$ need be computed. 
\subsection{Testing the Model Statistics}\label{Subsec:StatsTests}
\begin{figure}\centering
\subfigure{\includegraphics[width=0.8\linewidth]{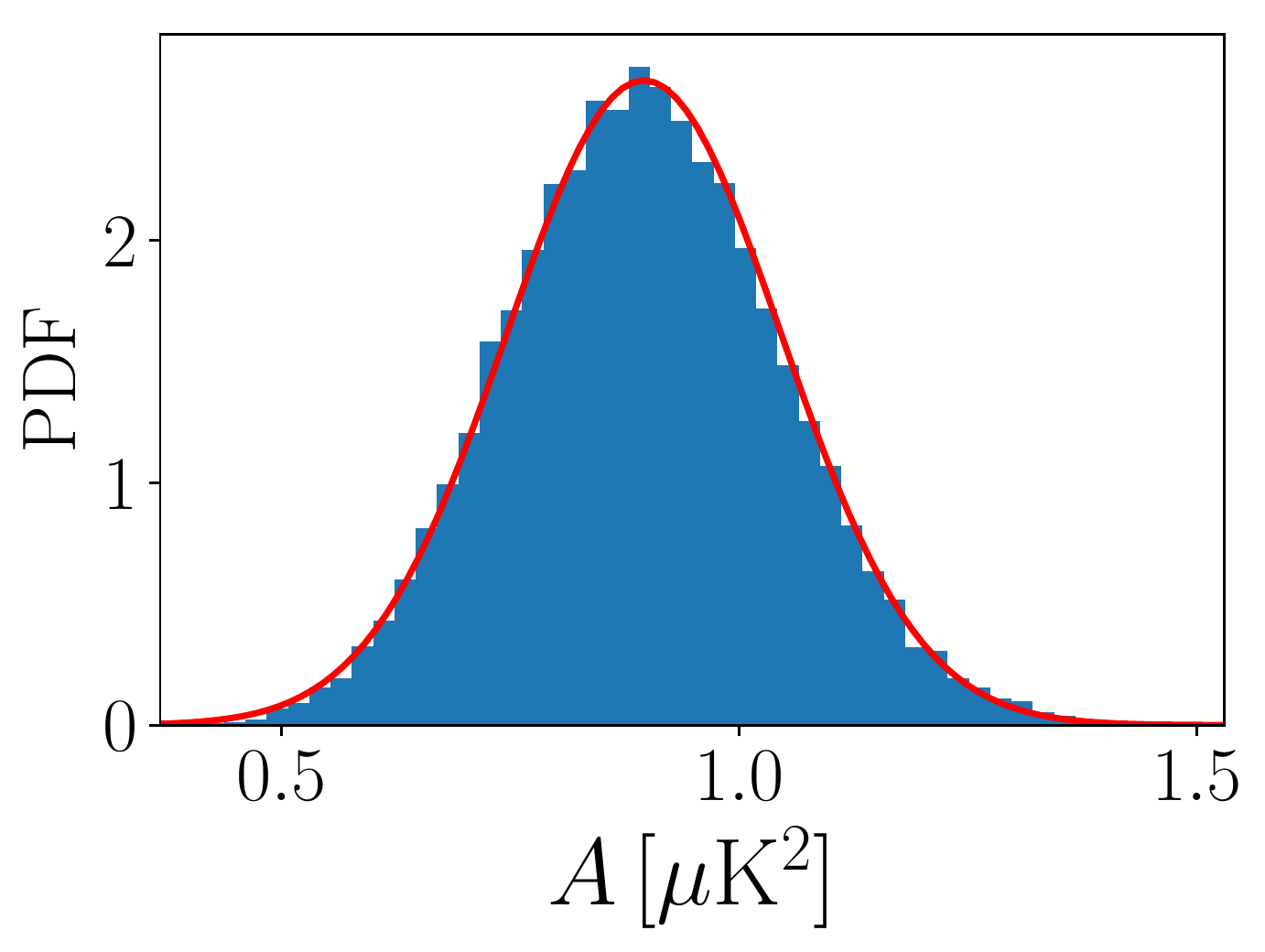}}
\subfigure{
\includegraphics[width=0.8\linewidth]{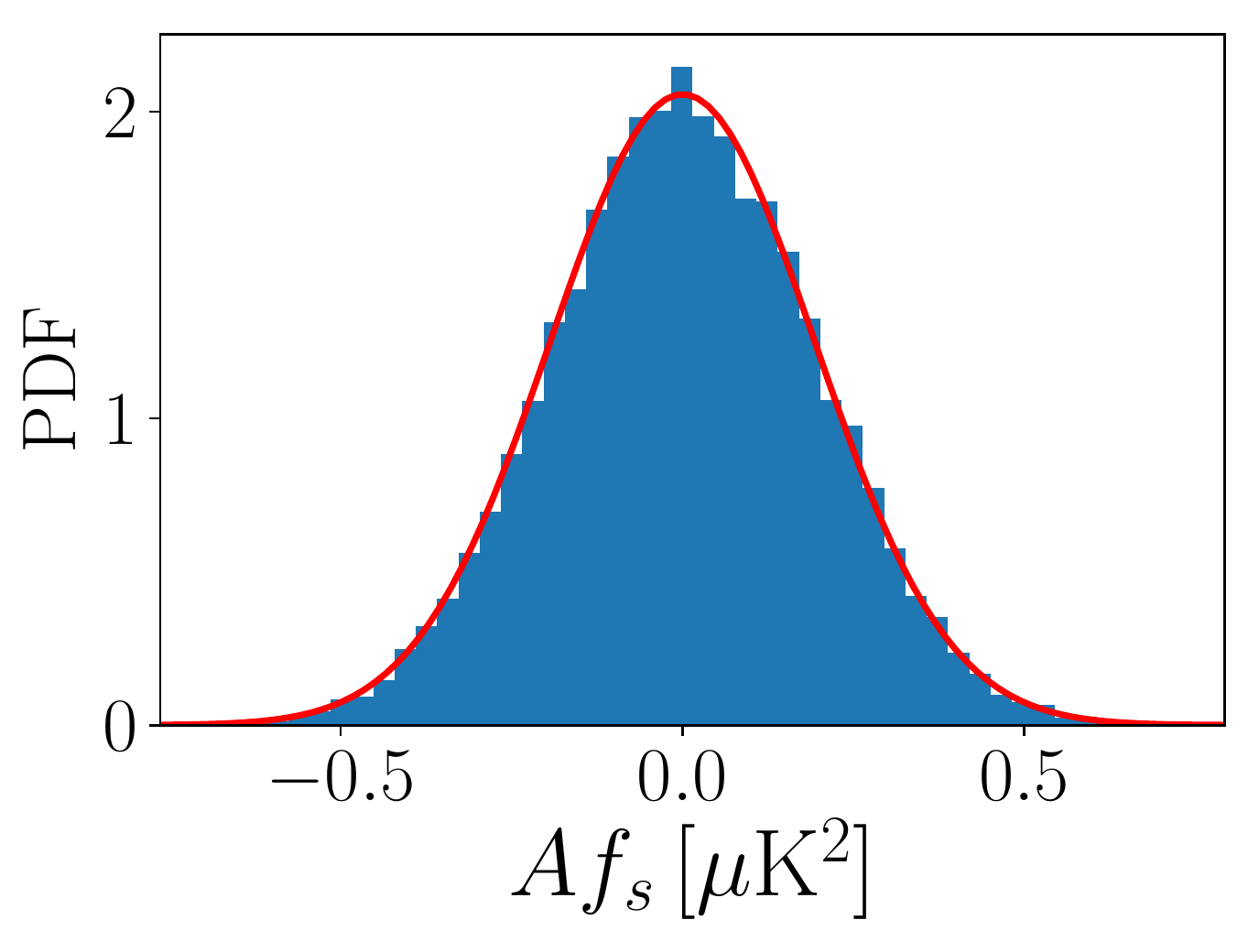}}
\subfigure{
\includegraphics[width=0.8\linewidth]{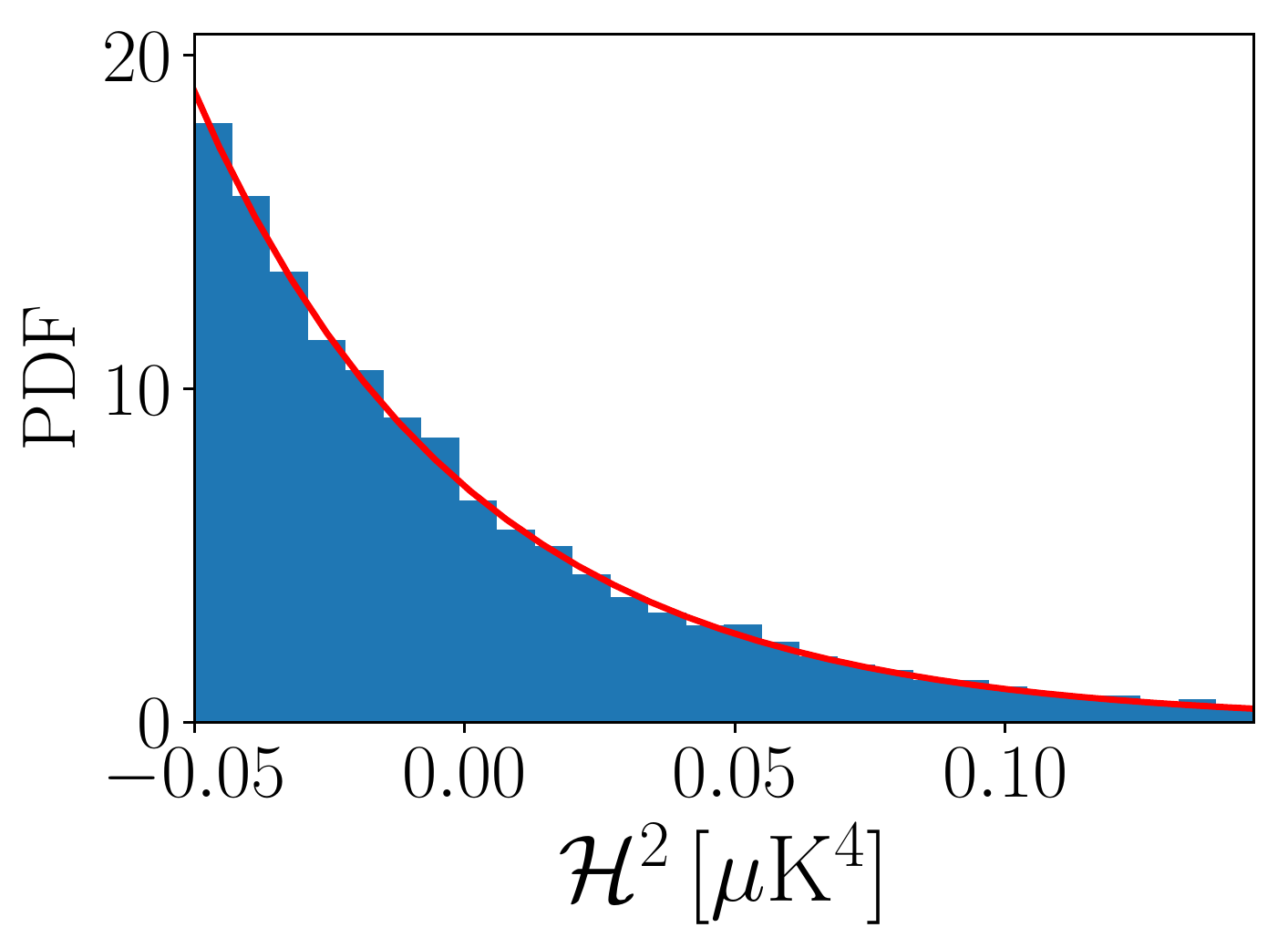}}
\caption{Histograms of the probability distribution function (PDF) for $A$, $Af_s$ and $\mathcal{H}^2$, as labelled, obtained from $5\times{}10^5$ random isotropic MC simulations of a specific tile within the BICEP patch. The data is fit with Gaussians for $A$ and $Af_s$ (the latter with zero-mean) with variances estimated from the data. For $\mathcal{H}^2$, the expected PDF (Eq.\,\ref{Eq:Chi2Prob}) is plotted which closely represents the fit.}\label{Fig:hists}
\end{figure}

To test our assumptions regarding the expected distributions of the monopole and hexadecapole parameters (appendix \ref{Appen:AnalyticPDFs}), we apply the hexadecapole estimators (Eq.\,\ref{Eq:estimators}) to $5\times{}10^5$ isotropic simulations of a representative tile from the 5\% sky region, using CMB-S4 noise and lensing parameters, as shown in Fig.~\ref{Fig:hists}.

$A$ and $Af_s$ are well represented by Gaussian curves (shown in red) with variances drawn from the data and zero-mean for $Af_s$. Furthermore, the analytic PDF for $\mathcal{H}^2$ (Eq.\,\ref{Eq:Chi2Prob}) well represents this dataset, using the mean of $\sigma_{Af_s}$ and $\sigma_{Af_c}$ as the only distribution shape parameter. This hence supports our assumptions of Gaussianity in the parameters $A$, $Af_s$ and $Af_c$, and we note that the analytic PDF may be used interchangeably with the statistical percentile described in the text.

\section{Bias from Noise Approximations}\label{Appen:NoiseBias}
Throughout the above, we have assumed that the effects of noise may be incorporated solely via single-tile Gaussian realisations of the 1D noise power spectrum (Equation \ref{Eq:noiseModel}). Here we consider the biases resulting from this assumption by reapplying our analysis pipeline to the case where noise is generated instead from a full-sky map, which will include correlations on larger scales than the tile widths. 

Using \texttt{HEALPix}, we created full-sky Q and U Stokes maps from a random realisation of the noise power spectra (via the \texttt{synfast} procedure). These were cut-out and transformed into Fourier space to compute a B-mode power spectrum for each tile as for the lensing modes, again utilising zero-padding. Comparison with the input $C_l$ showed that the correct spectra were indeed reproduced. The computed $B_\mathrm{noise}(\vec{l})$ map was combined with the lensing and dust maps to give the `data', which is a good approximation to a real experimental dataset. The remaining analysis proceeds as before, with the MC simulations being generated from the isotropic 1D power spectrum of the combined map. 

\begin{figure}
  \centering
  \includegraphics[width=\linewidth]{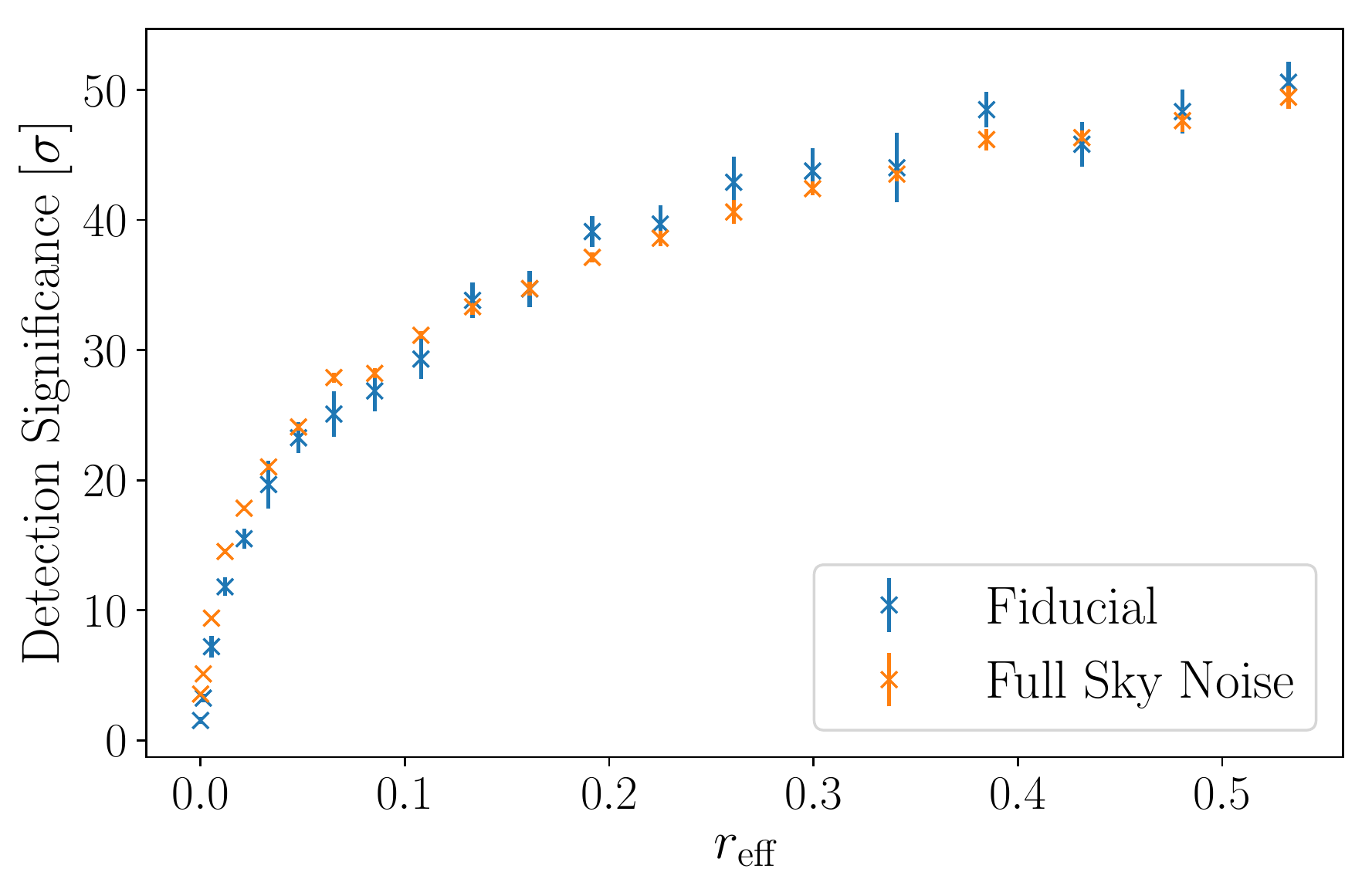} 
  \caption{Null test plot, as Fig.~\ref{Fig:NullPlot}, for the CMB-S4 experiment, with error-bars obtained from averaging over 10 computations of each data-point as before. Hexadecapole detection significances are given for two techniques used to simulate noise; (a) creating single-tile Gaussian random field realisations of the 1D noise power spectrum (Eq.\,\ref{Eq:noiseModel}) (blue, as used throughout this paper), and (b) using cut-out sections of a full-sky noise map generated from the same noise spectrum (orange). Significances are plotted as a function of the effective tensor-to-scalar ratio, $r_\mathrm{eff}$, and we observe no notable differences between the two datasets.}\label{Fig:NoiseBiases}
\end{figure}

Fig.~\ref{Fig:NoiseBiases} displays the analogue of Fig.~\ref{Fig:NullPlot} for CMB-S4 noise and lensing parameters, displaying the mean significance and $r_\mathrm{eff}$ for analyses using both the fiducial (single-tile) noise generating technique (blue) and the full-sky noise maps (orange). Notably, the $r_\mathrm{eff}-\mathcal{S}$ curve is very similar for both datasets across the range of dust amplitudes tested, thus we conclude that there is no significant bias obtained by our generating noise purely on single-tile scales. This motivates the use of the fiducial method in our analysis, which is considerably less computationally expensive, especially for tests where the noise parameters are varied (e.g.~Sec.~\ref{Subsec:NoiseParamsXi}). 

% This was achieved using a full-sky map of the lensed CMB scalar spectrum from the FFP10 simulations,\footnote{Available from NERSC: \url{https://nim.nersc.gov/}} (scaled by $\sqrt{f_\mathrm{lens}}$) which was cut out into tiles as in section \ref{Subsec:Cutouts} utilising the same technique of zero-padding and the \texttt{flipperPol} `hybrid' algorithm to determine a B-mode Fourier space map for each tile separately. Comparison of the 1D power spectra of these tiles with the previously used \texttt{CAMB} $C_l^\mathrm{lens}$ spectra show a slight ($\sim20\%$) increase in power due to different cosmological parameters in use, thus the measured Fourier maps were rescaled via a 1D transformation for consistency with our previous analysis.

% The generated $B_\mathrm{lens}(\vec{l})$ map is then combined with the simulated dust Fourier map and a Gaussian realisation of the noise spectra to produce the `data' power maps. The remaining analysis proceeds as before, with the MC simulations being generated from the isotropic 1D power spectrum. Our MC variances do not take into account lensing anisotropies, but if a significant hexadecapole signature was detected in the lensed maps, we would expect the significance, $\mathcal{S}$ of a detection of patch-hexadecapole, $\Xi$, to be significantly boosted, with a plateau in $\mathcal{S}$ as $f_\mathrm{dust}\rightarrow0$.

% Figure \ref{Fig:LensingBiases} displays the analogue of figure \ref{Fig:NullPlot} for SO and CMB-S4 noise parameters, displaying the mean significance and $r_\mathrm{eff}$ for analyses using both the isotropic lensing (red) and full FFP10 lensing (blue). At low $r_\mathrm{eff}$, $\mathcal{S}$ is slightly boosted by the inclusion of anisotropic lensing modes, but there is no net effect at high dust levels (as expected), and no clear plateau in $\mathcal{S}$ as $r_\mathrm{eff}\rightarrow0$. 

% The bias of anisotropic lensing modes is hence small, and the non-zero intercept ($\sim1.5\sigma$) caused by the extra hexadecapole power could be ameliorated by the use of MC simulations containing only noise and anisotropic lensing modes. A full analysis would include anisotropic lensing in the MC simulations used to compute the variances on the hexadecapole parameter $H^2$, but from figure \ref{Fig:LensingBiases}, we observe that the overall effect is small, thus may be safely neglected, avoiding a large increase in computation time.

\section{Distortions due to Map Projection}\label{Appen:ProjectionDistortions}
During the conversion of full-sky maps into small flat-sky tiles, we must project sections of a full \texttt{HEALPix} map onto a flat grid, which can give significant distortions. These are seen most clearly at high galactic latitudes, since the polar co-ordinates become singular there, leading to unwanted stretching effects. This can affect both the real- and power-space data, and may have a significant impact on the derived hexadecapole quantities. The effect could be negated by altering the projection axis for each cut-out such that the desired tile always lies in equatorial regions, where it would experience the least distortion, but this was not implemented to avoid significant additional computational expense.

To explore this effect, we simulated a full-sky map with known $C_l^{EE}$ and $C_l^{BB}$ spectra, using \texttt{HEALPix}'s \texttt{synfast} method. The spectra were derived from a dust-plus-contamination spectrum with dust amplitude appropriate for a tile in the BICEP region, and $C_l^{EE}$ was set to $2C_l^{BB}$, following \cite{planck2014-XXX,planck2016-XLVIII}.

\begin{figure}
\subfigure[Monopole amplitude $\log_{10}(A\,{[\mathrm{K}^2]})$]{\includegraphics[width=\linewidth]{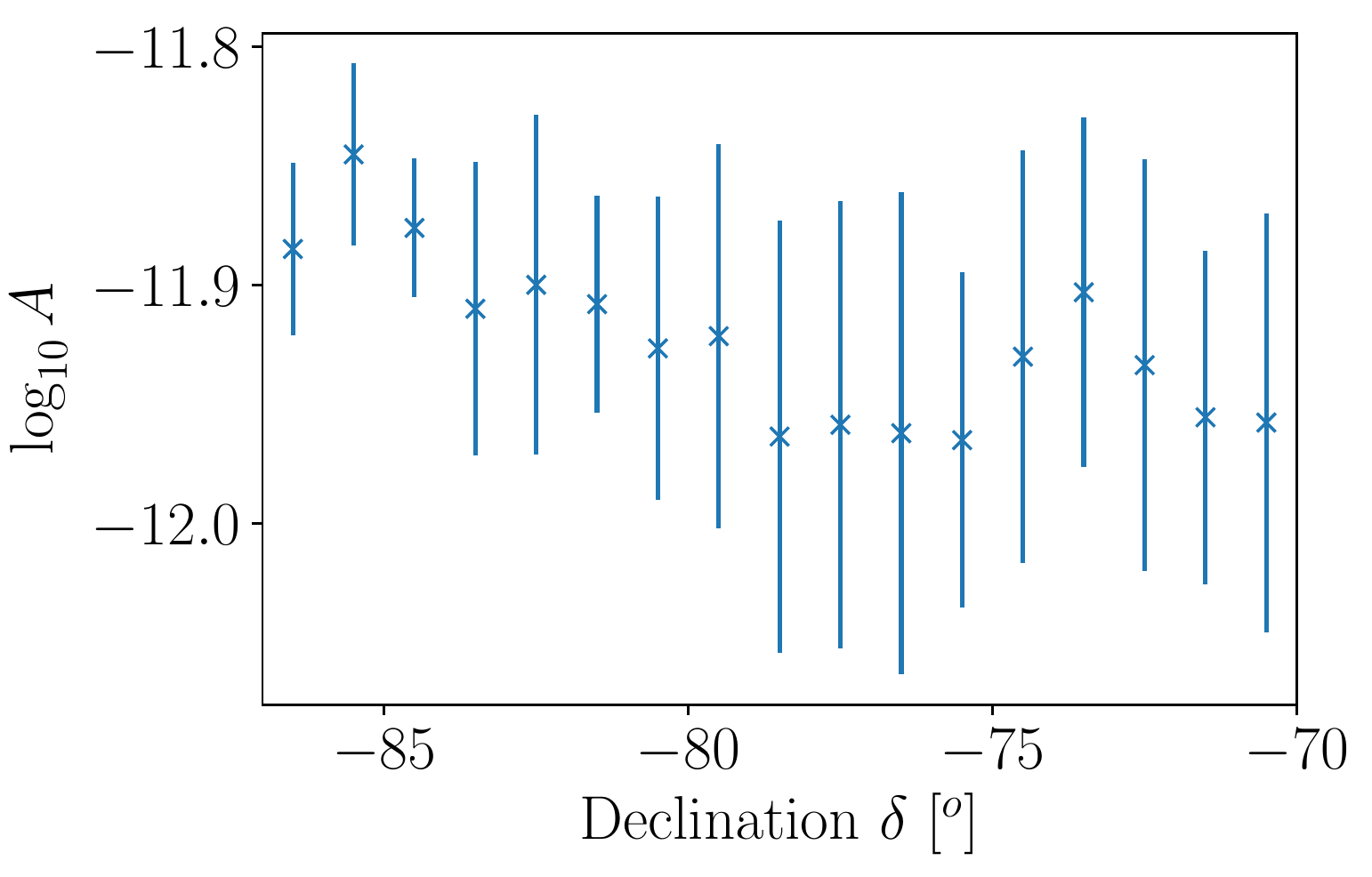}}
\subfigure[Hexadecapole Power $\log_{10}(H^2\,{[\mathrm{K}^4]})$]{\includegraphics[width=\linewidth]{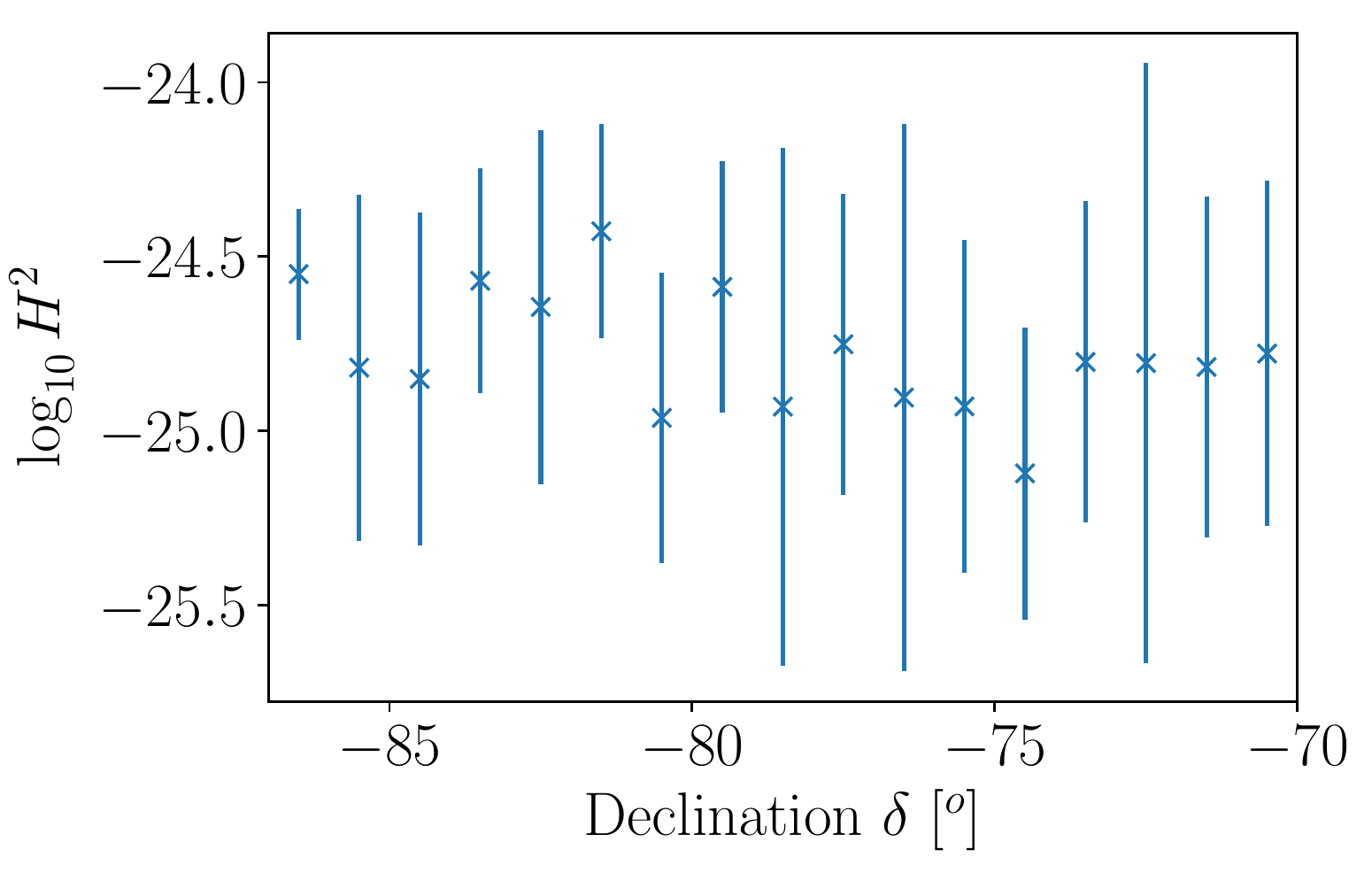}}
\caption{Monopole and hexadecapole powers reproduced from $3\degree$ tiles generated from cutting out a full-sky map of known power spectra, satisfying $C_l^{BB}/C_l^{EE} = 0.5$. Values are given as a function of the declination of the map-centre, $\delta$, and error-bars are from the standard deviations of parameter values across different tiles at that latitude. Any deviation from uniformity with respect to $\delta$ indicates unwanted effects from projection-derived distortions.}\label{Fig:EngelenDistortions}
\end{figure}

Focussing on the Southern Galactic pole (over an RA range of $100\degree$), this was cut out into $3\degree$ width tiles with separation 1$\degree$, and the power spectrum was computed for each tile. In particular, we considered the observables $A$ and $H^2$ as a function of the declination, $\delta$, of the tile-centre. Fig.~\ref{Fig:EngelenDistortions} shows the computed values for $\delta<-70\degree$, which should be constant with respect to $\delta$ in the absence of any distortions. Error-bars show the variation in the parameters for different tiles at this declination. We conclude that there is no significant biases to $\mathcal{H}^2$ from any latitude, but a possible slight enhancement in $A$ at $\delta<-85\degree$.

An additional test is to ensure that there is no significant E- to B-mode leakage deriving from the projection onto small tiles at high Galactic latitudes. This is investigated in a similar way, instead using a simulated full-sky map with only T- and E-modes present (generated in \texttt{HEALPix} from a \texttt{CAMB} power spectrum of the unlensed scalar CMB). For each $3\degree$ tile in the same region as before, the ratio of B- to E-mode power was computed, denoted $\mathrm{R}_\mathrm{BE} = \mean{C_l^{BB}}/{\mean{C_l^{EE}}}$.

\begin{figure}
\centering
\includegraphics[width=\linewidth]{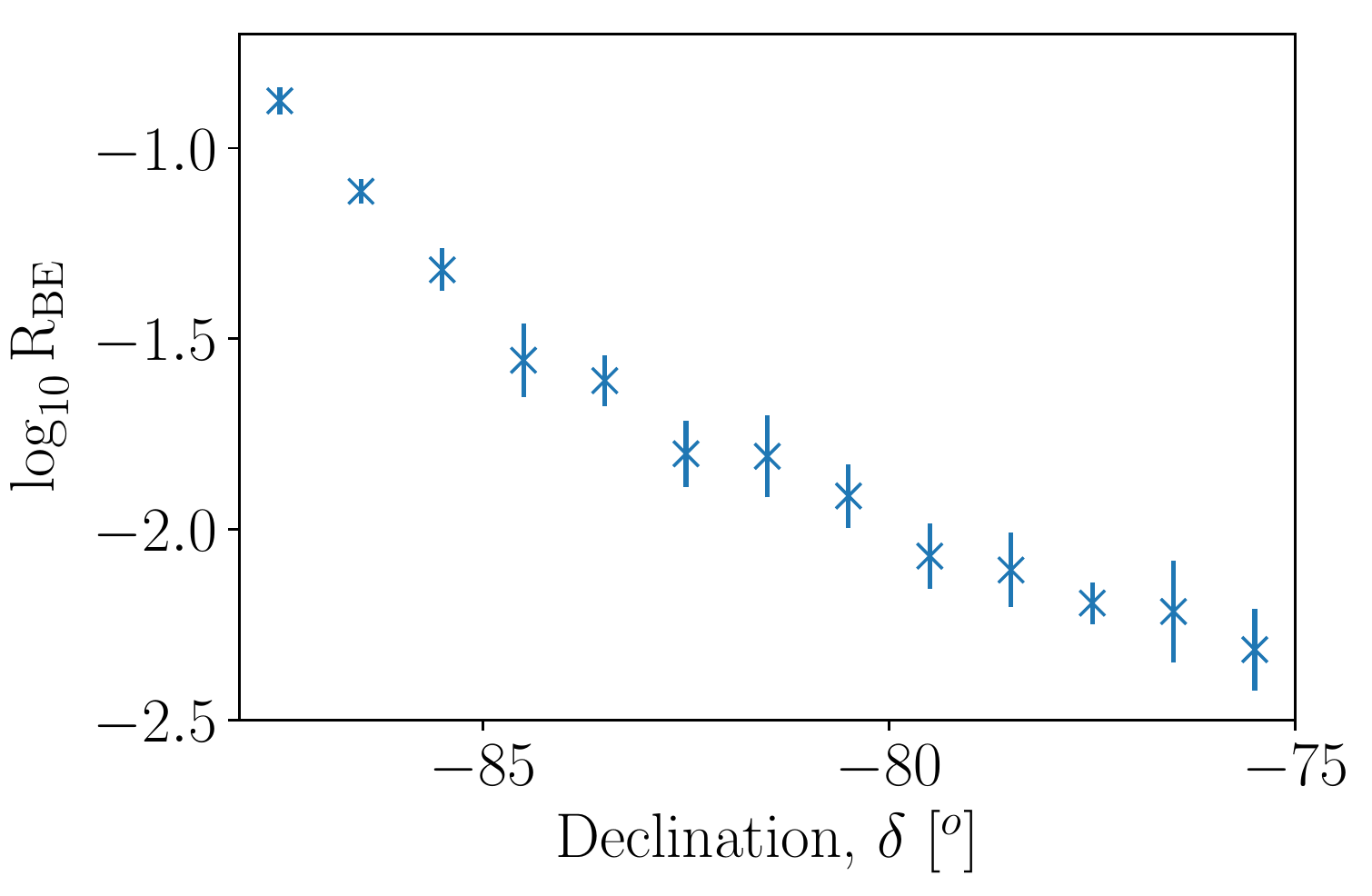}
\caption{Ratio of projected mean B- to E-mode power, $\mathrm{R_{BE}}$, for 350 $3\degree$ width tiles close to the South Galactic pole with centres at the stated declinations. This uses a simulated full-sky CMB map of unlensed scalar modes, created using \texttt{CAMB} and \texttt{HEALPix}, thus we expect no B-mode component in an ideal projection.}\label{Fig:EngelenLeakage}
\end{figure}

From Fig.~\ref{Fig:EngelenLeakage}, it is clear that there is non-zero E- to B-mode leakage at all tested latitudes, increasing as $\delta\rightarrow-90\degree$. However, $\mathrm{R_{BE}}<10^{-1.5}\approx3\%$ for $\delta<-85\degree$, and, since we expect comparable powers in both modes for dust, this is not an important source of error in this paper. Combining both tests, we see that there are significant biases only for $|\delta|>85\degree$, thus these regions have been excluded from all analyses.

\section{Reduction of Bias from Pixellation}\label{Appen:Pixellation}
In zero-padding, we insert extra zeros to the sides of all real-space cut-out tiles to increase the map width by a factor $R_\mathrm{pad}$. This effectively boosts the resolution in Fourier space by interpolating between pixels (via convolution with a 2D sinc function). The number of pixels in the data is thus increased which reduces pixellation errors and allows smaller $l$ to be probed, increasing the significance of any detection of anisotropy. It is important to note that following zero-padding the Fourier-space pixels are no longer independent, thus when Gaussian realisations of known power spectra are created, they are generated in Fourier space using an unpadded template before zero-padding is added in real-space, to ensure that they have the same correlation properties as the data. In addition, we note that zero-padding cannot be applied in the high-noise limit (e.g.~for noise parameters appropriate for BICEP2) since the Fourier maps are dominated by high-amplitude pixels at large $l$, which give significant leakage into other (non-local) pixels due to the sinc interpolation.

We now consider the effects of rotating the power-map prior to applying the hexadecapole estimators. Na\"ively, if the map is rotated by angle $\chi$ (i.e.~transforming $\phi_{\vec{l}}\rightarrow\phi_{\vec{l}}+\chi$), we expect the output hexadecapole parameters ($Af_s'$ and $Af_c'$) to be related to their unrotated forms ($Af_s$ and $Af_c$) via
\begin{eqnarray}\label{Eq:AfsRotations}
Af_s &=& Af_s'\cos(4\chi) - Af_c'\sin(4\chi)\nonumber\\
Af_c &=& Af_s'\sin(4\chi) + Af_c'\cos(4\chi).
\end{eqnarray}
(The $4\chi$ dependence comes from the hexadecapolar nature of the estimators.) However, this also has the effect of changing the orientation of the Fourier-space data with respect to the pixel shape, resulting in an \textit{additional} oscillation in $Af_s$ and $Af_c$ with respect to $\chi$. Here, we average the corrected (derotated) estimates over 20 linearly-spaced values of $\chi\in[0,\pi/2)$, which significantly diminishes the pixellation effects. 

\section{Determining the angle sign via cross-correlation}
\label{App:dedustAngle}
We now discuss how we can determine the sign of the quantities $\sin{2\alpha}$ and $\cos{2\alpha}$ used in the dedusting procedure in Sec.~\ref{Sec:Dedusting}. Firstly, note that in the flat-sky approximation for each tile (appropriate for single-tile scales), we may use the canonical relation 
\begin{eqnarray}\label{Eq:Bhat}
B(\vec{l})=\frac{1}{\sqrt{2}}\left[U(\vec{l})\cos{2\phi_{\vec{l}}}-Q(\vec{l})\sin{2\phi_{\vec{l}}}\right]
\end{eqnarray}
\citep[e.g.][Eq.\,30]{2016ARA&A..54..227K} to obtain an estimate of the single-tile B-mode Fourier dust spectrum, $\widehat{B}$, as 
\begin{eqnarray}\label{Eq:DedustingBModel}
\left.\eta\widehat{B}(\vec{l})\right|_\mathrm{tile}=\frac{\pm{}\eta{}I(\vec{l})R_\mathrm{tile}}{2\sqrt{\overline{f_c}+1}}\left[\overline{f_s}\cos{2\phi_{\vec{l}}}-(\overline{f_c}+1)\sin{2\phi_{\vec{l}}}\right],
\end{eqnarray}
(using Eqs.~\ref{Eq:QUdedust}\,\&\,\ref{Eq:sin2alphaDedust}) where $R_\mathrm{tile}$ is the scaling ratio (Eq.\,\ref{Eq:DedustingScaling}, constant for each tile) and $\eta>0$ is some unknown factor. $\phi_{\vec{l}}$ is the angle of $\vec{l}$ relative to the positive RA axis as before. We may compute the cross-spectrum between this estimate $\widehat{B}(\vec{l})$ and the (known) true B-mode map $B(\vec{l})$ ($C_l^{B\widehat{B}}$) which is expected to be positive if we have chosen the correct sign in Eq.~\ref{Eq:DedustingBModel}. Measuring the sign of $C_l^{B\widehat{B}}$ (by calculating the average value of the cross-spectrum for $l\in{[100,500]}$) and applying this same sign to both $\sin{2\alpha}$ and $\cos{2\alpha}$ expressions, we resolve our spatially-dependent sign ambiguity and obtain complete knowledge about the angle $2\alpha$. 
%%%%%%%%%%%%%%%%%%%%%%%%%%%%%%%%%%%%%%%%%%%%%%%%%%

% Don't change these lines
\bsp	% typesetting comment
\label{lastpage}
\end{document}